\RequirePackage{snapshot}
\documentclass[a4paper,12pt,oneside]{article}
\usepackage[warn]{mathtext}
\usepackage{amsmath}
\usepackage{fancyref}

\usepackage[figuresright]{rotating}
\usepackage{amsfonts,amssymb,amsmath}
\usepackage{boxedminipage}
\usepackage{makeidx}
\usepackage{url}

\usepackage[T2A]{fontenc}
\usepackage{latexsym}

\usepackage[english]{babel}
\usepackage{graphics}
\usepackage{epsfig}
\usepackage[table]{xcolor}
\usepackage{color,soul}
\usepackage{todonotes}
\usepackage{cite}
\graphicspath{{img/}} 

\title{High Energy Nuclear Optics of polarized nucleons and nuclei: research at complex Nuclotron M- NICA}

\author{Vladimir Baryshevsky}

\begin{document}


\maketitle

\begin{abstract}
\noindent
Refracton of particles (nucleons, nuclei, $\gamma$-quanta) in matter with polarized protons (nuclei) results in revealing coherent quasi-optical phenomenon of nuclear spin precession of particles (nuclei) in the pseudomagnetic field of matter with polarized spins and the phenomenon of birefringence of particles (nuclei) with spin $S \ge 1$.
These phenomena can be observed and studied at complex NuclotronM-NICA. 
The similar effects for $\gamma$-quanta could be observed at LINAC accelerator.
Quasi-optical coherent phenomena of spin rotation and dichroism
are not caused by strong interactions only, 
the T-odd P-odd, T-odd P-even, T-even P-odd interactions also contribute.
Limits for the value of these contributions at energies available at complex NuclotronM-NICA can be obtained by investigating all these phenomena.
When studying polarized particles collisions, it is necessary to consider possible influences of quasi-optical phenomena of spin rotation and spin dichroism caused by nuclear precession and birefringence.
\end{abstract}

\section{Introduction}
\label{sec:light}

The phenomena of interference, diffraction and refraction of light
are well known even to lycee and college students. A great variety
of their applications is described in school and university
manuals and popular science books
\cite{vfel_Born,B2,B3,B4,rins_78}. Centuries--long argument about
the nature of light:  whether light is a wave or a particle,
finally led to the creation of quantum mechanics and extension of
the wave conception to the behavior of any particles of matter. As
a result, optical concepts and notions were also introduced for
describing interaction of particles with matter, nuclei, and one
another \cite{Landau_3,rins_1,th_11}. In particular, widely used nowadays
is diffraction of electrons and neutrons  by crystals, which are
in fact natural diffraction gratings. Neutron interferometers were
designed \cite{rins_79}. It was found out that  scattering of
particles by nuclei (and by one another) is in many cases similar
to scattering of light by a drop of water (the optical model of
the nucleus).

A study of interaction between light and matter has shown that besides frequency and propagation direction, light waves are characterized by polarization.

The first experiment, which observed a phenomenon caused by the polarization of light,
was carried out in 1669 by E.~Bartholin,  who discovered the double refraction of a light ray
by Iceland spar (calcite). Today it is common knowledge that in the birefringence effect, the
stationary states of light in a medium are the states with linear polarization parallel
or perpendicular to the optic axis of a crystal. These states have different refractive indices
and move at different velocities in a crystal. As a result, for example, circularly
polarized light in crystals turns into linearly polarized and vice versa \cite{vfel_Born}.

\begin{figure}[h]
	\epsfxsize = 12 cm \centerline{\epsfbox{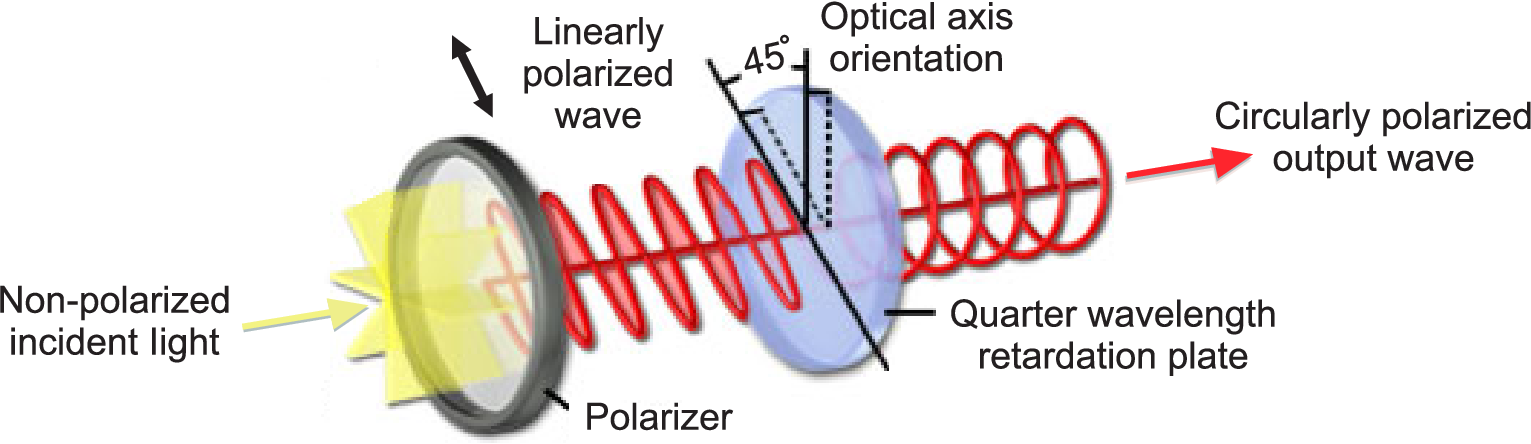}}
	\caption{Light optics: birefringence effect}
	\label{fig:light3}
\end{figure}

Another series of experiments was performed by D.F.~Arago in 1811 and J.B.~Biot in 1812. They
discovered the phenomenon of optical activity, in which the light polarization plane rotates  as
the light passes through a medium. In 1817, A. Fresnel established that in an optically active medium
rotating the polarization plane, the stationary states are the waves with right-hand and left-hand
circular polarizations, which, as he found out in 1823, move in a medium at different velocities
(i.e., propagate with different refractive indices), thus causing the polarization plane to rotate. Let us also
recall the effect of light polarization plane rotation in matter placed in a magnetic field, which
was discovered by Faraday, and the birefringence effect in matter placed in an electric field
(the Kerr effect).

\begin{figure}[h]
	\epsfxsize = 12 cm \centerline{\epsfbox{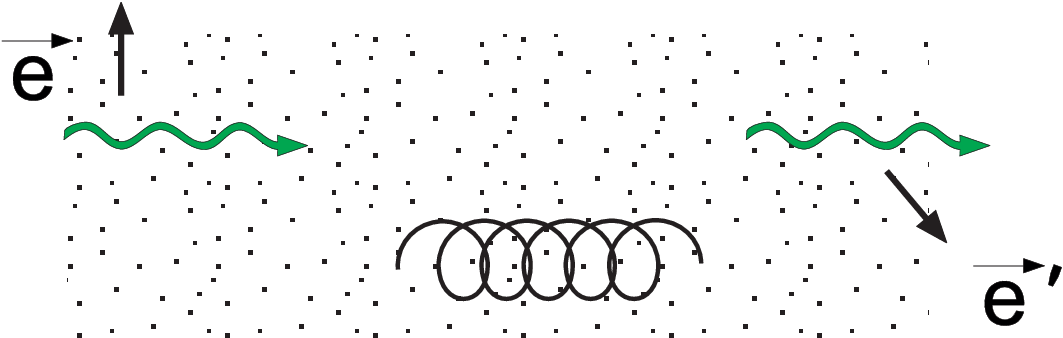}}
	\caption{Light optics: natural optical activity}
	\label{fig:light4}
\end{figure}

The above-mentioned phenomena and various other effects caused by
the presence of polarization of light and optical anisotropy of
matter have become the subjects of intensive studies and found wide applications.
The microscopic mechanism leading to the appearance of
optical anisotropy of matter is, in the final analysis, due to the
dependence of the process of electromagnetic wave scattering
by an atom (or molecule) on the wave polarization (i.e., on the photon spin)
and to bounds imposed on electrons in atoms and molecules. Beyond the optical
spectrum, when the photon frequency appears to be much greater
than the characteristic atomic frequencies, such bounds become
negligible, and the electrons can be treated as free electrons. As a
result, the effects caused by optical
anisotropy of matter, which are studied in optics, rapidly diminish, becoming practically
unobservable when the wavelengths are smaller than $10^{-8}$ cm.

Moreover, there is a widespread belief that it is only possible to speak of the refraction of
light and to use the concept of the refraction index of light
in matter because the wavelength of light
($\lambda \approx 10^{-4}$ cm) is much greater than the distance
between the atoms of matter $R_a$ ($R_a\approx 10^{-8}$ cm) since
only in this case ($\lambda \gg R_a$) matter may be treated as a
certain continuous medium. As a consequence, in a short-wave
spectral range where the photon wavelength is much smaller than
the distance between the atoms of matter, the effects similar to
the Faraday effects and birefringence, which are due to refraction,
should not occur. However, such a conclusion  turned out to be incorrect. The
existence of the refraction phenomena does not appear to be
associated with the relation between the wavelength $\lambda$ and the
distance between atoms (between scatterers). Even at high photon
energies when the wavelength is much smaller than $R_a$, the
effects due to  refraction of waves in matter can be quite appreciable.
Thus, for example, when a beam of linearly polarized
$\gamma$-quanta with the energies greater than tens of kiloelectronvolts
(wavelengths smaller than $10^{-9}$ cm) passes through matter with
polarized electrons, there appears rotation of the polarization plane of
$\gamma$-quanta, which is  kinematically analogous to the Faraday effect
(\cite{51,52,54,55,56}).  Moreover, with the growth of the energy of $\gamma$-quantum
(the decrease in the $\gamma$-quantum wavelength) the effect increases, attaining its maximum
in the megaelectronvolt energy range.
Unlike the Faraday effect, which is due to
the bounds imposed on electrons in atoms, for
$\gamma$-quanta electrons  may be treated as free electrons. The effect of
polarization plane rotation in this case is due to the
quantum--electrodynamic radiative corrections to the process of
scattering of $\gamma$-quanta by an electron, which are lacking in classical
electrodynamics.

Analogously, the propagation in matter of the de Broglie
waves, which describe  motion of massive particles, can be
characterized by the refractive index \cite{Goldberger,rins_1}. In this case, the
refractive index also characterizes particle motion in matter,
even at high energies, for which the de Broglie
wavelength $\hbar/m v$ ($m$ is the particle mass, $v$ is its
velocity, in the case of relativistic velocities $m$ stands for
the relativistic mass $m\gamma$, $\gamma$ is the Lorentz
factor) is small in comparison with the distance between the atoms
(scatterers). Furthermore, it turns out that for particles with nonzero spin,
there exist the phenomena analogous to light polarization plane
rotation and birefringence.  In this case such phenomena of quasi-optical
activity of matter ("optical" anisotropy of matter) are
due not only to electromagnetic but also to strong and weak
interactions.

The investigations in this field were initiated  by
V. Baryshevsky and M.Podgoretsky  \cite{24}, who predicted the existence of the phenomenon of quasi-optical spin
rotation of the neutron moving in matter with polarized nuclei, which is caused by strong interactions,
and introduced the concept of a nuclear pseudomagnetic field (neutron spin precession
in a pseudomagnetic field of matter with polarized nuclei). The concept of a nuclear
pseudomagnetic field and the phenomenon of neutron spin precession in matter with
polarized nuclei were experimentally verified by Abragam's group in France (1972)
(see \cite{Abragam_PRL_1973} and references therein) and Forte in Italy (1973) \cite{Forte_1973}.

Here we would also mention the paper by F. Curtis Michel
\cite{87}, who predicted the existence of spin "optical rotation"
due to parity nonconserving weak interactions (the phenomenon was
experimentally revealed \cite{92}
and is used for studying parity nonconserving weak
interactions between neutrons and nuclei).

Further analysis showed that the effects associated with the optical
activity of matter, which we consider in optics, are, in fact, the
particular case of coherent phenomena emerging when polarized
particles pass through matter with nonpolarized and polarized
electrons and nuclei \cite{164,201,Nuclear_optics}.
It was found out, in particular, that at high energies of particles
(tens, hundreds and thousands of gigaelectronvolts), the effects of
"optical anisotropy" are quite significant and they may become
the basis of unique methods for the investigation of the structure
of elementary particles and their interactions.


\section
{Gamma optics}
\label{sec:gamma}

{
	Preceding consideration of different quasi-optical phenomena in the high-energy range I'd like to mention that so far, the following statements,  similar to those recently expressed by a reviewer of a distinguished journal still appear:
	\textit{''... and the whole classical concept of a refractive index, which underpins this calculation, requires a beam of light coherent over some spatial extent much larger than the distance between the scatterers so that the (forward) scattered radiation interferes constructively to produce a phase-shifted beam - it is not a obvious that the gamma-ray beam at, say, MAMI satisfies this ...''}
	The above misinterpretation compels me to remind that
	%
	numerous  studies
	made it possible to establish a close relation between the elastic
	coherent zero-angle scattering amplitude $f(0)$ and the refractive
	index $n$ (see, for example, \cite{Goldberger,22,rins_1,rins_3,rins_80,rins_81})
	and to develop experimental methods for investigating refraction of polarized particles in matter.
	It turned out that in the case when
	matter is composed of randomly located scatterers and the
	condition $|n-1|\leq 1$ is fulfilled, the refractive index has the form
	\begin{equation}
		\label{I.1} n=1+\frac{2\pi\rho}{k^{2}}f(0)\,,
	\end{equation}
	where $\rho$ is the density of scatterers (the number of
	scatterers per cubic centimeter of matter); $k$ is the wave number
	of the incident wave.}

It was also found out that the possibility to introduce the
refractive index is not associated with the ratio of the radiation wavelength to the
distance between scatterers. 
(\ref{I.1}) 
also  describes
the refraction of short--wave radiation with a wavelength much shorter
than the distance between the scatterers. This can be explained
by the fact that the refractive index appears due to the interference
between an incident wave and secondary rescattered waves, which
always occurs in elastic coherent forward scattering. Moreover,
{(\ref{I.1})} describes not only scattering of photons but also
scattering of particles of different nature (neutrons, electrons
and others).

It is well known that an optically anisotropic medium is
characterized by the presence of several refractive indices. For
instance, in the case of the Faraday effect, the refractive indices
$n_{+}$ and $n_{-}$ are different for light with right-hand and left-hand circular polarizations.
In view of (\ref{I.1}), from this follows that the amplitude $f_{+}$ of
elastic coherent forward scattering of a right-hand polarized
photon  differs from a similar amplitude $f_{-}$ for a left-hand polarized
photon.

Thus, for the effect of light polarization plane rotation in a medium to occur, the elastic coherent zero--angle scattering
amplitude should be dependent on the photon polarization state, or, which is the same,
on its spin state.

As spin dependence of the scattering process is typical of
particle collisions for all particles with nonzero spin, it should be supposed that the quasioptical 
phenomena analogous to the Faraday effect (birefringence) will
occur for all such particles and various interactions.


It is well known that the phenomena caused by the optical
anisotropy of matter (the Faraday effect, birefringence, natural
rotation of the light polarization plane) are eventually due to
the influence that the forces acting on electrons in atoms have on
the interaction of electromagnetic waves with matter.

Beyond the optical spectrum when the frequency of electromagnetic
waves becomes much greater than the average energy of electrons in
atoms and molecules, the interaction between radiation and matter
is reduced to the interaction of a photon with free electrons. As
a result, the structure of atoms and molecules becomes
non-essential, and hence the phenomena caused by optical
anisotropy of matter should disappear. For example, in the case of
the Faraday effect, a simple theory based on the normal Zeeman
effect gives the following expression for the light polarization
plane rotation angle $\vartheta$ per 1 cm path length \cite{4}:
\begin{equation}
	\label{12.1} \vartheta=\frac{\omega}{c}\frac{\partial
		n(\omega)}{\partial\omega}\frac{eB}{2mc}\,,
\end{equation}
where $n(\omega)$ - is the refractive index of matter in the
absence of a magnetic field; $e$ is the electron charge; $m$ is
its mass.

In a high--frequency range the expression for $n(\omega)$ can be written as \cite{4}
\[
n=1-\frac{2\pi e^{2}N}{m\omega^{2}}\,,
\]
where $N$ is the number of electrons per 1 cm$^3$ of matter, $m$ is the electron mass.
Thus we have for the rotation angle $\vartheta$ of a
high--energy quantum:
\begin{equation}
	\label{12.2} \vartheta=\frac{2\pi e^{3}N}{m^{2}c^{2}\omega^{2}}\,.
\end{equation}
From (\ref{12.2}) we obtain that in the range of $\gamma$-quantum
energies of 100\,keV, the angle $\vartheta\approx 10^{-7}$\,rad/cm
for $B=10^{5}$ Gs=10 T, $ N\approx 10^{23}$ and decreases rapidly
with increasing quantum frequency.
Similarly, all other magneto-optical effects—for instance, the inverse Faraday effect \cite{reverse_Faradey}—as well as electro-optical effects studied in optics, decay rapidly and become negligibly small for gamma-quanta. 

However, in 1965, it was shown by  V.~Baryshevsky and V.~Lyuboshitz
\cite{51} that just at high energies of $\gamma$-quanta, there
becomes possible another mechanism of  photon  polarization plane
rotation in a target with polarized electrons, which is  due to
spin dependence of the amplitude of the elastic Compton forward
scattering of a $\gamma$-quantum by an electron. The calculations
showed that  the rotation angle reaches its maximum value of
$5.32\cdot10^{-3}$ rad/cm in the  range of quantum  energies from
500 to 700~keV \cite{52}. This new phenomenon was experimentally
observed by V.M. Lobashev, L.A.~Popeko, L.M.~Smotritskii, A.P.~Serebrov, E.A.~Kolomenskii \cite{54,55}, by P.~Bock and P.~Luksch
\cite{56} and was registered as a scientific discovery in the USSR State registry under the number 360.


\begin{figure}[h]
	\epsfxsize = 10 cm \centerline{\epsfbox{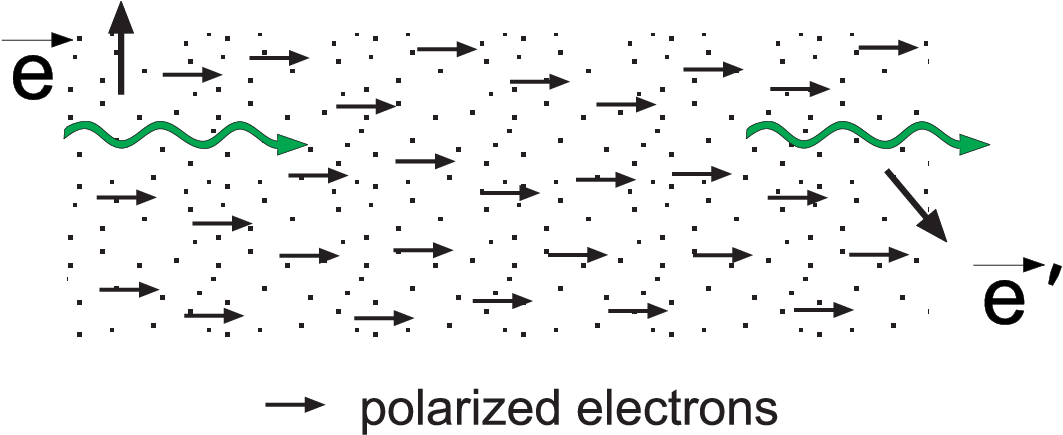}}
	\caption{Gamma-quanta polarization plane rotation}
	\label{fig:gamma1}
\end{figure}

Let us now consider processes caused by $\gamma$-quanta interaction with protons and nuclei of matter.
%
%
In accordance with the pioneering results obtained by M.~Gell-Mann, M.L.~Goldberger and W.E.~Thirring \cite{57} the amplitude of forward elastic scattering  of a photon by a particle with spin $s=1/2$ can be written as follows:

\begin{equation}
	f_{\mu \nu}(\omega)=f_{1}(\omega)\left( \vec{e}^*_{\mu} \vec{e}_{\nu}  \right) + i
	f_{2}(\omega) \vec{\sigma}\left[ \vec{e}^*_{\mu} \vec{e}_{\nu}  \right],
	\label{eq:1}  
\end{equation}
where $\vec{e}_{\mu}$ and $\vec{e}_{\mu}$ are the photon polarization vectors in states $\mu$ and $\nu$, $\vec{\sigma}$ is the Pauli spin matrix.

Another notation is often used now when considering $\gamma$-quanta interaction with protons (nuclei), namely: $f_{1}(\omega)=f_{0}(\omega)$
and $f_{2}(\omega)=g_{0}(\omega)$. 
Therefore, for further consideration and analysis of (\ref{eq:1}) we will use it in the following form (see, for example \cite{VG_3,VG_4,VG_5,VG_6,VG_7})
\begin{equation}
	f_{\mu \nu}(\omega)=f_{0}(\omega)\left( \vec{e}^*_{\mu} \vec{e}_{\nu}  \right) + i
	g_{0}(\omega) \vec{\sigma}\left[ \vec{e}^*_{\mu} \vec{e}_{\nu}  \right].
	\label{eq:2}  
\end{equation}

According to analysis \cite{VG_3,VG_4,VG_5,VG_6,VG_7} equations (\ref{eq:1}) and (\ref{eq:2}) can be used to define electromagnetic polarizabilities and spin polarizabilities as the lowest order coefficients in an $\omega$-dependent development of the scattering amplitudes.
It was shown that the real part of amplitude $g_0 (\omega)$, which determines the spin-dependent part of amplitude $f_{\mu \nu}(\omega)$ at small frequencies can be expressed as:
\begin{equation}
	g_0(\omega)= -\frac{e^2 \varkappa^2}{8\pi m^2} \omega +\gamma_0 \omega^3\, ,
	\label{eq:3}  
\end{equation}
where $\gamma_0$ is the spin polarizability, in the selected system of units electric charge $e$ meets the relation $\frac{e^2}{4 \pi}=\frac{1}{137.04}$, $\varkappa$ is the anomalous magnetic moment expressed in nuclear magneton. For a proton $\varkappa$=1.79, $\gamma_0 \sim -10^{-4}$ fm$^4$.

When frequency grows, expression (\ref{eq:3}) for $g_0(\omega)$, which includes constant $\gamma_0$ becomes invalid. 
Therefore, investigation of  $g_0(\omega)$ dependence on frequency $\omega$ for high-energy $\gamma$-quanta (energy range from few to dozens GeV or even higher) 
is of interest.

\subsection{Quasi-optical phenomenon of $\gamma$-quanta polarization plane rotation in matter with polarized proton (nuclei)}
\label{sec:rotation}

Let us consider the passage of a beam of  $\gamma$-quanta through
a medium with polarized protons (nuclei, electrons). 
%
%
%

Pioneering publications considering application of $\gamma$-quanta for investigation of the internal structure of a nucleon appeared in 50-th of the XXth century \cite{VG_1,57}.
%
Those papers initiated numerous theoretical and experimental studies of this interesting possibility
\cite{VG_8,VG_9,VG_9b,VG_3,
VG_4,VG_5,VG_6,VG_7,VG_9c,VG_9d,VG_9e,VG_9f,VG_9i,VG_9_1,
VG_n_10,VG_n_11,VG_n_12,VG_n_13,VG_n_14,VG_n_15,VG_n_16,VG_n_17,VG_n_18,VG_n_19,
VG_n_21,VG_n_22,51,A1}.   
The first experiment on Compton scattering by the proton to measure the polarizabilities was carried out in 1958 (see \cite{VG_4,VG_5}).

Fast advancing of experimental methods  makes both single and double polarization experiments possible with a polarized solid target and polarized high-energy $\gamma$-quanta 
\cite{VG_8,VG_9,VG_9b,VG_3,
VG_4,VG_5,VG_6,VG_7,VG_9c,VG_9d,VG_9e,VG_9f,VG_9i,VG_9_1,
VG_n_10,VG_n_11,VG_n_12,VG_n_13,VG_n_14,VG_n_15,VG_n_16,VG_n_17,VG_n_18,VG_n_19,
VG_n_21,VG_n_22}. 
Polarized targets with high degree of nuclei polarization were created, namely: $NH_3$, $ND_3$ and $^6LiD$, as well as a butanol polarized proton (deuteron) target \cite{VG_n_13,VG_n_14,VG_n_15,VG_n_16,VG_n_17,VG_n_21,VG_n_22}.

Let us intently consider the passage of a $\gamma$-quanta beam through
a medium with polarized electrons (nuclei) \cite{arxive.2411.04960}. 
%
%
If the photon in right-- and
left--circularly polarized states has different  refractive
indices $n_{1}$ and $n_{2}$, then \cite{51,A2}:
\begin{equation}
\label{12.3} \Delta n=n_2-n_1=\frac{2\pi
	Nc^{2}}{\omega^{2}}[f_{-}(0)-f_{+}(0)]\,,
	\end{equation}
	where $N$ is the number of electrons (nuclei) per unit volume; $f_{+}(0)$
	and $f_{-}(0)$ are the spin--non-flip amplitudes of elastic
	zero--angle scattering of right-- and left-- circularly polarized
	photons by polarized electrons (nuclei), respectively;  $\omega$ is the
	photon frequency.
	
	The scattering amplitude for the Compton forward scattering by a particle with
	spin $1/2$  can be written as follows \cite{57}:
	\begin{equation}
\label{12.4}
f_{\mu\nu}=f_{1}(\omega)(\vec{e}\,^{*}_{\mu}\vec{e}_{\nu})+i
f_{2}(\omega)\vec{\sigma}[\vec{e}\,^{*}_{\mu}\vec{e}_{\nu}]\,,
\end{equation}
where $\vec{e}_{\mu}$ and $\vec{e}_{\nu}$ are the photon
polarization vectors in states $\mu$ and $\nu$; $\vec{\sigma}/2$
is the particle spin operator. For the state with the right--hand
circular polarization
\begin{equation}
\label{12.5} \vec{e}_{+}=-(\vec{e}_{1}+i \, \vec{e}_{2})/\sqrt{2}\,,
\end{equation}
while for the state with the left--hand circular polarization
\begin{equation}
\label{12.6} \vec{e}_{-}=(\vec{e}_{1}-i \, \vec{e}_{2})/\sqrt{2}\,,
\end{equation}
Here $\vec{e}_{2}=[\vec{e}_{1}\vec{n}]$, where $\vec{n}$ is the
unit vector pointing in the propagation direction of the beam
of  $\gamma$-quanta.

In view of the above, it is easy to demonstrate
that
\begin{equation}
\label{12.7} f_{+}=f_{1}(\omega) - f_{2}(\omega)(\vec{p}\vec{n}),
f_{-}=f_{1}(\omega)+f_{2}(\omega)(\vec{p}\vec{n})\,,
\end{equation}
where $\vec{p}$ is the polarization vector of particles.

Suppose that photons in a vacuum are linearly polarized along the
direction  $\vec{e}_{1}$. Then, according to \cite{51} in the  medium the polarization vector
$\vec{e}\,^{\,\prime}_{1}$ is :
\begin{eqnarray}
\label{12.8}
\vec{e}\,^{\,\prime}_{1}&=&\left[\left(\frac{\vec{e}_{1}+i\vec{e}_{2}}{2}\right)\exp\left(-i\frac{\omega}{2c}\Delta n l \right)\right.\nonumber\\
&+&\left.\left(\frac{\vec{e}_{1}-i\vec{e}_{2}}{2}\right)\exp\left(i\frac{\omega}{2c}\Delta n l \right)\right]\exp\left(i\frac{n_{1}+n_{2}}{2c}\omega l \right)\nonumber\\
&=&\exp\left(i{\omega}\frac{n_{1}+n_{2}}{c} l \right)\left[\vec{e}_{1}\cos\left(\frac{2\pi Nc}{\omega}(\vec{p}\vec{n})f_{2}(\omega) l \right)\right.\nonumber\\
&+&\left.\vec{e}_{2}\sin\left(\frac{2\pi
	Nc}{\omega}(\vec{p}\vec{n})f_{2}(\omega) l \right)\right]\,,
	\end{eqnarray}
	where $l$ is the 
	path passed by $\gamma$-quanta in matter.
	
	
	Full rotation of the
	polarization vector takes place in the length
	\begin{equation}
\label{12.9} d=\frac{4\pi c}{\omega|\Delta n|}\,.
\end{equation}
One can easily see  that the positive sign of
$f_{2}(\omega)(\vec{p}\vec{n})$ corresponds to the right--hand
rotation, while the negative sign of this quantity corresponds to the left--hand rotation.

In the general case, $\texttt{Im}f_{2}(\omega)\neq 0$, i.e., the
coefficients of absorption are different for the states with left and
right circular polarizations. As in this case
(\ref{12.8}) includes the trigonometric functions for complex
arguments, the dependence of photon polarization on  distance
$l$ becomes more complicated. If the photons in a vacuum  are still
polarized along the direction  $\vec{e}_1$, the following formulas are
applicable to the Stokes parameters in a medium \cite{3,23}
\begin{equation}
\label{12.10} \varepsilon_{1}=r\cos 2\varphi;~~
\varepsilon_{2}=(1-r^{2})^{1/2};~~\varepsilon_{3}= r \sin 2\varphi\,,
\end{equation}
where
\begin{eqnarray*}
\varphi & = &\frac{2\pi N
	c}{\omega}(\vec{p}\vec{n})\texttt{Re}f_{2}(\omega)\,l\,;\\
r & = &\cosh\left(\frac{4\pi N
	c}{\omega}(\vec{p}\vec{n})\texttt{Im}f_{2}(\omega)\, l \right)=(\varepsilon_{1}^{2}+\varepsilon_{3}^{2})^{1/2}
	\end{eqnarray*}
	is the degree of linear polarization; $\varepsilon_{2}$ is the
	degree of circular polarization. At $l=0$, we have  $\varepsilon_{1}=1$,
	$\varepsilon_{3}=0$, $\varepsilon_{2}=0$.
	
	It is seen that when the imaginary part of the function
	$f_{2}(\omega)$ is nonzero, the  linear polarization of the  photon in a
	medium transforms into an elliptical one, and $\varphi$ is the angle of rotation
	of the ellipse's major axis  relative to the initial direction
	$\vec{e}_{1}$.
	
	It follows from the above that
	the full rotation of the ellipse's major axis occurs in the length
	\begin{equation}
\label{12.11} d=\left(\frac{N
	c}{\omega}(\vec{p}\vec{n})\,\texttt{Re}f_{2}(\omega)\right)^{-1}\,.
	\end{equation}
	Note that at $l\rightarrow\infty$, we have $|\varepsilon_{2}|=1$.
	This indicates the total absorption of photons with right-- or
	left--hand circular polarization.
	
	It immediately follows from  (\ref{12.10}) that the
	change in the polarization of $\gamma$-quanta  passing through a
	polarized target only depends on the function
	$f_{2}(\omega)$. As for the function $f_{1}(\omega)$ (see
	(\ref{12.4})), it has nothing to deal with the effect we are
	concerned with.
	
	
	From the
	optical theorem follows the below relation
	\begin{equation}
\label{12.12} \texttt{Im}f_{2}(\omega)=\frac{\omega}{4\pi
	c}\frac{\sigma_{\uparrow\downarrow}(\omega)-\sigma_{\uparrow\uparrow}(\omega)}{2}\,,
	\end{equation}
	where $\sigma\uparrow\uparrow$ and $\sigma\uparrow\downarrow$ are
	the values of the total Compton scattering  cross sections for
	parallel and antiparallel orientations of photon and electron (nucleus)
	spins, respectively. 
	

	To calculate the real part of $f_{2}(\omega)$, make use of the
	dispersion relation given in \cite{57}.
	\begin{equation}
\label{12.14} \texttt{Re}
f_{2}(\omega)=-\frac{2\omega}{\hbar c^{2}}(\Delta\mu)^{2}+\frac{2\omega^{3}}{\pi}\int\limits_{0}^{\infty}\frac{\texttt{Im}f_{2}(\omega^{\,\prime})}{\omega^{\,\prime
		2}(\omega^{\,\prime 2}-\omega^{2})}d\omega^{\,\prime}\,,
		\end{equation}
		where $\Delta\mu$  is the anomalous magnetic moment of the particle.
		

		Using expression (\ref{12.12}) for $\texttt{Im}f_{2}(\omega)$ one can express $\texttt{Re}f_{2}(\omega)$ as follows:
		\begin{equation}
\label{eq:4}
\texttt{Re}
f_{2}(\omega)=-\frac{2 \omega}{\hbar c^{2}}(\Delta\mu)^{2}+\frac{\omega^{3}}{4\pi^{2}c}\int\limits_{0}^{\infty}\frac{\sigma_{\uparrow\downarrow}(\omega^{\,\prime})
	-\sigma_{\uparrow\downarrow}(\omega^{\,\prime})}{\omega^{\,\prime}(\omega^{\,\prime
		2}-\omega^{2})}d\omega^{\,\prime}\,.
		\end{equation}
		According to expression (\ref{12.10}) the polarization plane rotation angle for $\gamma$-quanta moving through a polarized target is determined by $\texttt{Re}
		f_{2}(\omega)$. Expression (\ref{eq:4}) for $\texttt{Re}
		f_{2}(\omega)$ comprises two summands both conditioned by scattering: the first one is caused by anomalous magnetic moment $\Delta\mu$, while the second is associated with other scattering processes and reactions caused by $\gamma$-quanta interactions with protons (nuclei, electrons).
		
		In accordance with expressions (\ref{12.10}) and (\ref{eq:4})
		polarization plane rotation angle $\varphi$ for a $\gamma$-quantum, which passed in matter   path $l$, reads as follows:
		\begin{equation}
\label{eq:5}
\varphi=-\frac{4 \pi N}{\hbar c}(\Delta \mu)^2 (\vec{p} \vec{n}) l + \frac{2 \pi N c}{\omega} (\vec{p} \vec{n}) l \frac{2 \omega^3}{\pi} \int_{0}^{\infty} \frac{\texttt{Im} f_2(\omega^{\,\prime})}{\omega^{\,\prime \,2} (\omega^{\,\prime \, 2}-\omega^{ 2})} d \omega^{\,\prime}
\,.
\end{equation}
Expression (\ref{eq:5}) can be rewritten as follows:
\begin{equation}
\label{eq:6}
\varphi=-\frac{4 \pi N}{\hbar c}(\Delta \mu)^2 (\vec{p} \vec{n}) l + {2 \pi N c} (\vec{p} \vec{n}) l \gamma_0 (\omega) \omega^{2}
\,,
\end{equation}
where 
\begin{equation}
\gamma_0 (\omega)=\frac{1}{4 \pi^2 }  \int_{0}^{\infty} \frac{\sigma_{\downarrow \uparrow}(\omega^{\,\prime}) - \sigma_{\uparrow \uparrow}(\omega^{\,\prime})}{\omega^{\,\prime} (\omega^{\,\prime \,2}-\omega^{ 2})} d \omega^{\,\prime}.
\label{eq:gamma0}
\end{equation}

In the low energy range according to 
\cite{VG_4,VG_5,VG_6,VG_7,VG_9c,VG_9d,VG_9e,VG_9f,VG_9i,VG_9_1,
VG_n_10,VG_n_11,VG_n_12,VG_n_13,VG_n_14,VG_n_15,VG_n_16,VG_n_17,VG_n_18,VG_n_19}
the amplitude of Compton forward scattering can be expressed in the form (\ref{eq:3}) with $\gamma_0$ read as follows:
\begin{equation}
\label{eq:8}
\gamma_0=\frac{1}{4 \pi^2 }  \int_{\omega_{thr}}^{\infty} \frac{\sigma_{\downarrow \uparrow}(\omega^{\,\prime}) - \sigma_{\uparrow \uparrow}(\omega^{\,\prime})}{\omega^{\,\prime \,3} } d \omega^{\,\prime},
\end{equation}
where $\omega_{thr}$ is the pion photoproduction threshold. 
Contributions to the total cross-sections, caused by Compton scattering  and processes of electron-positron pair production  are not considered in (\ref{eq:8}).

Carried out experiments and theoretical analysis provide the following evaluation for proton spin polarizability $\gamma_0$:
$\gamma_0 \approx -1.34 \cdot 10^{-4}$\,fm$^4$\,$=-1.34 \cdot 10^{-56}$\,cm$^4$.
Therefore, in case if (\ref{eq:3}) is valid, polarization plane rotation angle $\varphi$ for a $\gamma$-quantum, which passed in polarized matter   path $l$, reads as follows:
\begin{equation}
\label{eq:9}
\varphi=-\frac{4 \pi N}{\hbar c}(\Delta \mu)^2 (\vec{p} \vec{n}) l + {2 \pi N} (\vec{p} \vec{n}) \gamma_0 k^2 l =\varphi(\Delta \mu)+\varphi(\gamma_0)
\,,
\end{equation}
where $k=\frac{\omega}{c}$ is the wavenumber of the $\gamma$-quantum.

The summand $\varphi(\Delta \mu)$, which includes anomalous magnetic moment, does not depend on energy, while  another one $\varphi(\gamma_0)$, which is determined by  spin polarizability $\gamma_0$,
depends on $\gamma$-quantum energy.
The latter grows proportionally to $k^2$ i.e. proportionally to the squared $\gamma$-quantum energy.

Let us now evaluate the value of polarization plane rotation angle $\varphi$ for a $\gamma$-quantum, which passes through a target with polarized nuclei.
Target $^{14}$NH$_3$, which is used for investigation of polarized $\gamma$-quanta  scattering  by polarized protons, can be considered as an example to evaluate rotation effect.
Note, that this target comprises both protons and nuclei $^{14}$N,
therefore, number of nuclei per cm$^3$ in this target is not equal to number of protons.
To evaluate the number of polarized protons per cm$^3$ let us use the density $\rho$ in units $[\frac{g}{cm^3}]$, which for solid NH$_3$ is as high as $\rho=0.85 [\frac{g}{cm^3}]$.
The weight of  NH$_3$ molecule is with high accuracy equal to $M=17m_p$, where $m_p$ is the proton mass. Therefore, the number of NH$_3$ molecules per cm$^3$ reads $N_{mol}=\frac{\rho}{17m_p}$. And, since each molecule comprises three protons, the number of protons can be expressed as 
$$
N=3 \cdot N_{mol}= \frac{\rho}{m_p} \cdot \frac{3}{17}= \frac{\rho}{m_p} \cdot f.
$$ 
Here $f$ is the dilution factor (see, for example, \cite{VG_n_13,VG_n_14,VG_n_21,VG_n_22}) and $\frac{\rho}{m_p}$ is the number of nucleons per cm$^3$.
The complicated internal structure of the target requires the target density to be reduced by the so called \cite{VG_n_13,VG_n_14,VG_n_21,VG_n_22} packing factor $\varkappa=0.6$ resulting in correction in the number of protons
per cm$^3$
as follows:
$$
N=\frac{\rho}{m_p} \cdot f \cdot \varkappa.
$$
Therefore, 
polarization plane rotation angle $\varphi$ is finally expressed as follows:
\begin{equation}
\label{eq:10}
\varphi=-\frac{4 \pi }{\hbar c}\frac{\rho}{m_p} f \varkappa p (\vec{n}_p \vec{n}) (\Delta \mu)^2  l + \frac{2 \pi \rho }{m_p} f \varkappa p (\vec{n}_p \vec{n})  \gamma_0 k^2 l 
\,,
\end{equation}
where $p$ is the proton polarization degree, $\vec{n}_p$ is the unit vector directed along proton polarization vector.
%

If $\gamma$-quanta momentum is directed along the polarization vector, then $\vec{n}_p \vec{n}=+1$, in case of antiparallel directed $\vec{n}_p$ and $\vec{n}$, product $\vec{n}_p \vec{n}=-1$.
Therefore, change of $\vec{n}_p$ direction with respect to $\vec{n}$ results in change of rotation direction (sign).
For $\vec{n}_p \uparrow \uparrow \vec{n}$ polarization plane rotation angle reads as
\begin{equation}
\label{eq:11}
\varphi=-\frac{4 \pi \rho \varkappa}{\hbar c \,m_p} f  p \, (\Delta \mu)^2  l + \frac{2 \pi \rho \varkappa}{m_p} f  p \, \gamma_0 k^2 l
\,.
\end{equation}
Let us evaluate rotation angle $\varphi$.
In case of NH$_3$ target with polarized protons the number of protons per cm$^3$ is $N=\frac{\rho \varkappa f}{m_p} \approx 5.4 \cdot 10^{22}$, therefore at  polarization degree $p=0.9$  the number of polarized protons in the target is as high as $N_p \approx 5 \cdot 10^{22}$.
Anomalous magnetic moment $\Delta \mu = 8.95 \cdot 10^{-24}$\,erg/Gs, therefore, contribution to rotation angle $\varphi(\Delta \mu)$ can be evaluated as 
\begin{equation}
\label{eq:12}
\varphi(\Delta \mu)=1.6 \cdot 10^{-6} \cdot l ~\textrm{rad}
\end{equation}
and for $l=$30~cm the angle $\varphi(\Delta \mu) \approx 5 \cdot 10^{-5}$\,rad. 
Contribution to rotation angle $\varphi(\Delta \mu)$ caused by anomalous magnetic moment does not depend on $\gamma$-quantum energy.

Let us now evaluate contribution to rotation angle $\varphi(\gamma_0)$, which is determined by  spin polarizability $\gamma_0$ and
depends on $\gamma$-quantum energy.
Using the second summand in (\ref{eq:11}) one can get for $\varphi(\gamma_0)$ in case of NH$_3$ target the following evaluation:
\begin{equation}
\label{eq:13}
\varphi(\gamma_0) \approx 3 \cdot 10^{-33} k^2\, l.
\end{equation}
It follows from (\ref{eq:13}) that for $\gamma$-quanta with energy 300~MeV ($k=1.6 \cdot 10^{13}$\,cm$^{-1}$) passing through the target of 30~cm thickness rotation angle $\varphi(\gamma_0)$ is as high as $\varphi(\gamma_0)=2 \cdot 10^{-5}$\,rad.
If $\gamma$-quanta energy is increased up to 1~GeV (such energies are available at the Bonn accelerator facility ELSA and at the Mainz accelerator MAMI
\cite{VG_n_13,VG_n_14,VG_n_21,VG_n_22}) 
this contribution could reach $\varphi(\gamma_0)=2 \cdot 10^{-4}$\,rad. 
Further increase of $\gamma$-quanta energies results in fast growth of rotation angle: for 3~GeV $\gamma$-quanta $\varphi(\gamma_0)=2 \cdot 10^{-3}$\,rad, 
for 10~GeV $\gamma$-quanta $\varphi(\gamma_0)=2 \cdot 10^{-2}$\,rad  
and for 100~GeV it appears to be $\varphi(\gamma_0)=1.8$\,rad  for the target of 30~cm thickness.

Experiments and theoretical analysis, in which the value $\gamma_0=-1.34 \cdot 10^{-56}$~cm$^4$ was obtained, were carried out for $\gamma$-quanta in the energy range $\le$1~GeV.
The above obtained evaluation for the angle of rotation  caused by the quasi-optical phenomenon of polarization plane rotation indicates a significant effect magnitude.  
This especially valid for the energy range GeV and higher.
In this energy range, the $\gamma$-quantum  wavelength is either comparable to or less than the electric radius of the proton, which is equal to $0.8 \cdot 10^{-13}$~cm.  
The amplitude of forward scattering, and thus $\gamma_0(\omega)$, may depend on the internal structure of the proton.

The question arises, what is the magnitude of $\gamma_0$ in this range?  
Does $\gamma_0$ begin to depend on the $\gamma$-quantum energy that leads to a gradual decrease of the rotation angle magnitude with energy growth?  
Such behavior of the rotation angle is observable for  $\gamma$-quanta passing through matter with polarized electrons \cite{52,Nuclear_optics,A2}.  
In this case, the drop occurs in the range of several MeV.

It seems very interesting not only to detect the quasi-optical effect of polarization plane rotation for $\gamma$-quanta in a target containing polarized protons but also to study the possible energy dependence of the spin polarizability $\gamma_0(\omega)$.

Let us evaluate the possibility of detecting the quasi-optical effect of polarization plane rotation for $\gamma$-quanta passing through a target with polarized protons (deuterons). 
For this purpose, let us estimate the number of $\gamma$-quanta required to detect the effect. 
As indicated above, in the $\gamma$-quanta energy range up to 10 GeV, the angle of polarization plane rotation $\varphi$ is much lower than 1. 
Due to polarization plane rotation a component of the polarization vector of the $\gamma$-quantum in the direction orthogonal to the polarization plane of the incident beam will appear. 
The magnitude of this polarization component can be measured, for example, by studying 
beam passage through a second polarized target containing polarized nuclei with spin $ \ge 1$. 
In such a target the phenomenon of birefringence and dichroism, which is sensitive to linear polarization, arises according to \cite{A1,A2,arxive.2411.04960}.  
Note that dichroism sensitive to linear polarization of $\gamma$-quanta also arises during $\gamma$-quanta passage through oriented crystals \cite{Cabibbo_33,Cabibbo_34,Cabibbo_35}.
To distinguish the signal from $\gamma$-quanta with polarization vector lying in the scattering plane against the background of all the scattered $\gamma$-quanta, the number of $\gamma$-quanta falling on the detector $N$ should  be greater than the value of $1/\varphi^2$.  

According to the above evaluations, the number of $\gamma$-quanta during the experiment for photons with energy 300 MeV is $N > \frac{1}{\varphi^2} \approx 2.5 \cdot 10^9$.  
For 1~GeV energy the number of quanta $N  \approx 2.5 \cdot 10^7$, at 3~GeV the number of quanta approaches $N  \approx 2.5 \cdot 10^5$ and at 10~GeV the number of quanta is as high as $N  \approx 2.5 \cdot 10^3$.  

At $\gamma$-quanta energies above 3~GeV, the observation conditions are further  improved. 
The number of quanta $N$ falling on the detector is equal to $N = \dot{N}T$, where $\dot{N}$ is the intensity of the photon beam ([$\dot{N}$ = ph/s]), $T$ is the observation time.

According to \cite{HIGS1}, modern accelerating facilities are capable of providing $\gamma$-ray flux as high as $\dot{N} \approx 10^{10}$~ph/s (the High Intensity Gamma-Ray Source (HIGS), operated by the Triangle Universities Nuclear Laboratory).  
This Compton gamma-ray source is capable of providing nearly mono-energetic, polarized gamma-ray beams with energies ranging from 1 to 100 MeV.  

The Laser Electron Photon beamline (LEPS, Japan) produces 1300 -- 2900 MeV photon beams with fluxes $\dot{N} \approx 10^6 - 10^7$~ph/s.

According to calculations carried out in \cite{Tikhomirov_2023},  
with the use of crystals at electron energy 10~GeV, one can obtain $10^{-3}$ $\gamma$-quanta with energy 3~GeV per electron. 
In this range of  $\gamma$-quanta energy and higher, according to \cite{Tikhomirov_2023}, the contribution from the Bethe-Heitler radiation mechanism becomes important. 
This means that even at a current of 1nA=$10^{10}$~e/s one can obtain $\dot{N} \approx 10^7$~ph/s.  
Therefore, in the range of $\gamma$-quanta energy about 3~GeV and above at SLAC accelerator complexes, it is possible to study the dependence of polarization plane rotation effect on the $\gamma$-quanta energy.


Thus, law (\ref{eq:3})  can be verified and the range, for which dependence of amplitude $\gamma_0(\omega)$ on $\gamma$-quanta energy should be taken into account, can be defined.

\begin{figure}[h]
\epsfxsize = 12 cm \centerline{\epsfbox{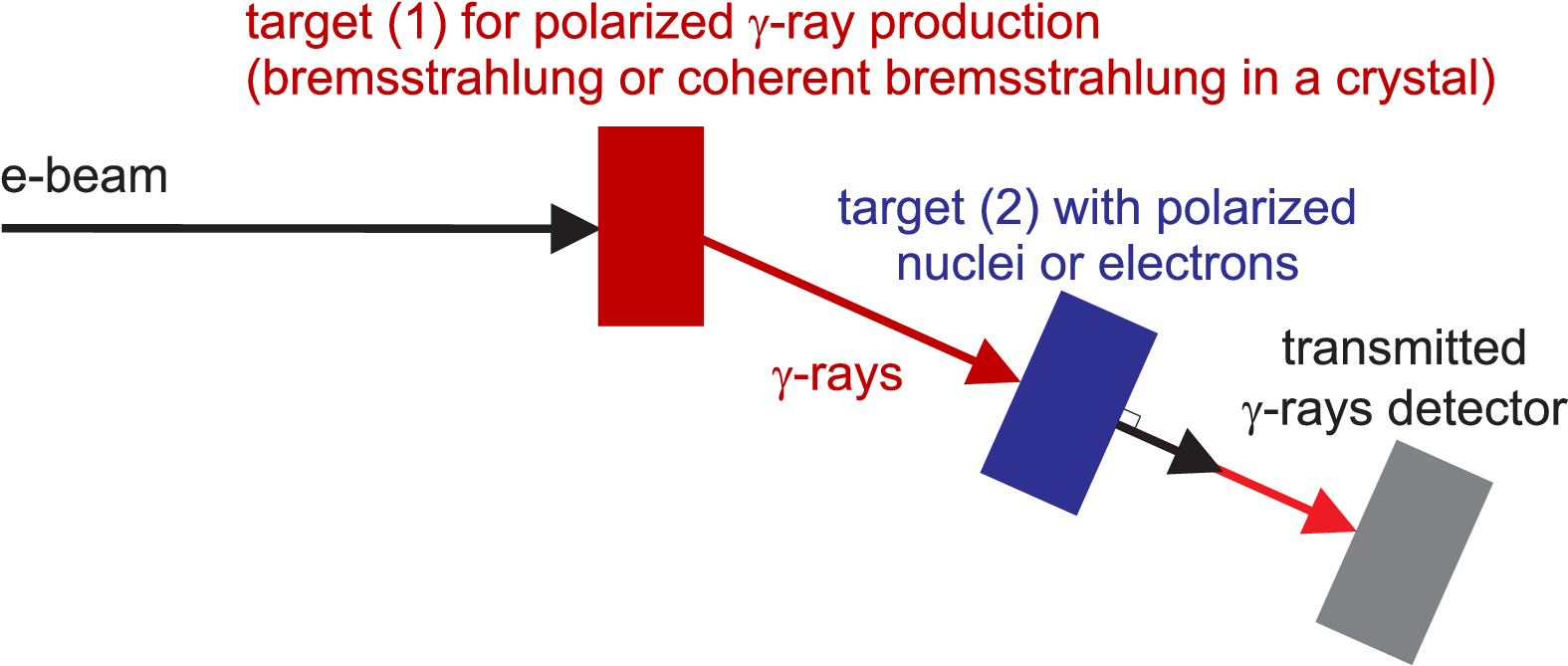}}
\caption{Layout of transmission experiment}
\label{fig:gamma3}
\end{figure}

Let us emphasize that investigation of quasi-optical effect of $\gamma$-quanta polarization plane rotation does not imply measurement of properties of scattered $\gamma$-quanta (scattered electromagnetic wave).
The considered effect is studied by investigation of coherent passing of $\gamma$-quanta (electromagnetic wave) through a target
(see Fig.\ref{fig:gamma3}). 
The number of $\gamma$-quanta passed through the target appears to be much higher as compared with the number of those scattered into some spatial angle.

	It is interesting to note that the phenomenon deceptively similar to inverse Faraday effect \cite{reverse_Faradey}, but caused by another mechanism \cite{51}, also appears at collisions of $\gamma$-quanta with the rest matter or a particle beam, namely: when $\gamma$-quanta with circular polarization fall onto nonpolarized matter, the latter acquires polarization due to  dependence of scattering cross-sections on orientation of electron (nuclear) spins. The same happens when circularly polarized $\gamma$-quanta collide with a beam of nonpolarized particles, for example, protons moving in an accelerator (or a storage ring): particle beam acquires polarization i.e. is magnetized (orientation of spins causes orienting magnetic moments).

\subsection{Quasi-optical birefringence effect for $\gamma$-quanta in matter with polarized nuclei}
\label{sec:birefringence}

It is interesting to spot that just as light in uniaxial and biaxial crystals experiences double refraction (birefringence), so the similar quasi-optical birefringence effect exists for  $\gamma$-quanta in matter containing polarized nuclei with spin {$S \ge 1$} \cite{A1,A2}. 
For example, for $^{14}$N and D spin is equal $S=1$, while $^7$Li has $S=3/2$.

According to the {above} (see (\ref{I.1}))
%
%
the refractive index of matter for $\gamma$-quanta is determined by the coherent elastic forward scattering amplitude $f(0)$ as follows:
\begin{equation*}
n=1+\frac{2 \pi \rho}{k^2} {f(0)}.
\end{equation*}

When $\gamma$-quanta interact with matter, which comprises polarized nulei with 
spin {$S \ge 1$},
amplitude $f(0)$ can be expressed as {\cite{A1,A2}}:
\begin{equation}
\label{eq:14}
f(0)=f_1(\omega) (\vec{e}^{\,\,\prime*}~\vec{e}) + i \, f_2(\omega) \, \vec{p} \, \left[\vec{e}^{\,\,\prime*}~\vec{e}\right] + {f_3 (\omega)} \, Q_{ik} {e}^{\,\,\prime*}_i  e_k + f_4(\omega) n_{\gamma i} n_{\gamma k} Q_{ik}.
\end{equation}
where $\vec{p}=\textrm{Tr} \hat{\rho} \vec{n}_S$ is the nuclear polarization vector, $\vec{n}_S= \hat{\vec{S}}/S$, $\hat{\vec{S}}$ is the nuclear spin operator, $\hat{\rho}$ is the spin density matrix of the target, $Q_{ik}=\textrm{Tr} \hat{\rho}\, \hat{Q}_{ik}$ is the polarization tensor of rank two, 
$$
\hat{Q}_{ik}=\frac{3}{2 S (2S-1)}\left\{  \hat{S}_i \hat{S}_k + \hat{S}_k \hat{S}_i - \frac{2}{3}S (S+1) \delta_{ik}     \right\} ;
$$
$\vec{n}_{\gamma}$ is the unit vector along the $\gamma$-quanta momentum.
Note that amplitudes $f_1$, $f_2$ and $f_3$ are expressed via polarizabilities $\alpha_0$, $\alpha_{\gamma}$ and $\alpha_T$ of nuclei introduced by A.M.~Baldin~\cite{VG_8,VG_9} as follows:
\begin{equation}
f_1=\left( \frac{\omega}{c} \right)^2 \alpha_0, ~~ f_2=\left( \frac{\omega}{c} \right)^2 \alpha_{\gamma},~~ f_3=\left( \frac{\omega}{c} \right)^2 \alpha_T.
\end{equation}


Let us assume that polarization vector $\vec{p}$ for a target is orthogonal to $\gamma$-quanta incidence direction: $\vec{p} \perp \vec{n}_{\gamma}$\,, and define direction of $\vec{n}_{\gamma}$  as axis $y$.
In this case from (\ref{eq:14}) one could observe the difference in refraction indices for a photon with linear polarization $\vec{e}_x$ and that with linear polarization $\vec{e}_z$.

Let linear polarization of a $\gamma$-quantum incident on a target is a superposition $\vec{e}= \alpha \vec{e}_x + \beta \vec{e}_z$\,.
As the $\gamma$-quantum moves deeper into target, its linear polarization  converts into  elliptical one that is in full similarity to optics.
Thus, circular polarization appears in the initially linearly polarized  beam of $\gamma$-quanta. Polarization degree is determined by $\textrm{Re}\,f_3(\omega)$. 
%

In case when a $\gamma$-quantum with circular polarization moves in the target,  it attains linear polarization, degree of which is determined by $\textrm{Re}\,f_3(\omega)$.
The detailed description of birefringence for $\gamma$-quanta is given in \cite{A1,A2}.


According to evaluations \cite{A1} in the vicinity of  giant resonance, the degree $\delta$ of attained circular polarization for a $\gamma$-quantum, which initially has linear polarization, when it passes through a target with $l=1$~cm thickness comprising $\textrm{Ta}$ nuclei, is as high as $\delta \approx 10^{-3} - 10^{-4}$.
For a target comprising polarized deuteron nuclei polarization degree could be evaluated as $\delta \approx 4 \cdot10^{-5}\, l$,
assuming that $\gamma$-quanta energy is about few MeV and deuteron tensor polarizability is evaluated by its  static polarizability $\alpha_T \approx 10^{-40}$\,cm$^3$ derived in {\cite{A3}}.

For other nuclei and $\gamma$-quanta energies as high as dozens and hundreds of MeV and higher, the birefringence effect should be evaluated separately.

\begin{figure}[h]
\epsfxsize = 8 cm \centerline{\epsfbox{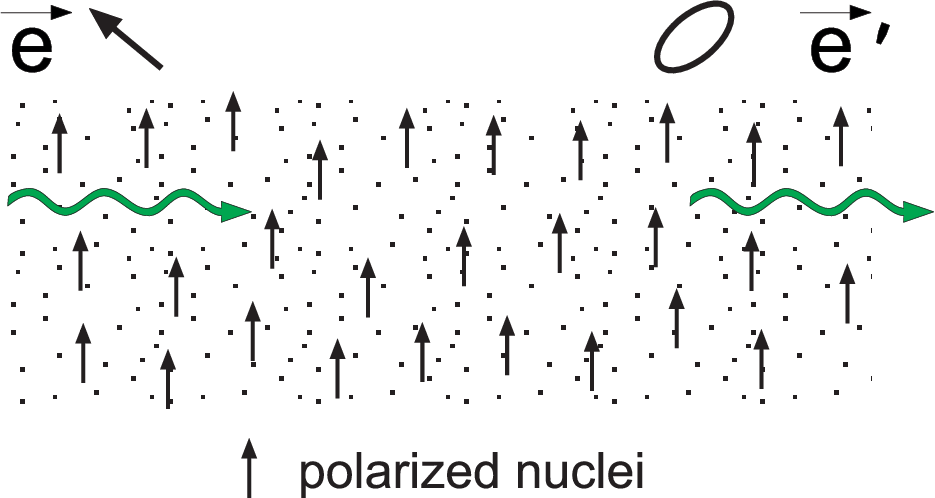}}
\caption{Birefringence of gamma quanta}
\label{fig:gamma2}
\end{figure}

Success in development of targets with polarized nuclei and beams of polarized high-energy $\gamma$-quanta  enables direct experimental observation of  quasi-optical phenomenon of $\gamma$-quanta polarization plane rotation in matter with polarized proton (nuclei).
Quasi-optical birefringence effect  for $\gamma$-quanta in matter with polarized nuclei having spin $S \ge 1$ could also be observed \cite{A1,Nuclear_optics,A2}. The latter effect is similar to double refraction (birefringence) of light in uniaxial and biaxial crystals (see Fig.\ref{fig:gamma2}).


The above discussed experiments can be carried at LINAC (JINR).
%

%
%

\section
{Quasi-optical phenomena for low-energy particles: 
} 
\label{sec:low-energy}

\subsection{Neutron Spin Precession in a Pseudomagnetic Field of Matter with Polarized Nuclei}

When a neutron moves in matter with polarized nuclei, the phenomenon of neutron spin precession (rotation) arises. 
The spin rotation is caused not merely by Larmor precession in a magnetic field, but also by
neutron spin interaction with the nuclear pseudomagnetic field caused by strong (weak) interaction of neutrons with polarized nuclei.

We shall start the description of the phenomenon of "optical" spin rotation of particles in matter with polarized nuclei from  consideration of
the passage of slow neutrons through  matter with polarized nuclei.
The possible contribution to neutron scattering from the interaction
between the neutron magnetic moment and  the magnetic moments of nuclei
(and electrons) will be temporarily ignored. In
this case the wave describing the process of elastic collision
between a neutron and a nucleus fixed at point $R_{i}$ has the form
\begin{equation}
	\label{6.1} \psi(\vec{r}) = e^{i \vec{k}\vec{r}}\chi_{n}\chi_{\mathrm{nuc}} +
	\hat{f}\frac{e^{ik|\vec{r}-\vec{R}_{i}|}}{|\vec{r}-\vec{R}_{i}|}e^{i\vec{k}\vec{R}_{i}}\chi_{n}\chi_{\mathrm{nuc}}\,,
\end{equation}
where $\chi_{n}$ is the spin wave function of the incident
neutron; $\chi_{\mathrm{nuc}}$ is the spin wave function of the nucleus.

For slow neutrons, the wavelength is much larger than the size of
the nuclei. That is why the scattering amplitude $\hat{f}$ does not depend on
the angle ($S$-scattering) and can be written as follows \cite{22}:

\begin{equation}
	\label{6.2} \hat{f} = \alpha+\beta\vec{\sigma}\vec{J}\,,
\end{equation}
where
\[
\alpha=\frac{J+1}{2J+1}a^{+}+\frac{J}{2J+1}a^{-}\,,
\]
\[\beta=\frac{a^{+}-a^{-}}{2J+1}\,,
\]
$a^{+}$ is the scattering amplitude
in the state with the total momentum $J+1/2$ of the neutron and the nucleus; $a^{-}$  is the same in state $J-1/2;
\vec{\sigma}=2\vec{S}$, $\vec{S}$ is the neutron spin operator;
$\vec{J}$ is the nuclear spin operator.

At scattering by a set of nuclei,
the wave function takes the form

\begin{equation}
	\label{6.3} \psi(\vec{r})=e^{i\vec{k}\vec{r}}\chi_{n}\prod_{m}\chi_{\mathrm{nuc}\,
		m}+
	\sum_{i}\hat{f}_{i}\frac{e^{ik|\vec{r}-\vec{R}_{i}|}}{|\vec{r}-\vec{R}_{i}|}e^{i\vec{k}\vec{R}_{i}}
	\chi_{n}\prod_{m}\chi_{\mathrm{nuc}\,
		m}\,,
\end{equation}
where $\prod_{m}\chi_{\mathrm{nuc} \,m}$ is the spin wave function of nuclei;
the nuclei are assumed not to interact with each other.

To find the coherent wave in this case, (\ref{6.3}) should be
averaged over the location of scatterers, as well as over their
spin state. Averaging of (\ref{6.3}) over the spin state of the
nuclei gives

\begin{equation}
	\label{6.4}
	\langle\psi(\vec{r})\rangle=e^{i\vec{k}\vec{r}}\chi_{n}+\sum\limits_{i}\langle\hat{f}\rangle\frac{e^{ik|\vec{r}-
			\vec{R}_{i}|}}{|\vec{r}-\vec{R}_{i}|}e^{i\vec{k}\vec{R}_{i}}\chi_{n}\,,
\end{equation}

where
\[
\langle\hat{f}\rangle=\alpha+\beta\vec{\sigma}\langle\vec{J}\rangle=\alpha+\beta
J\vec{\sigma}\vec{p}\,,
\]
$\vec{p}= \langle\vec{J}\rangle /J$ is the polarization vector of
the nuclei.

If the nuclei are chaotically distributed in a certain plane $z=z_{0}$, 
we obtain the
following expression for a coherent wave which has passed through the given
plane
%
%
\begin{equation}
	\label{6.5} \langle\psi(\vec{r})\rangle=\left(1+\frac{2\pi
		i\rho^{\prime}}{k_{z}}(\alpha+\beta J\vec{\sigma}
	\vec{p})\right)e^{i\vec{k}\vec{r}}\chi_{n}\,,
\end{equation}
Note that the distance between scatterers was not assumed to be less than the wavelengthin when deriving (\ref{6.5}).  
Expression (\ref{6.5}) is also valid in the case when the wavelength is much smaller than the distance between scatterers.

\begin{figure}[htbp]
	\epsfxsize = 10 cm \centerline{\epsfbox{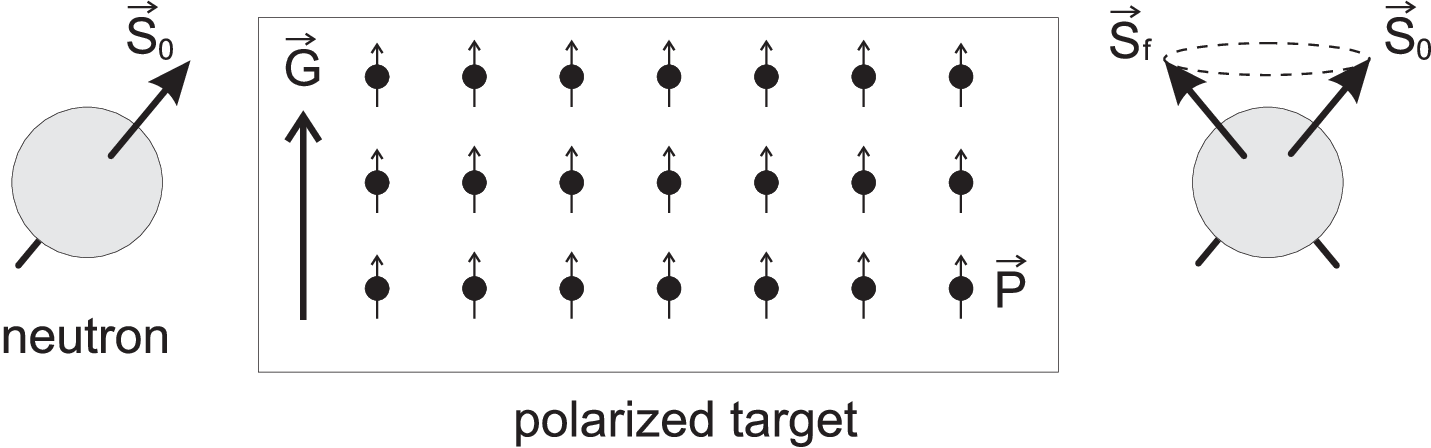}}
	\caption{Neutron spin precession} \label{presentation Figure 2}
\end{figure}

Recall (see, for instance, \cite{23}) that the operator of the
form
\begin{equation}
	\label{6.6} B= 1+\frac{i}{2}\delta\theta\vec{\sigma}\vec{\eta}_{p}
\end{equation}
is the operator of spin rotation by  an angle $\delta\theta
(\delta\theta \ll 1)$ about the axis characterized by a unit vector
$\vec{\eta}_{p}$. Comparing (\ref{6.6}) with (\ref{6.5}), one may
conclude that after the neutron has passed the polarized plane, the neutron spin
will rotate by the angle
\begin{equation}
	\label{6.7} \delta\theta= \frac{4\pi\rho^{\prime}}{k_{z}}J p \texttt{Re}\beta\,.
\end{equation}
If the wave passes through  $m$ planes, the total angle is
\begin{equation}
	\label{6.8} \delta\theta= \frac{4\pi\rho^{\prime}}{k_{z}}J p m \texttt{Re}\beta
\end{equation}
or, proceeding to a layer of finite thickness $l$, we obtain that as a
neutron passes through a polarized target, its spin rotates \cite{24} through an
angle (see Fig.\ref{presentation Figure 2})
\begin{equation}
	\label{6.9} \theta= \frac{4\pi\rho}{k_{z}}\texttt{Re}\beta J p l\,.
\end{equation}

The same result can be obtained in a different  way. Let us choose
the quantization axis parallel to vector $\vec{p}$. As a result, if
a neutron with spin parallel to vector $\vec{p}\left[\chi_{n}=
\left(\begin{smallmatrix}
	1\\
	0
\end{smallmatrix}\right)\right]$
falls on the plane, the coherent wave $\langle\psi(\vec{r})\rangle$
has the form
\begin{equation}
	\label{6.10} \langle\psi(\vec{r})\rangle_{+}= \left(1+ \frac{2\pi
		i\rho^{\prime}}{k_{z}}f_{+}\right)
	e^{i\vec{k}\vec{r}}\begin{pmatrix}
		1\\
		0
	\end{pmatrix}\,,
\end{equation}
where $f_{+} = \alpha +\beta J p$ is the amplitude of coherent
elastic zero--angle scattering of the neutron with spin parallel to
the  polarization vector of nuclei $\vec{p}$. For a neutron with
the opposite spin direction $\left[\chi_{n}=
\left(\begin{smallmatrix}
	0\\
	1
\end{smallmatrix}\right)\right]$, the coherent wave $\langle\psi(\vec{r})\rangle$
is given by the expression
\begin{equation}
	\label{6.11} \langle\psi(\vec{r})\rangle_{-}= \left(1+ \frac{2\pi
		i\rho^{\prime}}{k_{z}}f_{-}\right)
	e^{i\vec{k}\vec{r}}\begin{pmatrix}
		0\\
		1
	\end{pmatrix}\,,
\end{equation}
where $f_{-}=\alpha-\beta Jp$ is the corresponding coherent
amplitude of scattering of the neutron with spin antiparallel to
vector $\vec{p}$.

If a wave passes through a finite thickness layer of polarized
matter, then, recalling the conclusions that led us to the expression
describing the refractive index for a nonpolarized target, we obtain
that the refractive index for neutrons with spin parallel to
$\vec{p}$ is
\begin{equation}
	\label{6.12} n_{+}= 1+ \frac{2\pi\rho}{k_{z}^{2}} f_{+} = 1+
	\frac{2\pi\rho}{k_{z}^{2}} (\alpha+\beta Jp).
\end{equation}
For neutrons with the opposite polarization,
\begin{equation}
	\label{6.13} n_{-}= 1+ \frac{2\pi\rho}{k_{z}^{2}} f_{-} = 1+
	\frac{2\pi\rho}{k_{z}^{2}} (\alpha -\beta Jp)\,.
\end{equation}

In operator form the expression for $\hat{n}$ can be written as follows \cite{201,Nuclear_optics}  (see also (\ref{I.1})):
\[
\hat{n}=1+\frac{2\pi\rho}{k_z^2}(\alpha+\beta J\vec{\sigma}\vec{p})=1+\frac{2\pi\rho}{k_z^2}\hat{f}(0)\,.
\]
If $k_z$ decreases and the wavelength increases, then (with nuclei of different kinds present in the target) the following expression is true for $\hat{n}$ 
\begin{equation}
	\label{6.13a}
	\hat{n}^2=1+\frac{4\pi}{k_z^2}\sum\limits_i\rho_i\hat{f}_i(0)\,.
\end{equation}
The difference
\begin{equation}
	\label{6.13b}
	\Delta=n_{+} - n_{-} = \frac{2\pi\rho}{k_{z}^{2}} (f_{+}-f_{-})=\frac{4\pi\rho}{k_{z}^{2}}\beta J_{p}
\end{equation}
is determined by the difference of the corresponding coherent
scattering amplitudes and is nonzero only in a polarized medium.

Thus, in a polarized nuclear target, neutrons have two indices of refraction.

Let now neutrons with the polarization vector oriented at a certain
angle to the direction of $\vec{p}$ be incident on a polarized medium.
This neutron state can be considered as a superposition of
two states with polarizations along and opposite to the direction of vector $\vec{p}$.
The wave function of a particle before entering the target has the form
\begin{equation}
	\label{6.14} \psi(\vec{r})= e ^{i\vec{k}\vec{r}}\chi_{n},
	\chi_{n}=\begin{pmatrix}
		c_{1}\\
		c_{2}
	\end{pmatrix}\,,
\end{equation}
or
\begin{equation}
	\label{6.15} \psi(\vec{r})=c_{1} e
	^{i\vec{k}\vec{r}}\begin{pmatrix}
		1\\
		0
	\end{pmatrix}+ c_{2}e ^{i\vec{k}\vec{r}}\begin{pmatrix}
		0\\
		1
	\end{pmatrix}\,,
\end{equation}

To be more specific, let us study the case of refraction by the target whose nuclei
have the polarization vector $\vec{p}$ directed perpendicular to its surface. Choose the
direction  $\vec{p}$  as the $z$-axis. The state of the type
$\left(\begin{smallmatrix}
	1\\
	0
\end{smallmatrix}\right)$  has the refractive index $n_{+}$, while the state of the type $\left(\begin{smallmatrix}
	0\\
	1
\end{smallmatrix}\right)$  has the refractive index $n_{-}$.
If a neutron with spin parallel to the polarization vector $\vec{p}$ (spin state $\left(\begin{smallmatrix}
	1\\
	0
\end{smallmatrix}\right)$) were incident on the target, its motion in matter would be described by the wave function $\psi_{+}(r)=e^{i\vec{k}_{\perp}\vec{r}_{\perp}}e^{ik_{z}n_{+}z}\left(\begin{smallmatrix}
	1\\
	0
\end{smallmatrix}\right)$ (in this representation we neglect the influence of mirror reflection on the wave function of the neutron passing through matter, which is possible as $|n_{\pm}-1|\ll 1$ in a wide range of angles of incidence of thermal neutrons on the target).

If a neutron with spin antiparallel to $\vec{p}$ (spin state $\left(\begin{smallmatrix}
	0\\
	1
\end{smallmatrix}\right)$) is incident on the target, then in matter it is described by the wave function ${\psi_{-}(r)=e^{i\vec{k}_{\perp}\vec{r}_{\perp}} e^{ik_{z}n_{-}z}\left(\begin{smallmatrix}
		0\\
		1
	\end{smallmatrix}\right)}$. If a neutron with an arbitrary spin direction falls on the target, its wave function [see (\ref{6.15})] is the superposition of states $\left(\begin{smallmatrix}
	1\\
	0
\end{smallmatrix}\right)$ and $\left(\begin{smallmatrix}
	0\\
	1
\end{smallmatrix}\right)$. As a consequence, the wave function of neutrons in matter is also the superposition of these states, and can be written as
\begin{equation}
	\label{6.16} \psi(\vec{r})=\begin{pmatrix}
		c_{1}& \psi_{+}& (\vec{r})\\
		c_{2}& \psi_{-}& (\vec{r})
	\end{pmatrix}= c_{1}e ^{i\vec{k}_{\perp}\vec{r}_{\perp}}e^{ik_{z}n_{+}z}\begin{pmatrix}
		1\\
		0
	\end{pmatrix}+ c_{2}e ^{i\vec{k}_{\perp}\vec{r}_{\perp}}e^{ik_{z}n_{-}z}\begin{pmatrix}
		0\\
		1
	\end{pmatrix}\,.
\end{equation}

Now let us consider how the polarization of neutrons  changes as they  penetrate into the
interior of the target (with the growth of the target thickness). Suppose we have a detector
that transmits the particles with spin polarized along a certain direction in the detector
(the axis of the detector) and absorbs the particles with the opposite spin direction.
Such a detector is the analog of the Nicol prism \cite{vfel_Born} used in optics for analyzing
the polarization of light.
When polarized light is incident on the Nicol prism, one
component of the light polarization passes through it, while the component orthogonal
to the axis of the Nicol prism is absorbed.
In the case of neutrons, a target with polarized nuclei may act as a detector.
As the scattering cross section of a polarized neutron depends on whether the
neutron spin is oriented along the direction of the polarization vector of the
nucleus or opposite to it, neutron absorption in the  detector exhibits the same
dependence.
Suppose that the axis of the detector is parallel to the $z$-axis
along which the target nuclei are polarized. In this case the detector analyzes
those components of the neutron spin, which are directed along the $z$-axis and opposite to it.
From (\ref{6.16}) follows that the probability amplitude $A^{(+)}$ of finding the neutron with spin state $\left(
\begin{smallmatrix}
	1 \\
	0 \\
\end{smallmatrix}
\right)$,
i.e., of finding the neutron polarized parallel to the $z$-axis, is given by the expression
\[
A^{(+)}= (1\quad 0)\psi=c_1 e^{i\vec{k}_{\perp}\vec{r}_{\perp}}e^{ik_{z}n_{+}z}\,,
\] 
thus, the probability to find the neutron polarized parallel to the $z$-axis reads as follows:
\[
P^{(+)}_{z}=|(1\quad 0)\psi|^{2}=|c_{1}^{2}|e^{-2k_{z}\texttt{Im}n_{+}z}=|c_{1}|^{2}e^{-\rho\sigma_{+}z}\,.
\]
Similarly, the probability $P^{(-)}_{z}$ of finding the neutron polarized opposite to the $z$-axis is
\[
P_{z}^{(-)}=|(0\quad 1)\psi|^{2}=|c_{2}|^{2}e^{-2k_{z}\texttt{Im}n_{-}z}=|c_{2}|^{2}e^{-\rho\sigma_{-}z}\,,
\]
where $\sigma_{\pm}$ is the total cross section of scattering by the nucleus of the neutron polarized parallel (antiparallel) to the polarization vector of the nucleus.

Since in polarized matter $\texttt{Im}n_{+}\neq \texttt{Im} n_{-}$
($\sigma_{+}\neq \sigma_{-}$), one of the components of the
neutron spin wave function decays faster and at some depth the
rapidly damped component may be neglected. The beam will appear
polarized along the $z$-axis (along the direction of the
polarization vector of nuclei).

Let us now rotate the detector so that its polarization axis becomes perpendicular to the direction of the target polarization. Choose the direction of the polarization axis of the detector as the $x$-axis.

Now the detector analyzes
those components of the neutron spin, which are directed along and opposite to the $x$-axis.

To determine the probability $P_{x}^{(\pm)}$ of finding the component of the neutron spin parallel (antiparallel) to the direction of the $x$-axis, one should expand the wave function (\ref{6.16}) in terms of the spin wave functions $\chi_{x}^{\pm}$, which are the eigenfunctions of operator $\hat{S}_{x}$ of the spin projection onto the $x$-axis. They have the form
\[
\chi_{x}^{\pm}=\frac{1}{\sqrt{2}}
\begin{pmatrix}
	\phantom{+}1 \\
	\pm 1 \\
\end{pmatrix}\,.
\]
As a result, we find that the probabilities $P_{x}^{(\pm)}$ of neutron spin polarization along and opposite to the $x$-axis change with $z$ as \cite{24}:
\begin{eqnarray}
	\label{ins}
	P_{x}^{(+)}=\frac{1}{2}\left\{|c_{2}|^{2}e^{-2k_{z}\texttt{Im}n_{+}z}+|c_{2}|^{2}e^{-2k_{z}\texttt{Im}n_{-}z}\right.\nonumber\\
	\left.+2|c_{1}|c_{2}|e^{-k_{z}\texttt{Im}(n_{+}+n_{-})z}\cos\left[k_{z}\texttt{Re}(n_{+}-n_{-})z+\delta\right]\right\}\,,\nonumber\\
	P_{x}^{(-)}=\frac{1}{2}\left\{|c_{1}|^{2}e^{-2k_{z}\texttt{Im}n_{+}z}+|c_{2}|^{2}e^{-2k_{z}\texttt{Im}n_{-}z}\right.\nonumber\\
	\left.-2|c_{1}|c_{2}|e^{-k_{z}\texttt{Im}(n_{+}+n_{-})z}\cos\left[k_{z}\texttt{Re}(n_{+}-n_{-})z+\delta\right]\right\}\,,
\end{eqnarray}
where $\delta=\delta_{1}-\delta_{2}$ is the difference of the initial phases of states with neutron spin polarization  along and opposite to the $z$-axis ($c_{1}=|c_{1}|e^{i\delta_{1}}$; $c_{2}=|c_{2}|e^{i\delta_{2}}$).

If at $z=0$, the neutron is polarized along $x$, i.e.,
\[
c_{1}=c_{2}=\frac{1}{\sqrt{2}}\,, \quad \delta=0\,,
\]
then with growing $z$ the polarization opposite to $x$ appears and further change of the polarization  acquires the character of oscillations.

As the neutrons pass through the target, one of the components decays more strongly and the neutron beam eventually becomes polarized along or opposite to the $z$-axis. When a beam polarized along the $z$-axis is incident onto the target, no oscillations emerge: only damping occurs.

Using (\ref{6.16}), one can find the neutron polarization
vector
\begin{equation}
	\label{6.17} \vec{p}_{n}= \frac{\langle\psi|\vec{\sigma}|\psi\rangle}{\langle\psi|\psi\rangle}\,.
\end{equation}
As a result,
\begin{eqnarray}
	\label{6.18} p_{nx}&=&
	2\texttt{Re} c_{1}^{*}c_{2}\psi_{+}^{*}\psi_{-}\langle\psi|\psi\rangle^{-1}\,
	\nonumber \\
	p_{ny}&=&2 \texttt{Im} c_{1}^{*}c_{2}\psi_{+}^{*}\psi_{-}\langle\psi|\psi\rangle^{-1}\,,
	\nonumber \\
	p_{nz}&=&(|c_{1}\psi_{+}|^{2}-|c_{2}\psi_{-}|^{2})\langle\psi|\psi\rangle^{-1}\,.
\end{eqnarray}

Suppose that neutron spin in a vacuum is directed perpendicular to the
polarization vector of the nuclei. Choose this direction  as the
$x$-axis. In this case
\[
c_{1}=c_{2}=1/\sqrt{2}\,.
\]
Using relations (\ref{6.18}), we obtain
\begin{eqnarray}
	\label{6.19}
	p_{nx}&=& \cos[k_{z}\texttt{Re}(n_{+}-n_{-})z]e^{-k_{z}\texttt{Im}(n_{+}+n_{-})z}\langle\psi|\psi\rangle^{-1}\,, \nonumber \\
	p_{ny}&=&-\sin[k_{z}\texttt{Re}(n_{+}-n_{-})z]e^{-k_{z}\texttt{Im}(n_{+}+n_{-})z}\langle\psi|\psi\rangle^{-1}\,, \nonumber \\
	p_{nz}&=&\frac{1}{2}(e^{-2k_{z}\texttt{Im}
		n_{+}z}-e^{2k_{z}\texttt{Im}n_{-}z})\langle\psi|\psi\rangle^{-1} \,, \nonumber\\
	p_{nx}^{2}&+&p_{ny}^{2}~+~p_{nz}^{2}~=~1\,.
\end{eqnarray}

According to (\ref{6.19}), as the neutron penetrates into the interior of the target, its
polarization vector rotates about the polarization vector
of nuclei through the angle
\begin{equation}
	\label{6.20} \theta=
	k_{z}\texttt{Re}(n_{+}-n_{-})z=\frac{2\pi\rho}{k_{z}}\texttt{Re}(f_{+}-f_{-})z\,,
\end{equation}
which agrees perfectly with the result obtained above (see
(\ref{6.9})).

At the same time, as the neutrons pass through polarized matter, the transverse components $p_{nx}$ and $p_{ny}$ of the polarization vector  decay because neutron absorption depends on spin orientation, and finally the beam appears to be polarized along or opposite to the $z$-axis.

Thus, the dependence of the  absorption of neutrons in the target on the orientation of their spin results in the fact that the polarization vector $\vec{p}_{n}$ (recall that $|\vec{p}_{n}|=1$) not only rotates  about the $z$-axis (about the direction of the polarization vector of nuclei) but also undergoes additional rotation in the direction of the $z$-axis (the end point of the polarization vector moves along the unit sphere).

If the dependence of absorption on spin orientation can be neglected,
the polarization vector rotates about the direction of the polarization vector of nuclei only in the $(x, y)$ plane.

In terms of kinematics, this phenomenon is analogous to the light polarization
plane rotation in a magnetic field (the Faraday effect), while spin oscillations along and opposite to the direction of the $x$-axis  are analogous to the transitions $K^{0}\rightleftarrows \bar{K}^{0}$ occurring in  regeneration of neutral $K$-mesons (see e. g. \cite{rins_85}).

The conclusion about neutron spin rotation in a polarized target can
be derived from other considerations.

Since in a polarized nuclear target a neutron wave has two refractive indices, according to (\ref{6.3}), in
such a target it has two possible potential interaction energies
$U_{\pm}$:
\begin{equation}
	\label{6.26}
	U_{\pm}=\frac{\hbar^{2}k_{z}^{2}}{2m}(1-n_{\pm}^{2})=-\frac{2\pi\hbar^{2}}{m}\rho
	f_{\pm}(0)
\end{equation}
or in operator form
\begin{equation}
	\label{6.27}
	\hat{U}=\frac{\hbar^2k_z^2}{2m}(1-\hat{n}^2)=-\frac{2\pi\hbar^{2}\rho}{m}\hat{f}(0)=-\frac{2\pi\hbar^{2}\rho}{m}(\alpha+\beta J\vec{\sigma}
	\vec{p})\,.
\end{equation}

Recall now that the expression for the energy $W$ of interaction  between neutrons and a magnetic field
$\vec{B}$ \cite{Landau_3} is similar in form to the second term in (\ref{6.27}), i.e.,  $W=-\mu\vec{\sigma}\vec{B}$ ($\mu$ is the
neutron magnetic moment). In this case the interaction energy $W_{+}$ of
a particle with spin parallel to $\vec{B}$ is given by a
well-known expression $W_{+} =-\mu B$; the analogous quantity for a
particle with the opposite spin direction, by the expression
$W_{-} = \mu B$. The nonzero difference $W_{+}-W_{-} = - 2\mu B$ leads to the Larmor precession with
frequency $\omega_{L}=2\mu B/\hbar$ of neutron spin in a magnetic field $\vec{B}$.

Knowing the frequency $\omega_{L}$, one can find the magnetic field ${B=\hbar\omega_{L}/2\mu}$.

In perfect analogy with spin behavior in a magnetic field, the
non-zero difference $U_{+}-U_{-}$ leads to the neutron
spin precession about the direction of the nuclear polarization vector with the frequency \cite{24}:

\begin{equation}
	\label{6.28}
	\omega=\left|\texttt{Re}\frac{U_{+}-U_{-}}{\hbar}\right|=\frac{4\pi\hbar\rho}{m}\texttt{Re}\beta
	J p\,.
\end{equation}
During time $t$, spin will rotate by the angle $\vartheta=\omega t$. If a
polarized target has a layer of thickness $l$, the
neutron, which enters the area occupied by the target, passes through the layer in time $t=l/v_{z}$.

Therefore its spin rotates through the angle
\[
\vartheta=\omega\frac{l}{v_{z}}= \frac{m\omega l}{\hbar
	k_{z}}=\frac{2\pi\rho}{k_{z}}\texttt{Re} \beta Jpl\,,
\]
which is in complete agreement with the result obtained above.

Continuing by analogy with the magnetic field, to describe neutron
spin precession caused by nuclear
interaction (hereinafter referred to as "neutron nuclear precession"), it is
natural to introduce the effective magnetic field
\begin{equation}
	\label{6.29} B_{\mathrm{eff}} = \hbar\omega/(2\mu)\,,
\end{equation}
which causes precession with the same frequency $\omega$ as an
ordinary magnetic field.

As a result, we can write
\begin{equation}
	\label{ins_no6.1} \hat{U}=u_0-\mu\vec{\sigma}\vec{B}_{\mathrm{eff}}\,,
\end{equation}
where
\begin{equation}
	\label{ins_no6.2} u_0=-\frac{2\pi\hbar^2\rho}{m}\alpha,\qquad
	\vec{B}_{\mathrm{eff}}=\frac{2\pi\hbar^2\rho}{\mu m}\beta J\vec{p}\,.
\end{equation}

Note that in the  range of neutron energies where the scattering
amplitude is constant,  frequency $\omega$ is also constant and
characterizes the rotatory power of matter due to nuclear
interaction. This occurs when the neutron energy is small. With
increasing neutron velocity the spin precession frequency becomes
energy--dependent: in particular, it gives a sharp rise near each
of the resonances  and  can reverses sign when the neutron energy
changes near the resonance because the sign of the real part of
the scattering amplitude changes. It would be recalled in this
connection  (see e. g. \cite{22,Landau_3}) that in the vicinity of the
resonance, the scattering amplitude is
\[
f\sim \frac{1}{E-E_{0}-i\Gamma/2}\,,
\]
where $E$ is the particle energy; $E_{0}$ is the resonance energy,
$\Gamma$ is the width of the resonance level.

In view of (\ref{6.29}), at low energies the value of the effective
quasi-magnetic field of nuclear origin is a constant defining the
matter, while at higher energies it is energy--dependent. For a
polarized proton target, for instance, in the case of full
polarization, $\omega= 5\cdot 10^{8}$\,s$^{-1}$,
$B_{\mathrm{eff}}\approx 3\cdot 10^{4}$\,Gs $=3$ T and exceeds by
two orders of magnitude the ordinary magnetic field created by
polarized magnetic moments of protons. Under the same conditions,
for thermal neutrons $v = 2.2 \cdot 10^{5}$\,cm$\cdot$\,s$^{-1}$
and the full  rotation of spin occurs at length $L\approx
10^{-3}$\,cm.

{Nuclear spin precession phenomenon was observed in the experiments carried out by Abragam \cite{Abragam_PRL_1973} and Forte  \cite{Forte_1973,92}.}


\subsection{ Growth of Nuclear Spin Precession Frequency  of Antiprotons (Negative Hyperons) Under Deceleration in
	Matter with Polarized Nuclei }

The progress in development of the Facility for Low-Energy
Antiproton and Ion Research (FLAIR) has spurred the rapid
development of low--energy antiproton physics  \cite{x1,x2}.

The possibility to obtain  polarized antiprotons by a
spin-filtering method \cite{x3} opens up opportunities for
investigating a large number of  spin--dependent fundamental
processes arising when antiprotons pass through matter with
polarized nuclei. In particular, study of the phenomenon of
particle "optical" spin rotation in matter with polarized nuclei
enables investigation of the spin--dependent part of  the forward
scattering amplitude \cite{rins_98,A2,antiproton1,antiproton2}.

For low--energy neutrons, the phenomenon of "optical" spin
rotation (the phenomenon of nuclear precession of the neutron spin
in a nuclear pseudomagnetic field of a polarized target) was
predicted in \cite{24} and experimentally observed in \cite{Abragam_PRL_1973,Forte_1973}.
Based on this phenomenon,  the spin--dependent forward scattering
amplitude of thermal neutrons has been measured \cite{28,27}, and
similar new experiments  are being prepared
\cite{rins_7,rins_11,rins_8,rins_9}.

In contrast to neutrons, a charged particle undergoes Coulomb
interaction with the atoms of matter, which causes multiple
scattering and rapid deceleration of the charged particle due to
ionization energy losses. With decreasing  particle energy, the
influence of the Coulomb interaction on particle scattering by the
nucleus grows in significance. In particular, when the energy of a
positively charged particle diminishes, the Coulomb repulsion
suppresses nuclear interaction between the incident particle and
the target nucleus and hence, the phenomenon of spin rotation due
to nuclear interaction. Conversely, a negatively charged particle
(antiproton, hyperon) is attracted to the nucleus and, as a
result, participates in nuclear interaction even at low energies.
As a consequence of this, spin rotation of a negatively charged
particle in polarized matter does not disappear at very low
energies either.

The present section considers the influence of the Coulomb
interaction on the phenomenon of "optical" spin rotation of
negatively charged particles moving in matter with polarized
nuclei. It is shown that because the density of the antiproton
(negative hyperon) wave function on the nucleus increases, the
spin precession frequency grows as the particle decelerates. As a
result, spin rotation of negatively charged particles becomes
observable despite their rapid deceleration. This provides
information about the spin--dependent part of  the scattering
amplitude in the range of low energies, where scattering
experiments are practically impossible to perform.

\bigskip

\subsection{Forward Scattering Amplitude of Negatively Charged Particles}

According to \cite{rins_98,antiproton1,antiproton2,A2,24},  the spin rotation frequency
$\Omega_{\mathrm{nuc}}$ of a nonrelativistic particle passing
through a target with polarized nuclei can be expressed as
\begin{equation}
	\label{a1} \Omega_{\mathrm{nuc}}=\frac{\Delta
		\texttt{Re}U_{\mathrm{eff}}}{\hbar}=\frac{2\pi\hbar}{m}N \,P_t
	\Delta \texttt{Re}f(0)\,,
\end{equation}
where $\Delta \texttt{Re}U_{\mathrm{eff}}$ is the difference
between the real parts of the effective potential energy of
interaction between the particle and the polarized target for
oppositely directed particle spins, $m$ is the mass of the
particle, $N$ is the number of nuclei in 1 cm$^3$, $P_t$ is the
degree of polarization of the target nuclei, and $\Delta
\texttt{Re} f(0)$ is the difference between the real parts of the
amplitudes of coherent forward scattering for particles with
oppositely directed spins.

The scattering amplitude $f(0)$ is related to the $\mathcal{T}$-matrix  as
follows (see, e.g. \cite{Goldberger,Davydov}):
\begin{equation}
	\label{a2}
	f(0)=-\frac{m}{2\pi\hbar^2}\langle\Phi_a|\mathcal{T}|\Phi_a\rangle\,,
\end{equation}
where $|\Phi_a\rangle$ is the wave function describing the initial
state of the system "incident particle--atom (nucleus)". The wave
function $|\Phi_a\rangle$ is the eigenfunction of the Hamiltonian
$\hat{H}_0=H_p(\vec{r}_p)+H_A(\vec{\xi}\,,
\vec{r}_{\mathrm{nuc}})$, i.e.,
$\hat{H}_0|\Phi_a\rangle=E_a|\Phi_a\rangle$; $H_p(\vec{r}_p)$ is
the Hamiltonian of the particle incident onto the target;
$\vec{r}_p$ is the particle coordinate; $H_A(\vec{\xi}\,,
\vec{r}_{\mathrm{nuc}})$ is the atomic Hamiltonian; $\vec{\xi}$ is
the set of coordinates of the atomic electron;
$\vec{r}_{\mathrm{nuc}}$ is the set of coordinates describing the
atomic nuclei.

The Hamiltonian $H$ describing the particle--nucleus  interaction
can be written as:
\begin{equation}
	\label{a3} H=H_0+V_{\mathrm{Coul}} (\vec{r}_p\,,\vec{\xi}\,,
	\vec{r}_{\mathrm{nuc}})+V_{\mathrm{nuc}}(\vec{r}_p\,,
	\vec{r}_{\mathrm{nuc}})\,,
\end{equation}
where $V_{\mathrm{Coul}}$ is the energy of  Coulomb interaction
between the particle and the atom, and $V_{\mathrm{nuc}}$ is the
energy of nuclear interaction between the particle and the atomic
nucleus.

According to the quantum theory of reactions \cite{Goldberger,Davydov}, in the
case of two interactions, the matrix element of the operator $\mathcal{T}$,
which  describes the system transition from the initial state
$|\Phi_a\rangle$ into the final state $|\Phi_b\rangle$, can be
presented as a sum of two terms:
\begin{equation}
	\label{a9}
	\mathcal{T}_{ba}=\mathcal{T}_{ba}^{\mathrm{Coul}}+\mathcal{T}_{ba}^{\mathrm{nuc\,Coul}}=
	\langle\Phi_b|\mathcal{T}_{\mathrm{Coul}}|\Phi_a\rangle+\langle\varphi_b^{(-)}|\mathcal{T}_{\mathrm{nuc}}|\varphi_a^{(+)}\rangle\,,
\end{equation}
where the first term, $\mathcal{T}_{ba}^{\mathrm{Coul}}$, describes the
contribution to the T-matrix that comes from the Coulomb
scattering alone, the operator
\begin{equation}
	\label{a10}
	\mathcal{T}_{\mathrm{Coul}}=V_{\mathrm{Coul}}+V_{\mathrm{Coul}}(E_a-H_0+i\varepsilon)^{-1}\mathcal{T}_{\mathrm{Coul}}\,,
\end{equation}
and the second term describes the contribution to the T-matrix
that comes from nuclear scattering and accounts for the distortion
of waves incident onto the nucleus, $\varphi^{(\pm)}$, which is
caused by the Coulomb interaction. The operator
\begin{eqnarray}
	\label{a11}
	\mathcal{T}_{\mathrm{nuc}}&=&V_{\mathrm{nuc}}+V_{\mathrm{nuc}}(E_a-H_0- V_{\mathrm{Coul}}+i\varepsilon)^{-1}\mathcal{T}_{\mathrm{nuc}}\nonumber\\
	&=& V_{\mathrm{nuc}}+V_{\mathrm{nuc}}(E_a-H_0-
	V_{\mathrm{Coul}}-V_{\mathrm{nuc}}+i\varepsilon)^{-1}V_{\mathrm{nuc}}\,,
\end{eqnarray}
and  the wave functions $\varphi_{a}^{(\pm)}$ describe the
interaction between particles and atoms via the Coulomb
interaction alone ($V_{\mathrm{nuc}}=0$) \cite{Goldberger,Davydov,Landau_3}:
\begin{equation}
	\label{a6} \varphi_{a}^{(\pm)}=\Phi_a+(E_a-H_0\pm
	i\varepsilon)^{-1}V_{\mathrm{Coul}}\, \varphi_a^{(\pm)}\,,
\end{equation}
the wave function $\varphi_a^{(+)}$ at large distances has the
asymtotics of a diverging spherical wave, and the wave function
$\varphi_a^{(-)}$ at large distances has the asymtotics of a
converging spherical wave \cite{Goldberger,Davydov,Landau_3}.

Let us give a more detailed consideration of the matrix element
$\langle\varphi_b^{(-)}|\mathcal{T}_{\mathrm{nuc}}|\varphi_a^{(+)}\rangle$.
Because nuclear forces are short--range, the radius  of the domain
of  integration in this matrix element is of the order of the
nuclear radius (of the order of the radius of action of nuclear
forces in the case of the proton). The Coulomb interaction,
$V_{\mathrm{Coul}}$, in this domain is noticeably smaller than the
energy of nuclear interaction, $V_{\mathrm{nuc}}$. We can
therefore neglect the Coulomb energy in the first approximation in
the denominator of Eq. (\ref{a11}), as compared to
$V_{\mathrm{nuc}}$.


As a result, the operator $\mathcal{T}_{\mathrm{nuc}}$ is reduced to the
operator describing a purely nuclear interaction between the
incident particle and the nucleus. The effect of Coulomb forces on
nuclear interaction is described by wave functions
$\varphi_{ba}^{(\pm)}$  (distorted--wave approximation \cite{Davydov}).

In the range of antiproton energies of hundreds of
kiloelectronvolts and less, the de Broglie wavelength for
antiprotons is larger than the nuclear radius. Therefore, in Eq.
(\ref{a9}) for $\mathcal{T}_{ba}^{\mathrm{nuc}}$, one can remove the wave
functions $\varphi_{a(b)}^{\pm)}$ outside the sign of integration
over the coordinate of the antiproton center of mass, $\vec{R}_p$,
at the location point of the nuclear center of mass,
$\vec{R}_{\mathrm{nuc}}$. As a result, one may write the following
relationship:
\begin{equation}
	\label{a12}
	\mathcal{T}_{ba}^{\mathrm{nuc\,Coul}}=g_{ba}\mathcal{T}_{ba}^{\mathrm{nuc}}=
	\langle\varphi_b^{(-)}(\vec{R}_p=\vec{R}_{\mathrm{nuc}})|\varphi_a^{(+)}
	(\vec{R}_p=\vec{R}_{\mathrm{nuc}})\rangle \mathcal{T}_{ba}^{\mathrm{nuc}}\,,
\end{equation}
where $\mathcal{T}_{ba}^{\mathrm{nuc}}$ is the matrix element describing a
purely nuclear interaction (in the absence of Coulomb interaction)
between the incident particle and the nucleus. The factor
$g_{ba}=\langle\varphi_b^{(-)}(\vec{R}_p=\vec{R}_{\mathrm{nuc}})|\varphi_a^{(+)}
(\vec{R}_p=\vec{R}_{\mathrm{nuc}})\rangle$ appearing in 
	(\ref{a12}) defines the probability to find the antiproton (the
negative hyperon, e.g. $\Omega^{-}\, , \Sigma^{-}$) at the
location of the nucleus.

From (\ref{a9})-(\ref{a12}) one can derive the following
expression for the amplitude of coherent elastic zero--angle
scattering:
\begin{equation}
	\label{a13}
	f(0)=-\frac{m}{2\pi\hbar^2}g_{aa}\mathcal{T}_{aa}^{\mathrm{nuc}}=g_{aa}f_{\mathrm{nuc}}{(0)}\,,
\end{equation}
where $f_{\mathrm{nuc}}{(0)}$ is the amplitude of particle
scattering by the nucleus in the absence of Coulomb interaction,
and
$g_{aa}=\langle\varphi_b^{(-)}(\vec{R}_p=\vec{R}_{\mathrm{nuc}})|\varphi_a^{(+)}
(\vec{R}_p=\vec{R}_{\mathrm{nuc}})\rangle$ is the probability to
find the particle incident onto the nucleus at the location of the
nucleus.

Thus, the Coulomb interaction leads to a change in the value of
the amplitude of nuclear forward scattering. Let us estimate the
magnitude of this change.

According to \cite{Landau_3}, when a particle moves in the Coulomb field,
the probability $g_{aa}$ can be written in the form:
\begin{itemize}
	\item for the case of repulsion, i.e., scattering of similarly
	charged particles
	\begin{equation}
		\label{a14}
		g_{aa}^{\mathrm{rep}}=\frac{2\pi}{\kappa(e^{\frac{2\pi}{\kappa}}-1)}\,,\qquad
		\kappa=\frac{v}{Z\alpha c}\,,
	\end{equation}
	where $v$ is the particle velocity, $Z$ is the charge of the
	nucleus, $\alpha$ is the fine structure constant, $c$ is the speed
	of light; \item for the case of attraction
	\begin{equation}
		\label{a15}
		g_{aa}^{\mathrm{att}}=\frac{2\pi}{\kappa(1-e^{-\frac{2\pi}{\kappa}})}\,.
	\end{equation}
\end{itemize}

With decreasing particle energy (velocity), $\kappa$ diminishes,
and for such values of $\kappa$ when $\frac{2\pi}{\kappa}\geq1$,
one can write
\begin{equation}
	\label{a16} g_{aa}^{\mathrm{rep}}=\frac{2\pi\alpha Z
		c}{v}\,e^{-\frac{2\pi\alpha Z c}{v}}\,,\qquad
	g_{aa}^{\mathrm{att}}=\frac{2\pi\alpha Z c}{v}\,.
\end{equation}

According to Eq. (\ref{a16}), with decreasing energy of positively
charged particles, the amplitude $f(0)$ diminishes rapidly because
of repulsion. For negatively charged particles, the amplitude
grows with decreasing particle energy (velocity).

These results for the amplitude $f(0)$ generalize a similar,
well-known  relationship for taking account of the Coulomb
interaction effect on the cross section of inelastic processes,
$\sigma_r$, \cite{Landau_3}.

So in the range of low energies, the amplitude $f(0)$ of
antiproton (negative hyperon) scattering by the nucleus can be
presented in the form:
\begin{equation}
	\label{a18} f(0)=\frac{2\pi \alpha Z c}{v}
	f_{\mathrm{nuc}}=\frac{2\pi\alpha Z c}{v}\texttt{Re}\,
	f_{\mathrm{nuc}}+\frac{i}{2}\frac{\alpha
		Z}{\lambda_c}\sigma_{\mathrm{tot}}\,,
\end{equation}
where $\lambda_c=\frac{\hbar}{mc}$, $\sigma_{\mathrm{tot}}$ is the
total cross section of nuclear interaction between the particle
and the scatterer. In deriving Eq. (\ref{a18}), the optical
theorem ${\texttt{Im}\, f(0)=\frac{k}{4\pi}\sigma_{\mathrm{tot}}}$
was applied, where $k$ is the wave number of the incident
particle.

\bigskip

\subsection{Effective Potential Energy of Negatively Charged
	Particles in Matter}

Using the amplitude $f(0)$, one can write the expression for the
refractive index $\hat{n}$ of a spin particle in matter with
polarized nuclei  (see (\ref{I.1})), as well as the expression for the effective
potential energy $\hat{U}_{\mathrm{eff}}$ of interaction between
this particle and matter \cite{rins_98,antiproton1,antiproton2,A2,24}:
\begin{equation}
	\label{a19} \hat{n}^2=1+\frac{4\pi\,
		N}{k^2}\hat{f}(0)\qquad\mbox{and}\qquad
	\hat{U}_{\mathrm{eff}}=-\frac{2\pi\hbar^2}{m} N\hat{f}(0)\,,
\end{equation}
where $\hat{f}(0)=\frac{2\pi\alpha Z
	c}{v}\hat{f}_{\mathrm{nuc}}(0)$ is the amplitude of coherent
elastic forward scattering, which is the operator in the particle
spin space.

The amplitude $\hat{f}(0)$ depends on the vector polarization
$\vec{P}_t$ of the target nuclei and can be presented in the form:
\begin{equation}
	\label{a20} \hat{f}(0)=A_0+A_1(\hat{\vec{S}}\, \vec{P}_t)+
	A_2(\hat{\vec{S}}\, \vec{e})(\vec{e}\, \vec{P}_t)\,,
\end{equation}
where $A_0$ is the scattering amplitude independent  of the
incident particle spin, $\hat{\vec{S}}$ is the particle spin
operator, and  $\vec{e}$ is the unit vector in the direction of
the particle momentum. If the spin of the target nuclei $I\geq1$,
then the addition depending on the target tensor polarization also
appears \cite{rins_98,A2}.

Correspondingly, the effective potential energy
$\hat{U}_{\mathrm{eff}}$ of particle interaction with polarized
matter looks like
\begin{equation}
	\label{a21} \hat{U}_{\mathrm{eff}}=-\frac{2\pi\hbar^2}{m}
	N(A_0+A_1(\hat{\vec{S}}\, \vec{P}_t)+ A_2(\hat{\vec{S}}\,
	\vec{e})(\vec{e}\, \vec{P}_t))\,.
\end{equation}
Expression (\ref{a21}) can be written as
\begin{equation}
	\label{a22} \hat{U}_{\mathrm{eff}}={U}_{\mathrm{eff}}+
	\hat{V}_{\mathrm{eff}} (\vec{P}_t)\,,
\end{equation}
where
\begin{equation}
	\label{a23} {U}_{\mathrm{eff}}=-\frac{2\pi\hbar^2}{m} N
	A_0\,,\qquad
	\hat{V}_{\mathrm{eff}}(\vec{P}_t)=-\vec{\mu}\vec{G}=-\frac{\mu}{S}(\hat{\vec{S}}\,
	\vec{G})\,,
\end{equation}
with $\mu$ being the particle magnetic moment,
\begin{equation}
	\label{a25} \vec{G}=\frac{2\pi\hbar^2 S}{m\mu}
	N(A_1\vec{P}_t+A_2\vec{e}(\vec{e}\, \vec{P}_t))\,.
\end{equation}

Recall now that the energy of interaction between the magnetic
moment $\mu$ and the magnetic field $\vec{B}$ is as follows:
\begin{equation}
	\label{a26} V_{\mathrm{mag}}= -(\vec{\mu}\,
	\vec{B})=-\frac{\mu}{S}(\hat{\vec{S}}\vec{B})\,.
\end{equation}
Expressions (\ref{a23}) and (\ref{a25}) are identical. Therefore,
$\vec{G}$ can be interpreted as the effective pseudomagnetic field
acting on the spin of the particle moving in matter with polarized
nuclei and appears due to nuclear interaction between the incident
particles and the scatterers. Similarly to particle spin
precession in an external magnetic field, particle spin precesses
in field $\vec{G}$. This phenomenon was called the nuclear
precession of the particle spin, first described for slow neutrons
in \cite{24} and then observed in \cite{Abragam_PRL_1973,Forte_1973}.

From Eq. (\ref{a18}) follows that in the range of low energies,
$\hat{U}_{\mathrm{eff}}$ can be presented in the form:
\begin{equation}
	\label{a27} \hat{U}_{\mathrm{eff}}= \frac{2\pi \alpha Z
		c}{v}\hat{U}_{\mathrm{eff}}^{\mathrm{nuc}}\,,
\end{equation}
where $\hat{U}_{\mathrm{eff}}^{\mathrm{nuc}}$ has the same form as
that in Eq. (\ref{a21}) with the amplitudes $A_0$, $A_1$, and
$A_2$ replaced by $A_0^{\mathrm{nuc}}$, $A_1^{\mathrm{nuc}}$ and
$A_2^{\mathrm{nuc}}$ calculated ignoring the Coulomb interaction.

According to Eq. (\ref{a27}), with decreasing particle velocity,
$\hat{U}_{\mathrm{eff}}$ grows as well as the field $\vec{G}$, and
particle spin precession in this field is
$\Omega_{\mathrm{pr}}\sim \texttt{Re}\,G\sim\frac{1}{v}$.

Let us
take a somewhat different view of this issue. According to Eqs.
(\ref{a21}), (\ref{a23}), and (\ref{a27}), $U_{\mathrm{eff}}$
depends on the orientation of vectors $\vec{e}$ and $\vec{P}_t$.
As follows from Eq. (\ref{a21}), two simpler cases can be
distinguished: $\vec{e}\perp\vec{P}_t$ and
$\vec{e}\parallel\vec{P}_t$. Accordingly, we have two cases for
the effective interaction energy: $\hat{U}_{\mathrm{eff}}^{\perp}$
and $\hat{U}_{\mathrm{eff}}^{\parallel}$:
\begin{equation}
	\label{a29}
	\hat{U}_{\mathrm{eff}}^{\perp}=-\frac{2\pi\hbar^2}{m}\,
	N(A_0+A_1(\hat{\vec{S}} \vec{P}_t))\,,
\end{equation}
\begin{equation}
	\label{a30}
	\hat{U}_{\mathrm{eff}}^{\parallel}=-\frac{2\pi\hbar^2}{m}\,
	N(A_0+(A_1+A_2)(\hat{\vec{S}}\vec{P}_t))\,,
\end{equation}

Let us choose the quantization axis to be parallel to the
polarization vector $\vec{P}_t$. Hence for particle states with
magnetic quantum number $M_s$, one can write the below expressions
for $U_{\mathrm{eff}}^{\perp}(M_s)$ and
$U_{\mathrm{eff}}^{\parallel}(M_s)$, which follow from Eq.
(\ref{a29}):
\begin{equation}
	\label{a31} U_{\mathrm{eff}}^{\perp}(M_s)=-\frac{2\pi\hbar^2}{m}
	N(A_0+A_1M_s P_t)\,,
\end{equation}
\begin{equation}
	\label{a31b}
	U_{\mathrm{eff}}^{\parallel}(M_s)=-\frac{2\pi\hbar^2}{m}N(A_0+(A_1+A_2)M_sP_t)\,.
\end{equation}

In the case of antiprotons ($M_s=\pm\frac{1}{2}$), from Eqs.
(\ref{a31}), (\ref{a31b}) one can obtain two values of the
effective potential energy, depending on the antiproton spin
orientation. The difference of the real parts of these energies
defines the spin precession frequency of the antiproton in matter
with polarized nuclei:
\begin{equation}
	\label{a34} \Omega^{\perp}_{\mathrm{pr}} =
	\frac{\texttt{Re}(U_{\mathrm{eff}}^{\perp}(+\frac{1}{2})-U^{\perp}_{\mathrm{eff}}(-\frac{1}{2}))}{\hbar}
	=-\frac{2\pi\hbar}{m}\cdot\frac{2\pi \alpha Z c}{v} N P_t\,
	\texttt{Re}\, A_{1\,\mathrm{nuc}}\,,
\end{equation}
\begin{equation}
	\label{a34b}
	\Omega_{\mathrm{pr}}^{\parallel}=\frac{\texttt{Re}(U^{\parallel}_{\mathrm{eff}}(+\frac{1}{2})-U_{\mathrm{eff}}^{\parallel}(-\frac{1}{2}))}{\hbar}
	=-\frac{2\pi\hbar}{m}\cdot\frac{2\pi\alpha Z c}{v}
	NP_t\,\texttt{Re}(A_{1\mathrm{nuc}}+A_{2\,\mathrm{nuc}})
\end{equation}

\bigskip

\subsection{Spin Rotation Angle of Low--Energy Antiprotons in
	Polarized Matter}

Let us estimate the magnitude of the effect. Because in the range
of low energies, the de Brogile wavelength of a particle is much
larger than the nuclear radius, in making estimations we shall
concentrate on $S$-scattering (one should bear in mind, though,
that the analysis of the antiproton--proton interaction in
protonium has shown that at low energies,  $P$-waves also
contribute to antiproton--proton interaction \cite{x1}). In
$S$-wave scattering, the amplitude $A_2^{\mathrm{nuc}}$ equals
zero.

The amplitudes $A_0^{\mathrm{nuc}}$ and $A_1^{\mathrm{nuc}}$ can
be expressed in terms of the amplitudes $a^+$ and $a^-$, where
$a^+$ is the scattering amplitude in the state with total momentum
$I+\frac{1}{2}$, and $a^-$ is the same in the state with total
momentum $I-\frac{1}{2}$ ($I$ is the nuclear spin) \cite{Landau_3}:
\begin{equation}
	\label{a36} A_0^{\mathrm{nuc}}=\frac{I+1}{2I+1}a^++\frac{I}{2I+1}
	a^{-}\,,\quad A_1^{\mathrm{nuc}}=\frac{2I}{2I+1}(a^+-a^-)\,.
\end{equation}
As a consequence, one can write the following expression for
$\Omega_{\mathrm{pr}}$:
\begin{equation}
	\label{a37}
	\Omega_{\mathrm{pr}}=-\frac{2\pi\hbar}{m}\cdot\frac{2\pi \alpha Z
		c}{v} \, NP_t\frac{2I}{2I+1} \,\texttt{Re}(a^+-a^-)\,.
\end{equation}
When antiprotons pass through a hydrogen target ($I=\frac{1}{2}$),
we have:
\begin{equation}
	\label{a38} \Omega_{\mathrm{pr}}=\frac{\pi\hbar}{m}\cdot\frac{2\pi
		\alpha  c}{v} \, NP_t\,\texttt{Re}(a^+-a^-)\,.
\end{equation}
The factor $\frac{2\pi\alpha c}{v}$ makes (\ref{a38})
different from the equation for spin precession frequency of slow
neutrons moving in a target with polarized protons.

Recall that in the range of low energies, the amplitudes $a^+$ and
$a^-$ are often expressed in terms of the scattering lengths $b^+$
and $b^-$ \cite{Landau_3}:
\[
a^+=-b^+\qquad \mbox{and}\qquad a^-=-b^-\,.
\]

When the neutron is scattered by the proton, $b^+=5.39\cdot
10^{-13}$\,cm, $b^-=-2.37\cdot 10^{-12}$\,cm  \cite{Landau_3}. As a
result, in the case of $n-p$ scattering, for the amplitude
$A_0^{\mathrm{nuc}}$ (see Eq. (\ref{a36})) we have $A_0^{\mathrm{nuc}} \approx
-1.9\cdot 10^{-13}$\,cm, while for the amplitude
$A_1^{\mathrm{nuc}}$, we have $A_1^{\mathrm{nuc}}\approx 1.46\cdot
10^{-12}$\,cm. As is seen, $A_1\gg |A_0|$.

In the case of antiproton--proton interaction, the
spin--independent part of the scattering length, "$b$", is $b\sim
10^{-13}$\,cm \cite{x1}, which is comparable with
$|A_0^{\mathrm{nuc}}|$ for the case  of $n-p$ scattering. For
antiproton scattering by the proton, the magnitude of $A_1$ is
unknown in the considered range of low energies.

Let us assume for further estimations that
$\texttt{Re}\,A_1^{\mathrm{nuc}}$ in the case of $\bar{p}-p$
scattering is comparable with $\texttt{Re}\,A_1^{\mathrm{nuc}}$ in
the case of $n-p$ scattering, i.e.,
$\texttt{Re}\,A_1^{\mathrm{nuc}}$ is of the order of $10^{-12}$\,cm.

We can finally obtain the following estimate for the antiproton
spin precession frequency in matter with polarized protons:
\begin{equation}
	\label{a39} \Omega_{\mathrm{pr}}=\frac{2\pi\alpha
		c}{v}\cdot\frac{2\pi\hbar}{m} NP_t
	\texttt{Re}\,A_1\simeq\frac{2\pi\alpha c}{v} 6 \cdot 10^7
	\frac{N}{N_{l}} P_t\,,
\end{equation}
where $N_l$ is the number of atoms in 1 cm$^3$ of liquid hydrogen,
$N_{l}\simeq 4.25\cdot 10^{22}$. It will be recalled that
$\frac{2\pi \alpha c}{v}\gg 1$ for slow antiprotons with velocity
${v< 10^9}$\,cm/s.

Let us estimate now the spin rotation angle of the antiproton. An
antiproton moving in Hydrogen may be captured by a proton and
forms a bound state, protonium. We shall therefore estimate the
magnitude of the spin rotation angle during the characteristic
time $\tau$ necessary for the antiproton to be captured into a
bound state: $\tau\sim\frac{1}{N v\sigma_{\mathrm{pr}}}$, where
$\sigma_{\mathrm{pr}}$ is the cross section for protonium
formation.

The antiproton spin rotation angle $\vartheta$  during this time
can be estimated using formula:
\begin{equation}
	\label{a40}
	\vartheta\sim\Omega_{\mathrm{pr}}\tau=\frac{2\pi\hbar}{m}\cdot\frac{2\pi\alpha
		c}{v^2}\, P_t\,
	\frac{\texttt{Re}A_1}{\sigma_{\mathrm{pr}}}=\frac{2\pi^2\hbar}{E}\alpha
	c\frac{\texttt{Re}A_1}{\sigma_{\mathrm{pr}} }P_t\,,
\end{equation}
where $E=\frac{mv^2}{2}$ is the antiproton energy.

Presently,
antiproton beams with the energy of hundreds of electronvolts and lower \cite{x1,x2} are available.  According to \cite{11},
in the range of energies  higher than 10\,e\,V, the cross section  of
antiproton capture by hydrogen with the formation of protonium is $\sigma_{\mathrm{pr}}\leq
10^{-18}$\,cm$^2$.  As a result, the spin rotation angle for antiprotons with an energy of $10$\,eV can be
estimated as $\vartheta\simeq 6\cdot10^{-2} P_t$.
For antiproton energies of 20\,eV, the same is $\vartheta\sim 3\cdot 10^{-1}
P_t$.
When antiprotons are decelerated in a polarized gaseous target,
the degree of proton polarization is close to unity. As a
consequence, the rotation angle $\vartheta$ in such a target
reaches quite appreciable values $\vartheta\sim 10^{-1}$, giving
hope for experimental observation of the effect.

Thus, this effect  will be applicable for $\texttt{Re}\,A_1$
measurement in the region where the scattering experiments are
practically impossible to perform. One of the possibilities  to
study the spin rotation effect and to analyze the polarization
state of antiprotons (hyperons) under deceleration in matter
consists in use of  polarized targets with nuclei having a large
$Z$ (in fact, a thin film, in which case nuclei in a static
magnetic field can be polarized far more readily) and in
investigation of the intensity of the annihilation product
formation (angular distribution) as a function of  the spin
direction of the incident particle relative to the spin of target
nucleus.

Let us note in conclusion  that the deceleration of antiprotons in
Hydrogen finally results in the formation of a neutral atom ---
protonium \cite{x1}. Protonium undergoes interactions with a
nuclear pseudomagnetic field of polarized matter in a similar way
as a neutron.  This interaction leads to splitting and shifting of
energy levels of excited and ground states of protonium (and
similar atoms, such as $\bar{p}$\,d, $\bar{p}^3$He), as well as to
spin rotation and oscillations of the excited and ground states of
these atoms.


Thus, the above analysis shows that because the real part of the
coherent elastic forward scattering amplitude increases under
deceleration of $\bar{p}$ (hyperon), the effective potential
energy of interaction between the particle and matter  grows,
along with the pseudomagnetic field. As a result, the spin
rotation angle $\vartheta$ in the nuclear pseudomagnetic field of
matter reaches the value $\vartheta\sim10^{-1}$, which gives hope
to experimentally observe the real part of the amplitude in the
range where the scattering experiments are practically impossible
to perform.

\subsection{Mirror Reflection of Antiprotons from the Surface of
	Matter}

Before we start to consider mirror reflection of antiprotons from
the surface formed by the vacuum-matter boundary, let us make some
general remarks about mirror reflection of particles.

\begin{figure}[htbp]
	\begin{center}
		\resizebox{60mm}{!}
		{\includegraphics{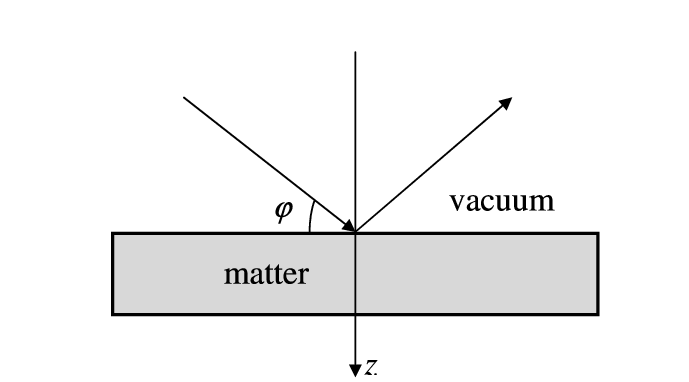}}
		\caption{Mirror reflection at the vacuum-matter boundary}
		\label{fig:fig1}
	\end{center}
\end{figure}

It is well known \cite{22,4} that the mirror-reflection coefficient
$R$ is defined by the refractive index $n$ of a wave in
matter. For such particles as, say,  neutrons (for light and
$\gamma$-quanta, whose polarization is orthogonal to the
mirror-reflection plane), the mirror-reflection coefficient has the
form \cite{22,4}
\begin{equation}
	R=|F|^2,
	\label{eq1}
\end{equation}
\begin{equation}
	F= - \frac{\sqrt{n^2-\cos^2\varphi}-\sin \varphi}{\sqrt{n^2-\cos^2\varphi}+\sin
		\varphi},
		\label{eq2}
\end{equation}
where $\varphi$ is the grazing angle (see Fig.\ref{fig:fig1}).

\noindent The refractive index of particles or $\gamma$-quanta can be
expressed as follows \cite{Goldberger,Davydov}  (see also (\ref{I.1})):
\begin{equation}
	n^2=1+\frac{4\pi\rho}{k^2}f(0),
	\nonumber
\end{equation}
where $\rho$ is the number of scatterers (nuclei, atoms) per
cubic centimeter of matter and $f(0)$ is the amplitude of
zero-angle coherent elastic scattering.
According to (\ref{eq1})--(\ref{eq3}), the mirror-reflection
coefficient is determined by the zero-angle scattering
amplitude. But at the same time, the wave scattered in the
mirror-reflection direction obviously makes a certain angle $\vartheta$
relative to the incident direction. Thus, the  amplitude of this wave is determined by the superposition of waves  scattered
at the angle $\vartheta$, i.e.,  instead of
$f(0)$, the amplitude  $F$ is now
determined by the amplitude  $f(\vartheta)$ of coherent elastic
scattering at the angle~$\vartheta$.

Indeed, let us consider elastic scattering of a wave
$e^{i\vec k\vec r} = e^{i\vec k_{\perp}\,\vec r_{\perp}}\,e^{ik_z z}$
by a set of scatterers located in the plane $z=z_0$.  Here $\vec
k_{\perp}$ is the wave vector's component perpendicular to the
$z$-axis (parallel to the plane where the scatterers are placed),
$k_z$ is the component of the particle's wave vector
that is parallel to the $z$-axis. Upon summation of the
spherical waves produced by the scatterers  in the plane $z=z_0$,
we obtain the following expression for the amplitude of a
mirror-reflected wave $F_1$ \cite{A2}:
\begin{equation}
	\label{eq4} F_1 =\frac{2\pi i \rho'}{k_z}f(\vec k'- \vec k)e^{2 i
		k_z z_0},
\end{equation}
where $|\vec k'| = |\vec k|$, $\vec k'$ is the wave vector  of the
scattered particle,  which has the components $\vec k_{\perp}' =
\vec k_{\perp}'$ and $k'_z=-k_z$, and $\rho'$ is the density of
scatterers in the considered plane (the number of scatters in
cm$^2$ of the plane).

If the layer $[0,z]$ contains $m$ number of planes, then the
amplitude of the wave reflected by these planes  can be written in
the form
\begin{equation}
	\label{eq5}
	F=\frac{2\pi i \rho'}{k_z} f(\vec k' - \vec k)\sum_{m}
	e^{2i k_z z_m},
\end{equation}

Passing to a continuous distribution of planes in the layer $[0,
z]$, i.e., replacing the summation in (\ref{eq5}) by integration,
we finally obtain \cite{A2}
\begin{equation}
	\label{eq6} F = - \frac{\pi\rho}{k_z^2}f(\vec k' - \vec k)= -
	\frac{4\pi\rho}{|\vec k'-\vec k|^2} f(\vec k '-\vec k).
\end{equation}

According to (\ref{eq6}), microscopic summation of waves scattered
at the vacuum-matter boundary yields the formation of a wave
reflected in the direction determined by the laws of classical
optics. Against (\ref{eq1}), its amplitude is defined not by the
amplitude $f(0)$, but by the amplitude of scattering at a nonzero
angle $\vartheta$ equal to a double grazing angle $\varphi$, i.e.,
$\vartheta = 2\varphi$ \cite{A2}. Equation (\ref{eq1}) is valid
for thermal and ultracold neutrons because of prevailing isotropic
S-scattering by nuclei ($f(\vartheta)= f(0)$). It is also valid
for photons, because at wavelengths much greater than the size of
the atom, dipole scattering, which is also isotropic under the
considered polarization, occurs.

According to (\ref{eq5}), the  amplitude $F$ of mirror reflection
rises with decreasing  $k_z=k \sin \varphi$, i.e., with decreasing
grazing angle $\varphi$. When $F$ reaches the values close to
unity, equation (\ref{eq6}) for $F$ is no longer valid. Mirror
reflection for this range of angles is considered, e.g., in
\cite{A2}.

In view of (\ref{eq6}), the amplitude $F$ is close to unity in the
range where the grazing angles $\varphi$ satisfy the condition
\begin{equation}
	\label{eq7} \sin^2 \varphi \sim \frac{\pi\rho |f|}{k^2}.
\end{equation}
If in estimating the angle $\varphi$ we use a typical value of the
amplitude of scattering via nuclear interaction, $|f|\sim
10^{-12}$\,cm,   then (\ref{eq7}) readily yields that for
antiproton energies from 100 to 1000 eV, the value of $F$ can be
close to unity only when the grazing angles are very small:
$\varphi \leq 10^{-4}\ \div 10^{-5}$. Further in our consideration
we shall demonstrate that owing to the interference of Coulomb and
nuclear interactions and the increase in the amplitude of nuclear
scattering of antiprotons with decreasing energy \cite{antiproton2}, the
coefficient of mirror reflection for antiprotons becomes
noticeable even at much larger grazing angles $\varphi$, making it
possible to use the phenomenon of mirror reflection to investigate
scattering of slow antiprotons by nuclei \cite{arxiv:1404.0197}.

According to (\ref{eq6}), the mirror-reflection coefficient $R$
can be written as follows:
\begin{equation}
	\label{eq8} R=|F|^2=\left|\frac{\pi\rho}{k_z^2} f(\vec k'-\vec
	k)\right|^2=\frac{\pi^2\rho^2}{k^4\sin^4\varphi} |f(\vec k'-\vec
	k)|^2.
\end{equation}

There are two interactions responsible for antiproton scattering
by nuclei (atoms): Coulomb and nuclear (here we neglect the
magnetic interaction of antiproton and electron spins).
Consequently, the scattering amplitude $f$ can be presented as a
sum of two amplitudes:
\begin{equation}
	\label{eq8} f= f_{Coul} + f_{N},
\end{equation}
where $ f_{Coul}$ is the amplitude of a purely Coulomb scattering
and $f_{N}$ is the amplitude related to nuclear interaction (it
contains the contribution from Coulomb interaction that affects
nuclear scattering \cite{antiproton2}). Let us note that the contribution
to the formation of a mirror-reflected wave comes from  elastic
scattering, in which the state of the target does not change. As
is known, the scattering amplitude in this case can be presented
as a product of the amplitude of elastic scattering by an
infinite-mass nucleus (atom)(the reduced mass equals the mass of
the incident particle) into the Debye-Waller factor $e^{-w(\vec k'
	-\vec k)}$ describing the effect produced by thermal oscillations
of nuclei (atoms) in matter  on the process of scattering
\cite{22}. In subsequent consideration, by the amplitude $f$
we shall  mean  scattering by an infinite-mass nucleus. As a
result, we have ($\varphi\ll 1$)
\begin{equation}
	\label{eq9} R=\frac{\pi^2\rho^2}{k^4
		\varphi^4}\left(|f_{Coul}(\vec k'-\vec k)|^2 + 2 \texttt{Re}
	f_{Coul}(\vec k'-\vec k) f_{N}^* + |f_{N}|^2\right) e^{-2w(\vec k'
		- \vec k)}.
\end{equation}

For further consideration we need to compare the amplitudes of
Coulomb and nuclear elastic scattering of antiprotons by nuclei
(atoms).

\subsection{The Amplitude of Antiproton Scattering by a Nucleus (Atom)
	at Low Energy}

The scattering amplitude relates to the $\mathcal{T}$-matrix as \cite{Goldberger,Davydov}
\begin{equation} \label{eq10} f_{ba}= -
	\frac{m}{2\pi\hbar^2}\langle\Phi_{b}|\mathcal{T}|\Phi_{a}\rangle,
\end{equation}
where $|\Phi_{a(b)}\rangle$ is the wave function describing the
initial (final) state of the system "incident particle--atom
(nucleus)". The wave functions $|\Phi_a\rangle$  are the
eigenfunctions of the Hamiltonian
${H}_0=H_p(\vec{r}_p)+H_A(\vec{\xi}\,, \vec{r}_{\mathrm{nuc}})$,
i.e., {${{H}_0|\Phi_a\rangle=E_a|\Phi_a\rangle}$};
$H_p(\vec{r}_p)$ is the Hamiltonian of the particle incident onto
the target; $\vec{r}_p$ is the particle coordinate;
$H_A(\vec{\xi}\,, \vec{r}_{\mathrm{nuc}})$ is the atomic (nuclear)
Hamiltonian; $\vec{\xi}$ is the set of coordinates of the atomic
electron; $\vec{r}_{\mathrm{nuc}}$ is the set of coordinates
describing the atomic nuclei.

The Hamiltonian $H$ describing the particle--nucleus  interaction reads as (\ref{a3}).

Using the same approach as applied for deriving (\ref{a9})-(\ref{a12})
one can obtain the following expression for
scattering amplitude \cite{arxiv:1404.0197}:
\begin{equation}
	\label{eq11} f_{ba}^N= - \frac{m}{2\pi\hbar} g_{ba}
	\mathcal{T}_{ba}^{nuc}=g_{ba}f _{ba}^{nuc},
\end{equation}
where $f _{ba}^{nuc}$ is the amplitude of particle scattering by
the nucleus in the absence of Coulomb interaction.

Thus, Coulomb interaction leads to a change in the value of the
amplitude of antiproton-nucleus scattering. Let us estimate the
magnitude of this change.


In what follows we shall be primarily concerned with elastic
scattering. In this case $|\vec k| = |\vec k'|$, and it follows
that from the expression given in \cite{Landau_3} for the wave functions
$\varphi_b^{(-)}$ and $\varphi_a^{(+)}$, which describe particle
scattering in the Coulomb field, we can derive  the below
relationships for $g_{ba}$:

\begin{itemize}
	\item for the case of repulsion, i.e., elastic scattering of
	similarly charged particles
	\begin{equation}
		\label{a14}
		g_{ba}^{\mathrm{rep}}=\frac{2\pi}{\kappa(e^{\frac{2\pi}{\kappa}}-1)}\,,\qquad
		\kappa=\frac{v}{Z\alpha c}\,,
	\end{equation}
	where $v$ is the particle velocity, $Z$ is the charge of the
	nucleus, $\alpha$ is the fine structure constant, $c$ is the speed
	of light; \item for the case of attraction
	\begin{equation}
		\label{a15}
		g_{ba}^{\mathrm{att}}=\frac{2\pi}{\kappa(1-e^{-\frac{2\pi}{\kappa}})}\,.
	\end{equation}
\end{itemize}

With decreasing particle energy (velocity), $\kappa$ diminishes,
and for such values of $\kappa$ when $\frac{2\pi}{\kappa}\geq1$,
one can write
\begin{equation}
	\label{a16} g_{ba}^{\mathrm{rep}}=\frac{2\pi\alpha Z
		c}{v}\,e^{-\frac{2\pi\alpha Z c}{v}}\,,\qquad
	g_{ba}^{\mathrm{att}}=\frac{2\pi\alpha Z c}{v}\,.
\end{equation}

According to (\ref{eq11}), with decreasing energy of positively
charged particles, the amplitude $f_{ba}^N$ diminishes rapidly
because of repulsion. For negatively charged particles, the
amplitude grows with decreasing particle energy (velocity).

These results for the amplitude $f^N_{ba}$ generalize a similar,
well-known  relationship for taking account of the Coulomb
interaction effect on the cross section of inelastic processes,
$\sigma_r$, \cite{Landau_3}.

So in the range of low energies, the amplitude $f(\vec k' -\vec
k)$ of antiproton (negative hyperon) scattering by a nucleus can
be presented in the form (for antiproton scattering in
ferromagnets, the magnetic interaction of antiprotons with
electrons in atoms should also be considered):
\begin{equation}
	\label{a18} f(\vec k' - \vec k)=f_{Coul}(\vec k' - \vec k) +
	f_N(\vec k' - \vec k),
\end{equation}
where
\begin{equation} \label{a18b}
	f_N(\vec k' - \vec
	k)=\frac{2\pi\alpha Z c}{v}f_{nuc}(\vec k'- \vec k).
\end{equation}

Using (\ref{a18b}) and the expression for $f_{Coul}$ in \cite{Landau_3},
we can obtain the following expression for the modulus of the
Coulomb (Rutherford) scattering amplitude in the range of small
scattering angles ($\vartheta \ll 1$, $\vartheta>\frac{1}{kR}$):
\begin{equation}
	\label{eq12a} |f_{Coul}(\vec k' -\vec k)|  = \frac{2 Z e^2}{m v^2
		\vartheta^2} = \frac{Z e^2}{2 m v^2 \varphi^2}
\end{equation}
and then write the ratio for these amplitudes as
\begin{equation}
	\label{eq12} \frac{|f_{N}|}{|f_{Coul}|}= 4\pi k \varphi^2
	|f_{nuc}|.
\end{equation}

As a result, using for the characteristic nuclear amplitude the
estimate {${|f_{nuc}|\approx 10^{-12}}$}\,cm,{ ${k\leq 10^{10}\div
		10^{11}}$}, $\varphi \sim 10^{-1}$, we can estimate the ratio
$\frac{|f_{nuc}|}{|f_{Coul}|}$ as
{${\frac{|f_{nuc}|}{|f_{Coul}|}\leq 3\cdot 10^{-4}\div 10^{-2}}$}.

In view of the above estimate, we can recast the expression for
the coefficient of reflection as follows:
\begin{equation}
	\label{eq13} R=R_{Coul}\left(1+2\frac{\texttt{Re} e^{i\beta}
		f_{nuc}^*}{|f_{Coul}|}\right) e^{-2w(\vec k' - \vec k)},
\end{equation}
where
$R_{Coul}=\frac{\pi^2\rho^2}{k^4\varphi^4}\left|f_{Coul}(\vec k' -
\vec k )\right|^2$ is the coefficient of mirror reflection due to
a purely Coulomb interaction,
$f_{Coul}=\left|f_{Coul}\right|e^{i\beta}$, and the term in
(\ref{eq9}) that is proportional to
$\frac{\left|f_{nuc}\right|^2}{\left|f_{Coul}\right|}$ is dropped
for its smallness. Let us recall that (\ref{eq13}) holds true for
such values of $R_{Coul}$ that are much less than unity.

Thus the coefficient of mirror reflection contains two
contributions: one comes from a purely Coulomb interaction and the
other is due to the  Coulomb-nuclear interference. This makes it
possible to obtain data about the amplitude of nuclear scattering
of antiprotons by the nucleus  from the experiments investigating
angular and energy dependence of $R$ on the grazing angle and the
energy of antiprotons.

\subsection{Spin Polarization of Antiprotons Reflected from the
	Vacuum-Matter Boundary}

As is well known, spin-orbit interaction  during scattering causes
the initially nonpolarized particle beam to become polarized
\cite{Landau_3}. In this case, the polarization vector of particles
appears to be orthogonal to the scattering plane, i.e., the plane
containing vectors $\vec k$ and $\vec k'$. If  the particle beam
had a nonzero polarization vector before the interaction, then a
left-right asymmetry  in the intensity of particle scattering is
observed.
For slow neutrons, the spin-orbit
interaction is caused by the interaction $V_{so}$ of the neutron
magnetic moment and the nuclear electric field \cite{23}
\begin{equation}
	\label{eq14} V^{neut}_{so}= i \frac{\mu_n \hbar}{m
		c}\vec{\sigma}[\vec E(\vec r)\vec\nabla_{\vec r}],
\end{equation}
where $\mu_n$ is the neutron magnetic moment,
$\vec{\sigma}=(\sigma_x, \, \sigma_y,\, \sigma_z)$ are the Pauli
spin matrices, $\vec E$ is the electric field  at  point $\vec r$
where the neutron is located, and $m$ is the neutron's mass. First
experiments to observe the effect of spin-orbit interaction on
neutron scattering were performed by C.G. Shull \cite{1404.0197_shull}. (For
further experiments see, e.g., \cite{A2}).

The presence of a charge in antiprotons appreciably affects  the
dependence of spin-orbit interaction on their magnetic moment. The
energy of spin-orbit interaction of antiprotons with nuclei is
determined by the electric field and has the form \cite{23}
\begin{equation}
	\label{eq15} V_{so}^{ap}=- \left(\mu'+\frac{e\hbar}{4m c}\right)
	\left(\vec{\sigma}\left[\vec E\frac{\hat{p}}{m}\right]\right),
\end{equation}
where $\mu'$ is the anomalous part of the antiproton's magnetic
moment, $e=-|e|$ is the antiproton charge, $|e|$ is the electron
charge, and $\hat{p}=- i\hbar \vec{\nabla}$ is the momentum
operator of the antiproton.

The energy dependence of the amplitude of spin-orbit scattering
also changes noticeably through the interference of Coulomb and
spin-orbit interactions. The energy dependence of the contribution
coming from the antiproton-nucleus strong interaction to the
amplitude of spin-orbit scattering changes, too.

The expression describing the amplitude of spin-orbit scattering
in a general form reads:
\begin{equation}
	\label{eq16} F_{so}= F_{so}\vec {\sigma}[\vec k'\times \vec
	k]=F_{so}\vec{\sigma}[\vec q\times\vec k],
\end{equation}
where $\vec q= \vec k'-\vec k$  is the  momentum transfer. It
should be noted that the general form of (\ref{eq16}) is clear
even without calculations and follows from the symmetry
considerations, being valid for all spin particles.

Thus, the amplitude of mirror reflection from matter with nonpolarized nuclei is a sum of three
amplitudes: the amplitude of Coulomb scattering (or the amplitude of magnetic scattering by  electrons in ferromagnets), the
particle-spin-independent amplitude of nuclear scattering, and
the amplitude of spin-orbit scattering. Hence, the
coefficient of mirror reflection contains the contributions coming from
these amplitudes and their interference. Let us
consider the contribution coming  to the coefficient of mirror
reflection from the interference of Coulomb and spin-orbit
interactions. Obviously, it is proportional to $ \vec \sigma [\vec k'\times\vec
k]$.

Let us choose the quantization axis to be parallel to the
reflecting plane (the plane containing vectors $\vec k'$ and $\vec
k$). It follows from (\ref{eq9}) and (\ref{eq16}) that due to Coulomb-nuclear interference, the reflection coefficients
$R_{\uparrow\uparrow}$ and $R_{\downarrow\uparrow}$ for antiprotons
with  spins parallel and antiparallel  to the direction of the axial vector $[\vec k\times \vec
k']$ will differ.   Let a nonpolarized antiproton beam be incident on the
target. Such beam is represented by a coherent sum  of two
beams with spins parallel and antiparallel to $[\vec k\times \vec
k']$. A mirror-reflected beam appears partially
polarized, and the degree to which the beam is polarized is
determined by the difference $R_{\uparrow\uparrow} - R_{\downarrow\uparrow}$
of mirror-reflection coefficients:
\begin{equation}
	\label{eq17}
	p=\frac{R_{\uparrow\uparrow}-
		R_{\downarrow\uparrow}}{R_{\uparrow\uparrow}+  R_{\uparrow\downarrow}}
\end{equation}
When estimating the magnitude of the effect, we should take into
account that in the Born approximation, the amplitude $F_{so}$ is
purely imaginary, while the amplitude $F_{Coul}$ is real, and so
it is important that the imaginary part of $F_{Coul}$ be
considered. It can be readily found,  since we know that in the
range of small scattering angles, $\texttt{Im} F_{Coul}$ can be
set equal to $\texttt{Im}F_{Coul}(0)$ ($\texttt{Im}F_{Coul}(0)$ is
the amplitude of Coulomb scattering by a nucleus (atom) at zero
angle).

According to the optical theorem,
$\texttt{Im}F_{Coul}(0)=\frac{k}{4\pi}\sigma_{tot}$. The total
cross section of scattering by a screened Coulomb potential,
$V_{Coul}=\frac{Z e^2}{r}e^{-\frac {r}{ R_A}}$, can be written as
\begin{equation}
	\label{eq18}
	\sigma= 16\pi \left(\frac{m Z e^2 R^2_A}{\hbar^2}\right)^2
	\frac{1}{1+ \frac{8 m E R_A^2}{\hbar^2}},
\end{equation}
where $E$ is the particle energy.

It follows from (\ref{eq18}) that for antiproton energies greater
than the characteristic energy $E_A =\frac{\hbar^2}{4 z m
	R^2_A}\approx 10^{-2}$ e V, the scattering cross section
$\sigma\sim
\frac{1}{E}\sim \frac{1}{v}$. 
As a result, $ \texttt{Im} F_{Coul}\simeq \mbox{const}$. The
amplitude of spin-orbit interaction contains the term
$\frac{2\pi\alpha Z c}{v}$ [see (\ref{eq11}), (\ref{a18b})] that
increases the amplitude of spin-orbit scattering of antiprotons by
nuclei (atoms), making it grow as $\sim \frac{1}{v}$ in the range
of small grazing angles (the momentum transfer in this case  is
required to be $q
> \frac{1}{R_A}$). As a result,  the contribution of spin-orbit
scattering to the coefficient of mirror reflection can be
estimated as \cite{arxiv:1404.0197}:
\[
R_{so}\sim \frac{\pi\rho p^2}{k^4\varphi^4}|\texttt{Im}
f_{Coul}|^2\frac{f_so}{\texttt{Im}f_{Coul}}\approx 10^{-1}\div
10^{-2}.
\]
Consequently, this process can be used to obtain  polarized
antiprotons in this low energy range.

Let us note here that when a polarized antiproton beam falls on
the surface, the particles' polarization vector rotates about the
quantization axis, i.e., about the direction $[\vec k'\times \vec
k]$. The phenomenon described here also occurs during the
diffraction reflection of antiprotons from a crystal's surface and
is caused by the interference between the spin-orbit amplitude and
the real part of the Coulomb amplitude in the case when crystal's
cells lack the center of symmetry. (Compare with a similar
phenomenon for slow neutrons \cite{A2}, in which case  the
effect occurs through the interference of the spin-independent
part of nuclear scattering and the amplitude of spin-orbit
scattering.)

Now, let us suppose that a particle beam is incident on the
boundary between vacuum and matter with polarized nuclei. The
elastic scattering amplitude $\hat f(\vec k' - \vec k)$ in this
case depends on spin orientations of the incident particle, $\vec
S$,  and the nucleus, $\vec J$, i.e., the scattering amplitude is
the operator in the spin space of particle and nucleus.
Investigating refraction and mirror reflection, we are interested
in coherent elastic scattering, in which the nuclear spin state
remains unchanged. The scattering amplitude $\hat f_N(\vec k' -
\vec k)$, describing such scattering, is obtained by averaging
the total amplitude $\hat f$ using the nuclear spin density matrix
$\hat{\rho}_J$: $\hat f(\vec k' - \vec k) = \mbox{Tr}\,
\hat{\rho}\hat{f}(\vec k' - \vec k)$. (The general expression for
the amplitude $\hat f $ of scattering of a particle with spin
$S=\frac {1}{2}$ by a nucleus  with spin $J=\frac{1}{2}$ is given,
e.g., in \cite{Landau_3}.)

Consequently, in this case the contribution from nuclear scattering to the amplitude
of a mirror-reflected wave can be written in the form:
\begin{equation}
	\label{eq19a} \hat F_{pol}=-\frac{\pi\rho}{k_z^2}\hat f_N (\vec
	k'- \vec k)
\end{equation}

As a result, the amplitude of a
mirror-reflected wave can be presented in the form
\begin{equation}
	\label{eq19} \hat F(\vec k'- \vec k) = F_{Coul}(\vec k' - \vec
	k)+\hat F_{so}(\vec k ' - \vec k) + \hat F_{pol}(\vec k'-\vec k),
\end{equation}
where $F_{Coul}$ is the amplitude of the
mirror-reflected wave that is due to Coulomb interaction, $\hat
F_{so} (\vec k' - \vec k) $ is the amplitude of the reflected wave
that is due to antiproton-nucleus (or antiproton-atom) spin-orbit
interaction, and { ${\hat F_{pol}(\vec k ' - \vec k)}$} is the
amplitude of antiproton scattering by a polarized nucleus (except
for those terms in the amplitude $\hat F_{pol}$ that describe
spin-orbit interactions).

Using (\ref{eq19}), we can find the intensity and polarization of
reflected particles. For example, the intensity of reflected
particles is related to spin orientation of incident particles by
the expression of the form
\begin{equation}
	\label{eq20} I_{ref}=I_0\,\mbox{Tr}\,F\rho_0 F^+ = I_0
	\,\mbox{Tr}\,F^+F \rho_0,
\end{equation}
where $\rho_0$ is the spin density matrix of the incident beam and
$I_0$ is the beam's intensity.

The polarization vector $\vec p$ of mirror-reflected particles has
the form
\begin{equation}
	\label{eq21} \vec p=\frac{1}{I_{ref}}\,\mbox{Tr}\, \rho_0
	F^+\,\frac{\hat{\vec S}}{S} F,
\end{equation}
where $\hat{\vec S}$ is the spin operator of particles; the spin
of antiprotons equals $1/2$, hence $\hat{\vec S}=\frac{1}{2}\vec
\sigma$.

It follows from (\ref{eq19}), (\ref{eq20}), and (\ref{eq21}) that
$I_{ref}$ and $\vec p$ depend on the interference of  the nuclear
amplitude $\hat F_{pol}(\vec k' - \vec k)$ and the Coulomb and
spin-orbit amplitudes.

The amplitude $\hat f_{N}$ that determines the reflection
amplitude $\hat F_{pol}$ has quite a complicated structure. For
slow antiprotons scattered at the angle $\vartheta\ll 1$ (the
grazing angle $\varphi\ll 1$), the amplitude $ \hat f_{N}$
coincides with a zero-angle scattering amplitude and has the form
\begin{equation}
	\label{eq22} \hat f_{N}= A_0+ A_1 (\vec S\, \vec p_{t}) + A_2
	(\vec S\,\vec e) (\vec e \,\vec p_t),
\end{equation}
where $\vec p_t$ is the polarization vector of the target. Using
(\ref{eq20}), (\ref{eq21}), and (\ref{eq22}), we can find the
intensity and polarization of reflected particles for each
particular case. These expressions yield that when nonpolarized
antiprotons are incident on a polarized target, the reflected
antiprotons appear to be polarized. If the incident antiproton
beam is polarized, then the spin of the mirror-reflected beam
undergoes rotation (the rotation angle is estimated at the order
of $10^{-1}\div 10^{-2}$), and the intensity of the reflected beam
depends on the mutual orientation between the spin of incident
particles and the target polarization. The degree of polarization
that the initially nonpolarized beam acquires through reflection
has the order of magnitude about $10^{-1}\div 10^{-2}$.


\section
{Quasi-optical Phenomena for High-energy  Particles in Polarized Matter}
\label{sec:high-energy}

For thermal neutrons $v = 2.2 \cdot 10^{5}$\,cm$\cdot$\,s$^{-1}$, thus, spin full turn occurs at length $L\approx 10^{-3}$\,cm.
Velocity grows with the energy growth approaching speed of light $c$. Therefore,
length $L$, at which spin full turn occurs,
increases $10^5$ times and can be as long as 100 cm.
With particle energy growth the nuclear spin phenomenon becomes available for
positively charged particles and nuclei (Coulomb repulsion does not prevent nuclear
interactions in this case).

Now we shall pass on to consideration of  the effects of "optical"
spin rotation arising when high--energy particles pass through
matter with polarized nuclei \cite{109,110,nim06_PR+}.

To be more specific, we shall first consider refraction of
relativistic particles with spin $1/2$  in matter. In the
beginning, let us consider scattering by a particular center. The
asymptotic expression for a particle wave function  far from the
scatterer can be represented in the form \cite{Goldberger,23}
\begin{equation}
	\label{18.1}
	\Psi=U_{E,\vec{k}}e^{ikz}+U^{\prime}_{E^{\prime}\vec{k}^{\prime}}\frac{e^{ik^{\prime}r}}{r}\,,
\end{equation}
where $U_{E\vec{k}}$ is the bispinor amplitude of the incident
plane wave; $U^{\prime}_{E^{\prime}\vec{k}^{\prime}}$ is the
bispinor describing the amplitude of the scattered wave; $E$ and
$\vec{k}(E^{\prime},\vec{k}^{\prime})$ are the energy and the wave
vector of the incident (scattered) wave.

According to \cite{23}, the bispinor amplitude is fully determined
by specifying a two--component quantity --- a three--dimensional
spinor  $W$, which is a nonrelativistic wave function in the
particle rest frame. For this reason the scattering amplitude,
i.e., the amplitude of a divergent spherical wave, in (\ref{18.1})
like that in the nonrelativistic case can be defined as a
two--dimensional matrix of $\hat{f}$ by the relation
$W^{\prime}=\hat{f}W$, where $W^{\prime}$ is the spinor
determining the bispinor
$U^{\prime}_{E^{\prime}\vec{k}^{\prime}}$. Thus determined
scattering operator is quite similar to the  scattering amplitude
in the nonrelativistic scattering theory.

As a result, deriving the expression for the refractive index by
analogy with a non-relativistic case 
we obtain the following expressions for the wave function of a
relativistic neutron (proton) in a medium
\begin{equation}
	\label{18.2} \Psi= \frac{1}{\sqrt{2E}}\left(\begin{array}{cc}
		\sqrt{E+m}e^{ik\hat{n}z} & W\\
		\sqrt{E-m}(\vec{\sigma}\vec{n})e^{ik\hat{n}z} & W
	\end{array}\right)\,,
\end{equation}
where
\begin{equation}
	\label{18.3} \hat{n}=1+\frac{2\pi\rho}{k^{2}}\hat{f}(0)
\end{equation}
is the operator refractive index, $\hat{f}(0)$ is the amplitude of
coherent elastic zero-angle scattering by a polarized scatterer;
$\vec{\sigma}$ is the vector made up of the Pauli matrices;
$\vec{n}=\vec{k}/\vec{k}$.

Using $\vec{\sigma}$, $\vec{J}$,
$\vec{n}$ ($\vec{J}$ is the nuclear spin operator), we may write
the amplitude of coherent elastic forward scattering 
of a particle with spin 1/2 ($\vec{S}=(1/2)\vec{\sigma}$)
by a
polarized nucleus in the general case of strong, electromagnetic,
and P--, T--violating  weak interactions as follows \cite{85,86,A2,Nuclear_optics}:



\begin{equation}
	\label{16.1} \hat{f}=\hat{f}_{\mu}+\hat{f}_{s}+\hat{f}_{w}\,,
\end{equation}
where $\hat{f}_{\mu}$ is the amplitude of coherent elastic
scattering by a nucleus through magnetic interaction. Its explicit
form depends on the observation conditions due to a long--range
character of the magnetic interaction. 
%
The scattering
amplitude caused by parity--conserving strong interactions:
\[
\hat{f}_{s}=A+A_{1}\vec{\sigma}\langle\vec{J}\rangle+A_{2}(\vec{\sigma}\vec{n})(\vec{n}\langle\vec{J}\rangle)
+A_{3}\vec{n}\vec{n}_{1}+\ldots
\]
and the contribution from 
weak
interactions to the scattering amplitude
\[
\hat{f}_{w}=B\vec{\sigma}\vec{n}+B_{1}\vec{\sigma}\left[\langle\vec{J}\rangle
\vec{n}\right]
+B_{2}\vec{\sigma}\vec{n}_{1}+B_{3}(\vec{\sigma}\vec{n})(\vec{n}\vec{n}_{1})+
B_{4}\vec{n}\langle\vec{J}\rangle+B_{5}\vec{\sigma}\left[\vec{n}\vec{n}_{1}\right]+\ldots\,,
\]
where $\langle\vec{J}\rangle=
\textrm{Tr}\,\rho_{\mathrm{nuc}}\vec{J}$ is the average nuclear
spin, $\rho_{\mathrm{nuc}}$ is the spin density matrix of the
nucleus, $\vec{n}_{1}$ is the vector with components
$n_{1i}=\sum\limits^{3}_{j=1}\langle Q_{ij}\rangle n_{j}$,
$\langle Q_{ij}\rangle=\textrm{Tr}\,\rho_{\mathrm{nuc}}Q_{ij}$ is
the target quadrupolarization tensor of rank two and
\[
Q_{ij}=\frac{1}{2J(2J-1)}{J_{i}J_{j}+J_{j}J_{i}-\frac{2}{3}J(J+1)\delta_{ij}}\,,
\]
the sign $+\ldots$ means the contributions of higher--order
multipoles to $\hat{f}$.


Where the terms containing $A$ are caused by strong P-,T-even interactions, 
those with $B$, $B_2$, 
$B_3$, $B_4$ --
by P-odd T-even, $B_1$ -- by P-odd T-odd and $B_5$ corresponds to P-even T-odd interactions.


{The terms proportional to $A_{1}$ and $A_{2}$ give rise to a
	phenomenon of nuclear precession of particle spin caused by a
	nuclear "pseudomagnetic" field. The terms
	proportional to $B_{i}$ also lead to spin precession. 
	%
	%
	The constant $B$ describes the phenomenon of
	P--violating particle spin rotation about the momentum direction,
	predicted by Michel \cite{87}, and P--odd spin dichroism
	\cite{92}. 
	The  constant $B_{1}$  \cite{85,86} describes
	the phenomenon of P-- and T--odd spin rotation and dichroism.
	It should be pointed out that
	unlike the term proportional to $B_{1}$, the contributions to
	$\hat{f}$ proportional to $B_{2}$, $B_{3}$, $B_{5}$
	\cite{85,86} include the second--rank nuclear spin operator.
	The terms proportional to constants $B_{2}$ and $B_{3}$ describe
	P--odd T-even particle spin rotation and dichroism; the term
	proportional to constant $B_{5}$ describes T--odd P--even particle
	spin rotation  and dichroism.} 

According to (\ref{18.2}), the spinor $W^{\prime}$ defining the
spin state of a particle  after passing the path
length $z$ in matter has the form
\begin{equation}
	\label{18.4} W^{\prime}=e^{ik\hat{n}z}W\,.
\end{equation}
Note that   $\hat{n}$ can be written as
\begin{equation}
	\label{18.5} \hat{n}=n_{0}+\frac{2\pi\rho}{k^{2}}(\vec{\sigma}
	\vec{g})\,,
\end{equation}
where $n_{0}$ is the $\vec{\sigma}$-independent part of $n$.


From  expression for $\hat{f}(0)$ we have
\begin{equation}
	\label{ins_18.5}
	n_{0}=1+\frac{2\pi\rho}{k^{2}}(A+A_{3}\vec{n}\vec{n}_{1}+B_{4}\vec{n}\langle\vec{J}\rangle+\ldots)\,.
\end{equation}
The explicit form of vector $\vec{g}$ immediately follows from
the expression for the amplitude:
\begin{equation}
	\label{ins1_18.5}
	\vec{g}=A_{1}\langle\vec{J}\rangle +A_{2}\vec{n}(\vec{n}\langle\vec{J}\rangle)+B\vec{n}+B_{1}[\langle\vec{J}\rangle\vec{n}]+\ldots\,.
\end{equation}
Suppose that particle absorption can be neglected, as a
consequence, $\vec{g}$ is a real vector. With the help of
(\ref{18.5}), represent (\ref{18.4}) as follows:
\begin{equation}
	\label{18.6}
	W^{\prime}=e^{ikn_{0}z}e^{i\frac{2\pi\rho}{k}(\vec{\sigma}\vec{\jmath}_{g})|\vec{g}|z}W\,,
\end{equation}
$\vec{\jmath}_{g}=\vec{g}/|\vec{g}|$.  Remember now that the
operator of spin rotation by an angle $\vartheta$ about a certain
axis, characterized by a unit vector $\vec{\jmath}$  ~has the form
\begin{equation}
	\label{18.7} \hat{T}=e^{i\frac{\vartheta}{2}\vec{\sigma}\vec{\jmath}}\,.
\end{equation}
Comparing (\ref{18.4}), (\ref{18.5}) and (\ref{18.6}), we obtain
that in the case in question the operator
\[
\exp\left\{i\frac{2\pi\rho}{k}|\vec{g}|(\vec{\sigma}\vec{\jmath}_{g})z\right\}
\]
acts as a spin rotation operator of a particle in its rest
frame.
The rotation angle is
\begin{equation}
	\label{18.8}
	\vartheta=\frac{4\pi\rho}{k}|\vec{g}|z=k(n_{\uparrow\uparrow}-n_{\downarrow\uparrow})z\,,
\end{equation}
where the quantization axis is selected along $\vec{\jmath}_{g}$.

\subsection{The effect of spin precession at high energies}

To be more specific, we shall further analyze the effect of spin
precession in a polarized target due to strong interactions. 
In (\ref{18.2}), (\ref{18.8}), for the angle of
particle spin rotation about the direction $\vec{\jmath}_{g}$ we obtain
\begin{equation}
	\label{18.9}
	\vartheta=\frac{2\pi\rho}{k}(f_{\uparrow\uparrow}-f_{\uparrow\downarrow})z=
	\frac{4\pi\rho}{k}|A_{1}\langle\vec{J}\rangle+A_{2}\vec{n}(\vec{n}\langle\vec{J}\rangle)|z\,.
\end{equation}
To answer a question about how the the spin precession
angle $\vartheta$ of a relativistic particle  in a polarized target
depends on the particle energy, let us remember that at scattering by a
potential, the Dirac equation for ultrarelativistic particles
reduces to the equation similar to a non-relativistic
Schr\"odinger equation, where the particle mass $M$ stands for its
relativistic mass \cite{23}, i.e., $M=\gamma m$, where $m$ is the
particle rest mass, $\gamma$ is its Lorentz factor. As the
amplitude of  particle scattering by a potential is proportional
to the particle mass, then the amplitude for a relativistic particle may be written as
\begin{equation}
	\label{18.10} f(E, 0)=\gamma f^{\prime}(E, 0)\,,
\end{equation}
where $E$ is the particle energy.

Such a relation also holds for the general case of scattering of a relativistic particle by the scatterer (for example, by a nucleus) \cite{Goldberger}.
According to \cite{Goldberger},
the forward scattering amplitude is related to the $\mathcal{T}$-matrix describing the collision of particles in the general case as follows
\begin{equation}
	\label{ins_18.10}
	f(E, 0)=-(2\pi)^{2}\,\frac{m\gamma}{\hbar^{2}}\,\mathcal{T}(E)\,.
\end{equation}
From this we have
\begin{equation}
	\label{insn_18.10}
	f^{\prime}(E, 0)=-(2\pi)^{2}\,\frac{m}{\hbar^{2}}\,\mathcal{T}(E)\,.
\end{equation}
Using (\ref{18.10}), we can rewrite (\ref{18.9}) as follows
\begin{equation}
	\label{18.12}
	\vartheta=2\pi\rho\lambda_{\mathrm{C}}(f^{\prime}_{\uparrow\uparrow}-f^{\prime}_{\downarrow\uparrow})z\,,
\end{equation}
where $\lambda_{\mathrm{C}}=\hbar/(mc)$ is the Compton  wavelength of the particle.

By means of the $\mathcal{T}$-matrix, the expression for the rotation angle $\vartheta$ may also be represented in the form:
\begin{equation}
	\label{ins_18.12}
	\vartheta=-\frac{(2\pi)^{3}\rho}{\hbar c}(\mathcal{T}_{\uparrow\uparrow}(E) - \mathcal{T}_{\downarrow\uparrow}(E))z\,.
\end{equation}
The particle path length $z$  in the target is $z = v t$, $v$ is
the particle velocity, $t$ is the time in which the particle
passed the path $z$. Hence, we also have
$\vartheta=\omega_{\mathrm{pr}} t$, where $\omega_{\mathrm{pr}}$
is the particle spin precession frequency in a polarized target
\[
\omega_{\mathrm{pr}} =-\frac{(2\pi)^{3}\rho}{\hbar c}\, (\mathcal{T}_{\uparrow\uparrow}(E)-   \mathcal{T}_{\downarrow\uparrow}(E))v\,.
\]
As we see, at relativistic energies, the dependence $\vartheta\sim
1/k$ disappears, and the entire possible dependence of the
rotation angle $\vartheta$ on the particle energy is contained in
the amplitude $f^{\prime}(E, 0)(\mathcal{T}(E))$.

Evaluate the possible rotation angle (see also \cite{nim06_PR+}).
To understand the magnitude of the effect, let us assume that the
difference
$f^{\prime}_{\uparrow\uparrow}-f^{\prime}_{\downarrow\uparrow}$ is
of the order of $10^{-12}\div 10^{-13}$ cm, i.e., of the order of
the difference between the spin amplitudes of neutron scattering
by a proton in a low--energy range. In this case we have for a
fully polarized target
\begin{equation}
	\label{18.14} \vartheta\sim 10^{-3}\div 10^{-4}z\,,
\end{equation}
where $\vartheta$ is in radians, i.e., $\vartheta\sim 10^{-2}\div
10^{-3}$\,rad for a particle passing through a polarized target of
length 10 cm.

The acquired $|p_x|$ values for a beam of polarized protons in a polarized external target at Nuclotron are given  in Table \ref{tab:slide20} for NH$_3$ and ND$_3$ targets. Thickness for both targets is as high as $L=30$~cm, proton energy is within 200~MeV -- 1~Gev range. 
Here $\Delta f =f_{\uparrow\uparrow}-f_{\downarrow\uparrow}$ and $\Delta \sigma = \sigma_{\uparrow\uparrow}-\sigma_{\downarrow\uparrow}$ denotes  difference in proton  total scattering cross-section values for spin parallel and antiparallel orientations.
\begin{table}[h]
	\caption{Effect evaluation for polarized protons in a polarized external targets at Nuclotron}
	\label{tab:slide20}
	\centering
	\setlength{\extrarowheight}{2pt}
	\renewcommand{\arraystretch}{1.6}
	\begin{tabular}{|c|c|c|c|c|c|c|c|}
		\hline
		{Target} & ${N_b}$ & $\sigma$, b & {L, cm} & $\frac{\Delta \sigma} {\sigma}=\frac{\sigma_{\uparrow\uparrow}-\sigma_{\downarrow\uparrow}} {\sigma}$ & $\frac{\text{Re}(\Delta f)}{\text{Im}(\Delta f)}$ & $|{p}_x| \approx \vartheta$ 
		\\ 		\hline
		NH$_3$ & $10^{10}$ & 0.5 & 30 & 0.02 & $ 0.5$ & $2 \cdot 10^{-3}$ 
		\\ 		\hline
		ND$_3$ & $10^{10}$ & 0.6 & 30 & 0.04 & $ 0.5$ & $4 \cdot 10^{-3}$ 
		\\  		\hline
	\end{tabular}
\end{table}


\subsection{Spin Rotation and Spin Dichroism} \label{sec:18.2}

With the increase of target thickness, the influence of scattering and
absorption on the polarization characteristics of a particle
transmitted through matter grows. Let, however, the target
thicknesses be such that the spin--dependent contributions to the
wave--function phase are small, i.e., inequalities $k \texttt{Re}g
z \ll 1$ and $k \texttt{Im}g z \ll 1$ (see (\ref{ins1_18.5})) are
valid. 
In this case 
we can make use of the above results
in order to find the polarization characteristics of
the beam. The spinor $W^{\prime}$ defining the
spin state of a particle  after passing the path
length $z$ in matter has the form
\begin{equation}
	\label{ins_18.14}
	W^{\prime}=e^{ik\hat{n}z}W\simeq e^{ikn_{0}z}(1+i\frac{2\pi\rho}{k}(\vec{\sigma}\vec{g})z)W\,.
\end{equation}
From (\ref{ins_18.14})
follows the expression for the number of particles $N$ transmitted
through the target in the same direction as the direction of the momentum of the particles $N_{0}$ incident on the target:
\begin{equation}
	\label{ins1_18.14}
	N=N_{0}e^{-\rho\sigma z}[1-\frac{4\pi\rho}{k}\vec{P}_{0}\texttt{Im}\vec{g}z]\,,
\end{equation}
where $\sigma$ is the spin--independent part of the total
scattering cross section determined by the imaginary part of
$n_{0}$ in  (\ref{ins_18.5})
\begin{equation}
	\label{ins12_18.14}
	\sigma=\frac{4\pi}{k} \texttt{Im}(A+A_{3}\vec{n}\vec{n}_{1}+\ldots)\,,
\end{equation}
$\vec{P}_{0}$ is the particle polarization vector before entering the target, $\vec{g}$ is defined by (\ref{ins1_18.5}).

According to (\ref{ins1_18.14}), the number of particles $N$
transmitted through the target  depends on the orientation of
$\vec{P}_{0}$: $N_{\uparrow\uparrow}\neq N_{\downarrow\uparrow}$,
where $N_{\uparrow\uparrow}$ describes $N$ for
$\vec{P}_{0}\uparrow\uparrow \texttt{Im} \vec{g}$,
$N_{\downarrow\uparrow}$ denotes $N$ for
$\vec{P}_{0}\downarrow\uparrow \texttt{Im} \vec{g}$.

So spin dichroism occurs because the absorption coefficient of incident particles in the target depends on the orientation of their spin.

Using (\ref{ins_18.14}) for the spin wave function $W^{\prime}$, we
may obtain the following expression for the polarization vector
$\vec{P}$ of the particles transmitted through a polarized target:
\begin{equation}
	\label{ins2_18.14} \vec{P}=\frac{\langle
		W^{\prime}|\vec{\sigma}|W\rangle}{\langle
		W^{\prime}|W\rangle}=\vec{P}_{0}+\frac{4\pi\rho z}{k}
	\texttt{Im}((\vec{P}_{0}\vec{g})\vec{P}_{0}-\vec{g}) z
	+\frac{4\pi\rho z}{k}[\vec{P}_{0}\times \texttt{Re}\vec{g}]\,.
\end{equation}
In view of (\ref{ins2_18.14}), the polarization vector of
high--energy particles, like that of low--energy particles,
undergoes rotation about the direction of $\texttt{Re}\vec{g}$. In
a similar manner as in the case of low energies, the contributions
associated with strong and P--, T--odd weak interactions can be
distinguished by measuring the magnitudes of $N$ and $\vec{P}$ for
different orientations of the polarization vector $\vec{P}_{0}$ of
the particles incident on the target.

It is worthy of mention that for  nuclei with spin $J\geq 1$, in
the cross-section $\sigma$ the term $\texttt{Im}
A_{3}\vec{n}\vec{n}_{1}$  determined by the target
quadrupolarization is different from zero.

Suppose that a nonpolarized beam of particles ($\vec{P}_{0}=0$) is
incident on a target (e.g. a beam of protons is incident on a
deuterium target). This situation is typical for the present--day
experiments with protons (antiprotons) and a deuterium gas target
in a storage ring or with beams in a collider, for example, NICA.

Let us choose the quantization axis along the direction $\vec{n}$ of
the beam momentum at the target location. We have (see (\ref{ins12_18.14}))
\begin{equation}
	\label{ins3_18.14}
	\sigma=\frac{4\pi}{k}(\texttt{Im}A + \texttt{Im}A_{3}Q_{zz}^{D})\,,
\end{equation}
where $Q_{zz}^{D}$ is the quadrupolarization tensor of the deuteron of the target.

Let the deuterons be in spin state with a magnetic quantum number
$M \pm 1$.

In this case  $Q_{zz}^{D}=\frac{1}{3}$.

But if the deuteron is in spin state  $M=0$, then $Q_{zz}^{D}=-\frac{2}{3}$.
Consequently, the below equalities may be written for the corresponding total interaction cross sections:
\begin{eqnarray}
	\sigma_{\pm 1}=\sigma_{\mathrm{non}}+\frac{4\pi}{3k} \texttt{Im} A_{3}\,,\nonumber\\
	\sigma_{0}=\sigma_{\mathrm{non}}-\frac{4\pi}{k}\frac{2}{3}\texttt{Im} A_{3}\,,
	\label{ins4_18.14}
\end{eqnarray}
$\sigma_{\mathrm{non}}$ is the total scattering cross section of a nonpolarized proton by a nonpolarized deuteron,
i.e.,
\begin{equation}
	\label{ins5_18.14}
	\texttt{Im} A_{3}=\frac{k}{{4\pi}}(\sigma_{\pm 1}-\sigma_{0})\,.
\end{equation}
According to (\ref{ins1_18.14}), the number $N(\pm 1)$ of protons
(antiprotons)  transmitted through a polarized target with deuterons in spin state $(M= \pm 1)$ is
\begin{equation}
	\label{ins6_18.14}
	N(\pm 1)= N_{0}e^{-\rho\sigma_{\pm 1}z}\,,
\end{equation}
while in the case when $M=0$, it is
\begin{equation}
	\label{ins7_18.14}
	N(0)=N_{0}e^{-\rho\sigma_{0} z}\,.
\end{equation}

In view of (\ref{ins1_18.14}), (\ref{ins6_18.14}),
(\ref{ins7_18.14}), the absorption of protons (antiprotons)
transmitted through the target will be different for different
orientations of the tensor polarization for the deuteron
target. Suppose now that a beam of nonpolarized particles
(protons, antiprotons, deuterons, nuclei) moves in a storage ring.
It is obvious that because the absorption coefficient of particles
depends on the orientation of  tensor polarization for the
deuteron target, the number of particles moving in a storage ring
will also depend on this characteristic of the target, i.e., the
lifetime of a nonpolarized beam in a storage ring depends on  the orientation of  tensor polarization for the
deuteron target \cite{rins_128,rins_127}. 
As a result, measuring  the beam
lifetime in this case, one can determine $\texttt{Im}A_{3}$, i.e.,
a spin-dependent part of the total cross-section of proton
(antiproton, deuteron) scattering by a polarized  deuteron, which
is proportional to $Q_{zz}$ .

\subsection
{Proton (Antiproton) Spin Rotation  in a Thick Polarized Target and Spin Filtering of Particle Beams in a Nuclear Pseudomagnetic	Field} \label{sec:protonspin}

With further increase in the target thickness the influence of the
spin--dependent part of particle absorption in matter is enhanced.

In order to obtain equations describing  the evolution of intensity and polarization of a beam in the target,
we shall split vector $\vec{g}$ into real and imaginary parts:
\begin{equation}
	\label{ins_protonspin}
	\vec{g}=\vec{g}_{1}+i\vec{g}_{2}\,,
\end{equation}
where $\vec{g}_{1}= \texttt{Re} \vec{g}$ and $\vec{g}_{2}=
\texttt{Im} \vec{g}$.

Using (\ref{18.4}), (\ref{18.5}), one may obtain the following system of equations defining the relation between the number of
particles transmitted through the target $N(z)$ and their polarization $\vec{P}(z)$ \cite{nim06_PR+}:
\begin{equation}
	\label{protonspin_1.51}
	\frac{d\vec{P}(z)}{dz}=\frac{4\pi\rho}{k}[\vec{g}_{1}\times\vec{P}(z)]-\frac{4\pi\rho}{k}
	\left\{\vec{g}_{2}-\vec{P}(z)[\vec{g}_{2}\vec{P}(z)]\right\}\,.
\end{equation}
Acting analogously, we can obtain the equation for the change in the polarized beam intensity in a polarized nuclear target:
\begin{equation}
	\label{protonspin_1.52}
	\frac{d N(z)}{dz}=-2\texttt{Im}(n_{0})k N(z)-\frac{4\pi\rho}{k}[\vec{g}_{2}\vec{P}(z)]N(z)\,.
\end{equation}
(\ref{protonspin_1.51}) and (\ref{protonspin_1.52}) should be solved together under the initial conditions
$\vec{P}(0)=\vec{P}_{0}$, $N(0)=N_{0}$. According to (\ref{protonspin_1.51}),
the polarization of the incident particles passing through the polarized
nuclear target undergoes rotation through the angle
\begin{equation}
	\label{protonspin_1.53} \theta=\frac{4\pi\rho}{k}g_{1} z\,.
\end{equation}

Assume that the target polarization $\vec{P}_{t}=\langle\vec{J}\rangle/{J}$ is
parallel (this is the case of longitudinal polarization) or orthogonal (transversal polarization)
to the momentum of the incident particle $\vec{k}$. Then $\vec{g}_{1}\parallel\vec{g}_{2}\parallel\vec{P}_{t}$
and  (\ref{protonspin_1.51}) and (\ref{protonspin_1.52}) are reduced
to a simple form.

Consider two specific cases for which the initial polarization of the incident beam is (a) $\vec{P}_{0}\parallel \vec{P}_{t}$
or  (b) $\vec{P}_{0}\perp\vec{P}_{t}$.

Case (a) is a standard transmission experiment wherein we observe the process of
absorption in the polarized target without a change in the direction of the initial beam
polarization. The absorption is different for particles polarized parallel and antiparallel to the target polarization.
The number of particles $N$  changes according to
\begin{equation}
	\label{protonspin_1.55} N(z)=N_{0}\exp(-\sigma_{\pm}\rho z)\,,
\end{equation}
where
\begin{equation}
	\label{protonspin_1.56}
	\sigma_{\pm}=\frac{4\pi}{k}\left[\texttt{Im}(A+A_{3}\vec{n}\vec{n}_{1})\pm\texttt{Im}A_{1}J
	P_{t}\pm \texttt{Im} A_{2} J P_{t}\right]\,.
\end{equation}
In case (b) the coherent scattering by the polarized nuclei
results in spin rotation of the incident particles about the target polarization $\vec{P}_{t}$.

According to (\ref{18.9}), the spin rotation angle
\begin{equation}
	\label{ins.protonspin_1.56}
	\vartheta=\frac{4\pi\rho}{k} \texttt{Re} g=\frac{4\pi\rho}{k}\left[\texttt{Re}A_{1} J P_{t}+\texttt{Re}A_{2} J (\vec{n}\vec{P}_{t})\right]z
\end{equation}
is directly connected with the real part of the forward scattering
amplitudes. The values $\texttt{Re}A_{1}$ and $\texttt{Re}A_{2}$
can be determined separately by measuring spin rotation angles for
two cases when the target spin is parallel and antiparallel to the
beam direction $\vec{n}$. This means that by measuring the final
intensity and polarization of the beam in cases (a) and (b)  we
can directly reconstruct the spin dependent forward scattering
amplitude.

In particular, as shown in \cite{nim06_PR+} the measurement of the rotation
angle near the resonances or near the threshold values at which  inelastic
reaction channels open up enables, in particular, the apprehension of the nature
of resonances and the investigation of the nature of threshold phenomena.

Let us assume now that the the target polarization is directed at some angle
(which does not equal $\pi/2$ ) with respect to the incident particle momentum
and that the incident beam polarization is perpendicular to the plane formed
by  vectors $\vec{P}_{t}$ and $\vec{n}$ (see Fig.~\ref{optical_fig1}). In this case the effect of proton (antiproton)
spin rotation about vector $\vec{g}_{1}$ combined with absorption dichroism,
determined by vector $\vec{g}_{2}$, will cause the dependence of the total number
of particles transmitted through the target on $\texttt{Re} A_{1}$ and  $\texttt{Re} A_{2}$ \cite{protonspin_18c,nim06_PR+}:

\begin{figure}[htbp]
	\epsfxsize = 4 cm \centerline{\epsfbox{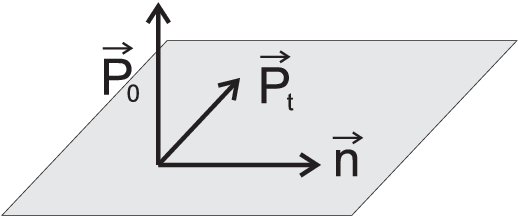}} \caption{Beam
		polarization $\vec{P}_0$ is perpendicular to the plane formed by
		vectors $\vec{P}_t$ and $\vec{n}$} \label{optical_fig1}
\end{figure}

\begin{equation}
	\label{ins.protonspin_1.56+}
	N(z)\sim(\texttt{Re}A_{1}\cdot\texttt{Im}A_{2}-\texttt{Re}A_{2}\cdot\texttt{Im}A_{1})\left[\vec{P}_{t}
	\times\vec{n}(\vec{n}\vec{P}_{t})\right]\vec{P}_{0}\,.
\end{equation}
Such behavior of $N(z)$ enables measuring spin--dependent
contributions to the amplitude $f(0)$ in the transmission
experiment without measuring the polarization of the beam
transmitted through the target \cite{protonspin_18c,nim06_PR+}.

This implies that if a vertically polarized beam rotates in the storage ring (the beam's spin is
orthogonal to the orbit plane), then the decrease in the beam intensity (the beam lifetime in the storage ring)
in the case when vectors $\vec{P}_{t}$ and $\vec{n}$ lie in the horizontal plane (in the orbit plane) enables
determining $\texttt{Re} A_{1}$ and  $\texttt{Re} A_{2}$.

Note that the contribution under study reverses sign when vector
$\vec{P}_{t}$  rotates about $\vec{n}$ through $\pi$ (or
$\vec{P}_0$ changes direction: $\vec{P}_0\rightarrow -\vec{P}_0$).
Measuring the difference between the beam's damping times for
these two orientations of $\vec{P}_{0}$ ($\vec{P}_{t}$),
one can find the contribution of $\texttt{Re} A_{1}$ and  $\texttt{Re} A_{2}$ to the real part of the amplitude $f(0)$.

Unlike a spin rotation experiment, this transmission experiment does not allow one to determine
$\texttt{Re}A_{1}$ and $\texttt{Re}A_{2}$ separately.

There is also an inverse process: if a nonpolarized beam of
particles is incident on the target, then after passing through
the target, the beam acquires polarization orthogonal to the plane
formed by $\vec{P}_{t}$ and $\vec{n}$. This phenomenon is most
interesting in the case of a storage ring. If $\vec{P}_{t}$ and
$\vec{n}$ lie in the orbit plane of a circulating beam (i.e., the
horizontal plane), then in the course of time the beam acquires
vertical polarization orthogonal to the orbit plane. By measuring
the arising vertical polarization it is also possible to determine
the spin--dependent part of the  coherent elastic zero-angle
scattering amplitude. In particular, in spin--filtering
experiments for obtaining polarized beams of antiprotons (protons)
\cite{24a,R,PAX,mil,nik1,nik2,nik,mil1,PAX1}, it is sufficient to
perform an experiment under the conditions when the polarization
vector $\vec{P}_t$ of a gas target is directed at a certain angle,
which is not equal to $0,\, \pi$ or $\pi/2$, with respect to the
particle momentum direction $\vec{n}$. If $\texttt{Re}A_{1,2}\sim
\texttt{Im}A_{1,2}$, the degree of arising polarization is
comparable to the anticipated degree of polarization of the
anti(proton) beam arising in the
method \cite{24a,R,PAX,mil,nik1,nik2,nik,mil1,PAX1} and enables one to measure the contribution proportional to $\texttt{Re}A_{1,2}$.

Worthy of mention is that in the case under consideration the beam
lifetime depends on the orientation of $\vec{P}_t$ in the
$\vec{P}_t,\vec{n}$-plane. In this case the measurement of the
beam lifetime also gives information about $\texttt{Re}A_{1,2}$.

Let us note that the terms similar to those proportional to $A_1$
and $A_2$ in the expression for the zero-angle scattering
amplitude also appear in the expression for the scattering
amplitude of particles with spin $S\ge 1$ (nuclei). Hence, the phenomena described above exist for
particles with spin  $S\ge 1$,  too
\cite{Nuclear_optics,A2}.

\subsection{Spin density matrix formalism}

Under the conditions when multiple scattering appears important,
quantum mechanical description of a beam of polarized particles
traveling through a polarized target utilizes the spin density
matrix $w$.

In our case, we treat the polarized target as a thermal reservoir
with an infinite set of degrees of freedom which we then average.
Therefore the density matrix for the system, $\hat{\rho}(b, T,
t)$, consisting of the polarized target plus polarized particle
beam, can be written as a direct product of the matrices
$\hat{\rho}_b (t)$  and $\hat{\rho}_T(t)$ ($\hat{\rho}_b(t)$
is the beam spin density matrix, $\hat{\rho}_T(t)$ is the target
spin density matrix):
\begin{equation}
	\label{protonspin_1.12} \hat{\rho} (b, T,
	t)=\hat{\rho}_b(t)\otimes \hat{\rho}_T (t)\,.
\end{equation}
In accordance with the general theory  \cite{23a} the master equation for the spin density matrix of above system can be expressed in
a form \cite{VG+nastya,nim06_PR+,rins_98,A2} as follows:
\begin{equation}
	\label{protonspin_1.13}
	\frac{d\hat{\rho}_b(t)}{dt}=-\frac{i}{\hbar}(\hat{H}\hat{\rho}_b-\hat{\rho}_b\hat{H})+
	\left(\frac{d\hat{\rho}}{dt}\right)_{\mathrm{sct}}
\end{equation}
with the Hamiltonian $\hat{H}$ including the interaction between
particles of the beam and the external electromagnetic fields. The
term $(d\hat{\rho}/dt)_{\mathrm{sct}}$  describes the change in
the density matrix due to collisions in the target.

A detailed account of general methods for obtaining the explicit
form of the collision integral is given in many course books
\cite{23a,Goldberger}. 
They can also be applied to the consideration of the
interaction between  polarized beams and a polarized target
\cite{nim06_PR+,VG+nastya}.

Let us consider the process of particle passage through a target
consisting of $N_{T}$ particles interacting with one another by
means of a certain potential $U$.

Suppose that an incident particle (proton, antiproton, deuteron,
nucleus) has a rest mass $m$ and spin $S_{b}$. The target consists
of $N_{T}$ bound particles with mass $M$ and spin $s$.

The Hamiltonian of the scattering system is written in the form:

\begin{equation}
	H_{T}=\sum_{\alpha=1}^{N_{T}}K_{\alpha}+U\,, \label{shir_1.1}
\end{equation}
where $K_{\alpha}$ is the kinetic energy operator of particle
$\alpha$, $U$ is the interaction energy of  $N_{T}$ particles of
the scatterer.

The solution of the Schr\"{o}dinger  equation
\begin{equation}
	H_{T}\Psi_{n}=W_{n}\Psi_{n} \label{shir_1.2}
\end{equation}
determines the possible values of the energy $W_{n}$ of the system
and the corresponding set of wave functions

\begin{equation}
	\Psi_{n}=\Psi_{n}(\vec{R}_{1},s_{1},\ldots,\vec{R}_{N},s_{N})\,,
	\label{shir_1.3}
\end{equation}
where $(\vec{R}_{\alpha},s_{\alpha})$ are the spatial and spin
coordinates of particle $\alpha$  in the target.

The operator of interaction between the incident particle and the
target is defined in terms of $V$: $\displaystyle
V=\sum_{\alpha=1}^{N_{T}}V_{\alpha}$.
Here $V_{\alpha}$ describes the interaction between the beam
particle  and the target particle $\alpha$.

The Hamiltonian of the
whole system has the form:
\begin{equation}
	H=K_{b}+H_{T}+V\,,
	\label{shir_1.4}
\end{equation}
where $K_{b}$ is
the kinetic energy operator of the beam particle $b$.

To describe the process of particle $b$ transmission through the
target, find the density matrix $\hat{\rho}(t)$ of the system
"incident particle + target". This density matrix satisfies the
quantum Liouville equation \cite{23a}:

\begin{equation}
	i\frac{\partial\hat{\rho}(t)}{\partial
		t}=[\hat{H},\hat{\rho}(t)]\,. \label{shir_1.5}
\end{equation}
The solution of this equation can formally be written using the
evolution operator $\hat{U}(t^{\prime},t)$

\begin{equation}
	\displaystyle
	\hat{\rho}(t^{\prime})=\hat{U}(t^{\prime},t)\hat{\rho}(t)\hat{U}^{\scriptscriptstyle+}(t^{\prime},t)\,,
	\label{shir_1.6}
\end{equation}
which is related to the explicitly time--independent Hamiltonian
of the system as
$\hat{U}(t^{\prime},t)=e^{-\frac{i}{\hbar}\hat{H}(t^{\prime}-t)}$.

Let us consider the target as a thermostat ($N_{T}\gg1$). Then the
statistical operator $\hat{\rho}$ of the system can be represented
as a direct product
$\hat{\rho}=\hat{\rho}_{b}\otimes\hat{\rho}_{T}$, where
$\hat{\rho}_{b}$ is the density matrix of the incident particle,
$\hat{\rho}_{T}$ is the equilibrium density matrix of the medium.
The equilibrium density matrix $\hat{\rho}_T$ of the target is diagonal in  the stationary states $n$
of the target: $\langle n|\hat{\rho}_T|n^{\prime}\rangle=\delta_{n
	n^{\prime}} \cdot \hat{\rho}_{n n^{\prime}}=\delta_{n n^{\prime}}
\hat{\rho}(n)$. (Note that $\hat{\rho}_T(n)$ is the operator in
the spin space of a nuclear target).  The spin density matrix
$\hat{\rho}_b$ includes the elements diagonal and nondiagonal with
respect to the momenta $\vec{k}$ and $\vec{k}^{\prime}$ of the
particle: $\hat{\rho}_b=\hat{\rho}_b(\vec{k},\vec{k}^{\prime})$.
However, the nondiagonal elements oscillate fast and after several
collisions with target nuclei, in order to describe the process of
multiple scattering, one can assume that the density matrix
$\hat{\rho}_b(\vec{k},\vec{k}^{\prime})$ is diagonal
\cite{protonspin_25,23a,cosy_bar99,nim06_PR+,VG+nastya}, i.e.,
$\hat{\rho}_b(\vec{k},\vec{k}^{\prime})=\delta_{\vec{k},\vec{k}^{\prime}}\hat{\rho}_b(\vec{k})$.
We shall note here that the particle spin density matrix
$\hat{\rho}_b (\vec{k})$ is nondiagonal in the spin states
describing a particle incident on the target.

The  time interval $\Delta t$, during which the density matrix
$\hat{\rho}_b(\vec{k}, \vec{k}^{\prime})$ is diagonalized,
satisfies the inequality $\Delta t\gg R/\bar{v}$, where $R$ is the
radius of action of the forces, $\bar{v}$ is the particle mean
velocity in matter. We shall further concern ourselves with the
behavior of the system over the time interval $\Delta
t\gg\frac{R}{\bar{v}}$.
This means that the evolution operator $U(t^{\prime},t)$ can be
replaced by the Heisenberg's $S$-matrix
$S\equiv\lim_{{t^{\prime}\rightarrow\infty},{t\rightarrow-\,\infty}}U(t^{\prime},t)$,
which relates the asymptotic states of the system before
scattering to those after scattering
\cite{protonspin_25,23a,cosy_bar99,nim06_PR+,VG+nastya}.
The matrix elements of the $S$-matrix for a scattering system
consisting of $N_{T}$ particles are defined as follows:
\begin{equation}
	\displaystyle
	S_{fa}=\delta_{fa}-i(2\pi)\delta(E_{f}-E_{a})\sum_{\alpha=1}^{N_{T}}\left(\mathcal{T}_{\alpha}\right)_{fa}\,,
	\label{shir_1.7}
\end{equation}
where $E_{a}$  and  $E_{f}$ are the total energies of the system
before and after scattering, respectively;
$E_a=\varepsilon_k+W_{n}$,
$E_f=\varepsilon_{k^{\prime}}+W_{n^{\prime}}$, $\varepsilon_{{k}}$
and $\varepsilon_{{k}^{\prime}}$ are the energies of the incident
particle before and after the collision; $\mathcal{T}_{\alpha}$ is
the scattering matrix of particle $b$ interacting with a free
particle $\alpha$.  We shall make use of the fact that in the case
of high-energy particles, the energies of the particles are much
greater than the binding energies of the scatterers in the target.
This enables using impulse approximation \cite{Goldberger}.

In the impulse approximation, the scattering matrix
$\mathcal{T}_{\alpha}$  coincides with the  matrix of scattering
of particle $b$ by a free particle $\alpha$
\cite{Goldberger}.
Expression 
(\ref{shir_1.6}) can be rewritten for the density
matrix of the incident particle, using the $S$-matrix  defined in
(\ref{shir_1.7}):
\begin{equation}
	\hat{\rho}_{b}(t+\Delta t)=\textrm{Tr}_{T}S\hat{\rho}(t)S^{+}\,,
	\label{shir_1.8}
\end{equation}
where $\textrm{Tr}_T$ means  taking the trace over the states of
the target.

Write  (\ref{shir_1.8}) for the elements of the density matrix
which are diagonal in the  momentum space:
\begin{eqnarray}
	& &\hat{\rho}_{b}(t+\Delta t,
	\vec{k})=\hat{\rho}_{b}(t,\vec{k})-i(2\pi)\frac{\Delta
		t}{2\pi}{Tr}_{T}\sum_{\alpha=1}^{N_T}
	\langle\vec{k},n|\hat{\mathcal{T}}_{\alpha}|\vec{k}, n\rangle\hat{\rho}(t, \vec{k},n)\nonumber\\
	& &+ i(2\pi)\frac{\Delta
		t}{2\pi}{Tr}_{T}\sum_{\alpha=1}^{N_T}\hat{\rho}(t, \vec{k},
	n) \langle\vec{k},
	n|\hat{\mathcal{T}}_{\alpha}^{\scriptscriptstyle+}|\vec{k},
	n\rangle
	\nonumber\\
	& &+(2\pi)^{2}\frac{\Delta
		t}{2\pi}{Tr}_{T}\sum_{\alpha,\beta=1}^{N_T}\sum_{\vec{k}'
		n'}
	\langle\vec{k}, n|\hat{\mathcal{T}}_{\alpha}|\vec{k}', n'\rangle\hat{\rho}(t, \vec{k}',n')\nonumber\\
	& &\times\delta(\varepsilon_{k'} -\varepsilon_{k}+W_{n'}-W_{n})
	\langle\vec{k}',
	n'|\hat{\mathcal{T}}_{\beta}^{\scriptscriptstyle+}|\vec{k},
	n\rangle\,. \label{shir_1.9}
\end{eqnarray}

Let us recall that $\hat{\rho}(t,
\vec{k}^{\prime},n)=\hat{\rho}_b(t,\vec{k})\otimes
\hat{\rho}_T(n)$. In the momentum space, the matrix elements for
the scattering matrix operator $\hat{\mathcal{T}}_\alpha$ have the
form \cite{23}:
\begin{equation}
	\langle\vec{k}',\vec{P}_{\alpha}'|\hat{\mathcal{T}}_{\alpha}|\vec{k},\vec{P}_{\alpha}\rangle=(2\pi)^{3}
	\delta(\vec{k}'+\vec{P}_{\alpha}'-\vec{k}-\vec{P}_{\alpha})
	\langle\vec{k}',\vec{P}_{\alpha}'|\hat{\mathcal{T}}_{\alpha}|\vec{k},\vec{P}_{\alpha}\rangle\,,
	\label{shir_1.10}
\end{equation}
$\vec{P}_{\alpha}$ denotes the momenta of the  $\alpha$-th
scatterer, $\hat{\mathcal{T}}_{\alpha}$ is the scattering matrix on
the momentum shell.
Recall that  $\hat{\mathcal{T}}_{\alpha}$ and
$\hat{\mathcal{T}}_{\alpha}$ remain the operators with respect to
spin variables.

The states with a definite value of the momentum $|\vec{k}\rangle$
are normalized according to equality
\[
\langle\vec{k}'|\vec{k}\rangle=\frac{(2\pi)^{3}}{V}\,\delta(\vec{k}'-\vec{k})\,,
\]
where $V$ is the normalization volume. Thus, substitution of
summation over all $\vec{k}$ by integration is made as follows
\[
\sum_{\vec{k}}\rightarrow \frac{V}{(2\pi)^{3}}d^{3}\vec{k}\,;
\]
the Kronecker symbol and the
Dirac function  $\delta$ are related as
\[
\delta_{\vec{k}\vec{k}'}\rightarrow\frac{(2\pi)^{3}}{V}\delta(\vec{k}-\vec{k}')\,.
\]

Scattering processes that make the contribution to (\ref{shir_1.9})
can be both elastic and inelastic. Elastic scattering occurs in
the absence of any excitation in the scatterer. Inelastic
scattering accompanied by a small excitation in the scatterer is
called "quasi-elastic" \, \cite{Goldberger}. For such scattering the
momentum $\vec{q}$ transferred to the $\alpha$-th scatterer is
$q\ll\sqrt{2M K_{\alpha}}$, $q\ll\sqrt{2M U_{\alpha}}$,
$\vec{q}=\vec{k}'-\vec{k}$; $\vec{k}$ is the momentum of the
incident particle before the collision with the scatterer,
$\vec{k}'$ is the momentum of the beam particle after the
collision.

When the momentum transferred to the scatterer $\alpha$ exceeds
the momentum of the target particle in the initial bound state
$q\gg\sqrt{2M K_{\alpha}}$, $q\gg\sqrt{2M U_{\alpha}}$, then the
so-called "quasi-free"  approximation  can be used. In this
approximation the excitation energy of the system in one collision
is exactly equal to the recoil energy $\vec{q}\,^2/2M$ of a free
particle of the target. Note that in summation over different
nuclei of the target, sums of the form given below appear in the
last term in (\ref{shir_1.9})
\begin{eqnarray}
	\displaystyle \sum_{\alpha,\beta=1}^{N_{T}}\int
	d^{3}\vec{R}_{1}...d^{3}\vec{R}_{N}
	e^{-i\vec{q}(\vec{R}_{\alpha}-\vec{R}_{\beta})}\hat{\rho}_{T}(\vec{R}_{1},...\vec{R}_{N})=N_{T}\hat{\rho}_{T}
	\nonumber\\
	\displaystyle+\sum_{\alpha\neq\beta}^{N_T}\int
	d^{3}\vec{R}_{1}...d^{3}\vec{R}_{N}
	e^{-i\vec{q}(\vec{R}_{\alpha}-\vec{R}_{\beta})}\hat{\rho}_{T}(\vec{R}_{1},...\vec{R}_{N})\,,
	\label{shir_1.11}
\end{eqnarray}
where $\hat{\rho}_T$ is the spin density matrix of the target
nucleus. In deriving (\ref{shir_1.11}) it is assumed that the
positions of target nuclei and their spin states are uncorrelated.

Upon averaging over the spin states of nuclei in the target,
the second term in (\ref{shir_1.11}) can be expressed in terms of
the so-called particle pair distribution function \cite{Goldberger}, and
vanishes when the transferred momentum $\vec{q}$ exceeds the
magnitude inverse to the correlation radius $r$, i.e.,
$\displaystyle q\gg r^{-1}$ \cite{Goldberger}. For a non-crystalline
target, the magnitude of the correlation radius $r$ is of the
order of the distance between nuclei. Consequently, the second
term can only contribute to the kinetic equation at very small
scattering angles $\vartheta_{sc}\lesssim{1}/{k r}$, so it will be
neglected in further consideration.

The resulting expression for the density matrix can be obtained
from (\ref{shir_1.9}) for
\[
\frac{\hat{\rho}_b(t+\Delta t)-\hat{\rho}_b(t)}{\Delta
	t}=\frac{\Delta\hat{\rho}_b(t)}{\Delta t}\,.
\]
The time interval $\Delta t$ should be chosen to be much greater
than the characteristic correlation time of the process (the time
interval during which the density matrix is diagonalized), but
still small enough to fulfil the condition that the difference
$\Delta\hat{\rho}_b$ is linear over $\Delta t$.

Thus, the equation  for the density matrix describing the
behavior of the polarized proton (antiproton, deuteron, nucleus)
beam in polarized matter can be written as:
\begin{eqnarray}
	& &\frac{d\hat{\rho}_{b}(\vec{k},t)}{dt}=- iVN_{T}
	{Tr}_{T}\left(\hat{\mathcal{T}}(\vec{k},\vec{k})
	\hat{\rho}(\vec{k},t)-\hat{\rho}(\vec{k},t)\hat{\mathcal{T}}^{\scriptscriptstyle+}(\vec{k},\vec{k})\right)
	\nonumber\\
	& &+\frac{V^{3}}{(2\pi)^{2}}N_{T}{Tr}_{T}\int d^{3}\vec{k}'
	\delta\left(\varepsilon_{k}-\varepsilon_{k'}-\frac{\vec{q}^{2}}{2M}\right)\nonumber\\
	& &\times
	\hat{\mathcal{T}}(\vec{k},0;\vec{k}',-\vec{q})\hat{\rho}(\vec{k}',t)
	\hat{\mathcal{T}}^{\scriptscriptstyle+}(\vec{k}',-\vec{q};\vec{k},0)\,.
	\label{shir_1.12}
\end{eqnarray}
$\hat{\rho}(\vec{k})$ denotes the following dependence:
$\displaystyle\hat{\rho}(\vec{k})=\hat{\rho}_{b}(\vec{k};\vec{S}_{b})\otimes\hat{\rho}_{T}(\vec{S}_{T})$,
where $\hat{\rho}_T(\vec{S}_{T})$ is the spin density matrix of
the target nucleus.

Let us introduce the scattering amplitude $\hat{F}$ with matrix
elements equal to \cite{23}:
\begin{equation} \displaystyle
	\hat{F}(\vec{k},\vec{k}')=-\frac{M_{r}}{2\pi}V^{2}\hat{\mathcal{T}}(\vec{k},0;\vec{k}',-\vec{q})\,,
	\label{shir_1.13}
\end{equation}
where
\[
M_{r}=\frac{mM}{m+M}
\]
is the reduced mass.

Then (\ref{shir_1.12}) can be transformed into a form:
\begin{eqnarray} \displaystyle
	\frac{d\hat{\rho}_{b}(\vec{k},t)}{dt}=\frac{2\pi i}{M_{r}}N_T
	{Tr}_{T}\left(\hat{F}(\vec{k},\vec{k})
	\hat{\rho}(\vec{k},t)-\hat{\rho}(\vec{k},t)\hat{F}^{\scriptscriptstyle+}(\vec{k},\vec{k})\right)
	\nonumber\\
	\displaystyle+N_T{Tr}_{T}\int d\Omega_{\vec{k}'}
	\frac{k'^{2}}{M_{r}^{2}\left(\frac{k'}{m}-\frac{(\vec{k}-\vec{k}')\vec{n}'}{M}\right)}
	\hat{F}(\vec{k},\vec{k}')\hat{\rho}(\vec{k}',t)
	\hat{F}^{\scriptscriptstyle+}(\vec{k}',\vec{k})\,,
	\label{shir_1.14}
\end{eqnarray}
where $N_T$  denotes the number of  particles in the target per
unit volume, $\vec{n}'$ is the unit vector in the direction of the
momentum $\vec{k}'$. The absolute value of vector $\vec{k}'$ is
determined from the equation:
\begin{equation} \displaystyle
	\varepsilon_{k}=\varepsilon_{k'}+\frac{(\vec{k}-\vec{k}')^{2}}{2M}\,.
	\label{shir_1.15}
\end{equation}
Note that when the condition $m\geq M$ for the masses of the
incident particles is fulfilled, the denominator in the integrand
of (\ref{shir_1.14}) tends to zero for the value of the scattering
angle of the incident particle $\theta$, which is defined by the
equality:
\[
\cos\theta=\frac{\sqrt{m^{2}-M^{2}}}{m}\,,
\]
with the absolute value of vector $\vec{k}'$
\[
\vec{k}'=k\sqrt{\frac{m-M}{m+M}}\,.
\]
Expression (\ref{shir_1.14}) is simplified when a particle (proton, deuteron,
antiproton) passes through a target with nuclei whose mass is much
larger than the mass of the incoming particle. In this case we can
neglect the effect of the energy loss of the incident particle
through scattering. So we can neglect the recoil energy
$\displaystyle{\vec{q}^2}/{2M}$ in the $\delta$-function
(\ref{shir_1.12}), (\ref{shir_1.15}). As a result, one obtains a
simple kinetic equation describing the time and spin evolution of
the incident particle as it passes through the target
\cite{nim06_PR+,VG+nastya}:
\begin{eqnarray}
	& &\frac{d\hat{\rho}_b(\vec{k},t)}{dt}=\frac{2\pi N_T}{m}
	{Tr}_{T}\left[\hat{F}(\vec{k},\vec{k})
	\hat{\rho}_b(\vec{k},z)-\hat{\rho}_b(\vec{k},z)\hat{F}^{\scriptscriptstyle+}(\vec{k},\vec{k})\right]\nonumber\\
	& &+N_T\frac{k}{m}{Tr}_{T}\int d\Omega_{\vec{k}'}
	\hat{F}(\vec{k},\vec{k}')\hat{\rho}(\vec{k}',t)
	\hat{F}^{\scriptscriptstyle+}(\vec{k}',\vec{k})\,,
	\label{shir_1.16}
\end{eqnarray}
where $|\vec{k}|=|\vec{k}'|$.

The first term on the right--hand side of (\ref{shir_1.16}), which
describes refraction of particle in the target, can be represented
as follows:
\begin{eqnarray}
	& &\hat {F}(0)\hat {\rho }(\vec {k},t) - \hat {\rho }(\vec
	{k},t)\hat
	{F}^ + (0) = \\
	& &\left[ \frac{1}{2}\left( {\hat {F}(0) + \hat
		{F}^{{\scriptscriptstyle +}} (0)}\right),\hat {\rho }(\vec {k},t)
	\right] + \left\{ \frac{1}{2}\left( {\hat {F}(0) - \hat
		{F}^{\scriptscriptstyle +} (0)} \right),\hat {\rho }(\vec {k},t)
	\right\}\,,\nonumber \label{shir_1.17}
\end{eqnarray}
\noindent where $\left[ ...\,,\,... \right]$ is the commutator, $\left\{\, ...\,,\,...
\right\}$ is the anticommutator.

The part proportional to the commutator leads to the rotation of
the polarization vector due to elastic coherent scattering (as a
result of the refraction effect \cite{nim06_PR+,VG+nastya}); the
anticommutator describes the reduction in the intensity and
polarization of the beam which has passed through the target. The
last term in (\ref{shir_1.16}) determines the effect of incoherent
scattering on the change of $\hat{\rho}_b$ (in the general case,
single and multiple scattering).

As stated above, (\ref{shir_1.16}) is not applicable to the
description of the process of proton (deuteron) transmission
through the target containing light nuclei (protons, deuterons).
To describe multiple scattering in this case, a more general
(\ref{shir_1.14}) should be solved.


As a result, it is possible to find the dependence of the intensity and
the polarization characteristics of the beam on the direction of
the particle scattering and on the distance $z$ traveled by the
particle in matter.

In a real experiment, the scattered particles
are registered  within a certain interval of finite momenta
because the collimator of the  detector has a finite angular
width. Therefore we should study  the
characteristics of the beam transmitted through the target in the
interval of solid angles $\Delta \Omega $ with respect to the
initial direction of the beam propagation.

In fact, due to the
axial symmetry of the collimator, $\Delta \Omega $ is determined
by the angular width $2\vartheta _{\mathrm{det}}$ of the detector collimator.

\begin{figure}[!h]
	\centering
	\includegraphics[scale=0.7]{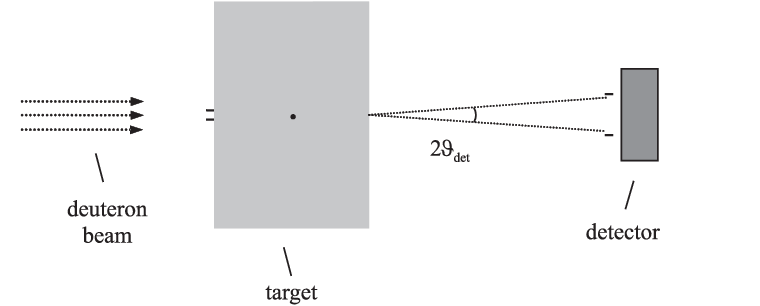}
	\caption{Scheme of transmitted beam detection  by the central
		detector}
	\label{shir_FigGr23}
\end{figure}

Equation (\ref{shir_1.16}) was analyzed in
\cite{nim06_PR+,cosy_bar99,VG+nastya}, where  two physical mechanisms of spin
rotation were indicated: one due to refraction of particles in a
polarized target ("optical" spin rotation) (the second term in
(\ref{shir_1.16}))
and the other, appearing as a result of spin rotation through incoherent scattering and caused by Coulomb--nuclear
interaction and incoherent scattering by nuclei \cite{nim06_PR+,B9,cosy_bar99,bar+shirv,A2}
(the third term in (\ref{shir_1.16}) (see \cite{bar+shirv,VG+nastya}).  As is shown in \cite{bar+shirv}, using different angular
resolutions $\Delta\Omega$ of the detector,  one can  study different contributions  to spin rotation.

Thus, multiple scattering does not cancel the effect of
spin rotation of charged particles in a polarized
target.


\section{The phenomenon of birefringence (spin oscillation and spin dichroism)
	of particles with spin $ S\ge 1$}
\label{sec:birefringence1}


In the previous sections, we have found out that refraction of
particles in matter with polarized nuclei leads to "optical" spin
rotation in a pseudomagnetic nuclear field. This effect is
kinematically analogous to the effect of light polarization plane
rotation in matter placed in a magnetic field.

However, it is known  \cite{B2,vfel_Born,B4,B3} that light
demonstrates another interesting phenomenon: the birefringence
effect in an optically anisotropic medium. This implies that as
light passes through such a medium, its linear polarization (which
is a vector) is converted into circular one and then the circular
polarization (which is a pseudovector) is converted into  linear
one.

The birefringence effect is accompanied by linear dichroism: the absorption coefficient of
light with linear polarization parallel to the optic axis of matter differs from that of light
with linear polarization orthogonal to the optic axis of matter.

It is appropriate to raise a question of whether similar birefringence and dichroism effects exist
for particles other than the photon.

The answer appears to be affirmative \cite{232,110}. According to
\cite{232,110}, a quasi-optical phenomenon of birefringence is
possible for particles with spin $S\geq 1$ which move in a
nonpolarized homogeneous isotropic medium.


Appearance of two refraction indices of deuteron can be easily explained

\begin{figure}[tbph]
	\epsfxsize=7.0 cm \centerline{\epsfbox{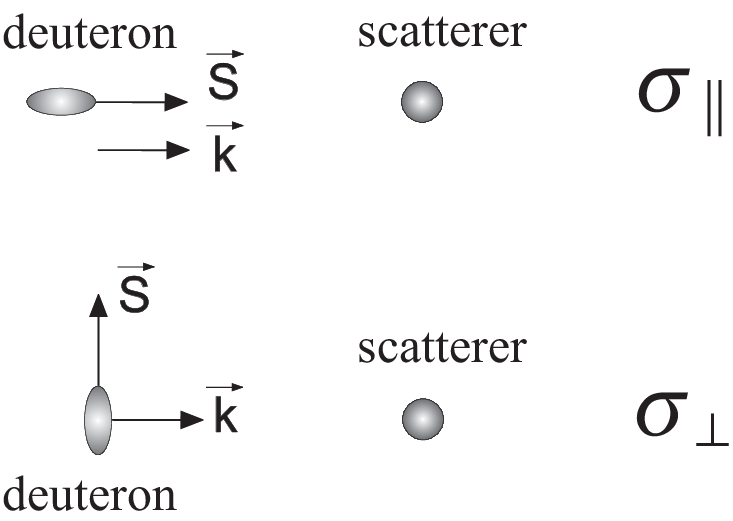}} 
	\caption{Explanation of deuteron birefringence effect}
	\label{fig:deuteron}
\end{figure}

As the ground state of deuteron is non-spherical, then it has different scattering
cross-sections depending on the angle between spin and momentum of scattering deuteron.

$$
\textrm{Im}{f_{\parallel}}(0)=\frac{k}{4 \pi} \sigma_{\parallel}~~\ne~~\textrm{Im}{f_{\perp}}(0)=\frac{k}{4 \pi} \sigma_{\perp}
$$

According to the dispersion relations
\[
Re~f(0) \sim \Phi  {\large ( Im~f(0)  )}
\]
and as a result
$$
Re~f_{\perp}(0) \ne Re~f_{\parallel}(0).
$$

In this case the vector
polarization of a particle is converted into tensor polarization
and vice versa.

This phenomenon also exhibits spin dichroism, which results in the fact that an initially nonpolarized deuteron beam
acquires tensor polarization after passing through a nonpolarized homogeneous isotropic medium.

\begin{figure}[h]
	\epsfxsize = 8 cm \centerline{\epsfbox{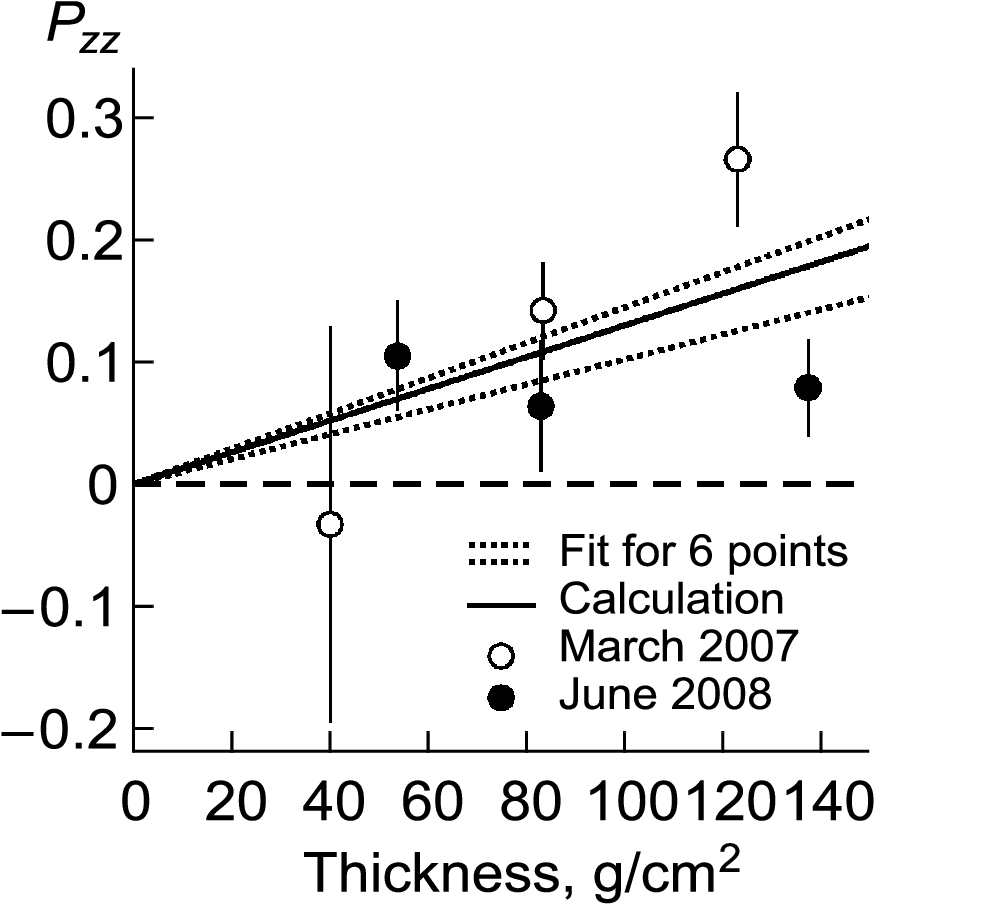}}
	\caption{A tensor polarization of deuterons as a function of the thickness of the carbon target 
		obtained in \cite{azhgirei2010} (Fig.11 therein)}
	\label{fig:slide28}
\end{figure}

For particles, the birefringence effect is due to intrinsic anisotropy of particles with spin $S\geq 1$. Currently,
the spin dichroism--related effect of acquisition  of tensor polarization by a nonpolarized deuteron beam which has passed
through a nonpolarized target is experimentally revealed for deuterons  of energy of 10--20\,MeV \cite{spinorb_ex1,spinorb_ex2,rins_63}
and deuterons with the momentum of 5\,GeV/\,s  transmitted through a carbon target \cite{VKB_dub,spinorb_ex3,azhgirei2010}.
Measured tensor polarization for deuterons with the momentum of 5\,GeV/\,s  transmitted through a carbon target as published in \cite{azhgirei2010} is shown in Fig.~\ref{fig:slide28}.

It is worthy of mention here that for neutrons (protons) and other
particles of spin $S=\frac{1}{2}$, the phenomenon similar to
birefringence of light is impossible. A particle with spin
$S=\frac{1}{2}$ possesses only one polarization characteristic
$\sim \langle\vec{S}\rangle$ --- the polarization vector which is
a pseudovector. No other vector and tensor quantities may be
constructed from the operators of spin
$\vec{S}=\frac{1}{2}\vec{\sigma}$  alone: all the products of
Pauli matrices of the type $\sigma_{i} \sigma_{n}$, etc., are
finally reduced to the Pauli matrix again \cite{Landau_3}.

The situation is different for particles with spin $S\geq 1$. In
this case the polarization state of  particles is described by
both the polarization vector and the polarization tensors. In
particular, for a particle with spin $S=1$ (e.g., deuteron), there
are two polarization characteristics: the polarization vector
$\sim\langle \vec{S}\rangle$ and the quadrupolarization tensor
(tensor of rank two) $\sim S_{i}S_{k}$ ($i$, $k$ correspond to the
axes of the Cartesian coordinate system $x$, $y$, $z$) \cite{Landau_3}.

It should be pointed out here that there is a difference between the polarization state of a photon (zero rest mass) and
that of a particle with a nonzero rest mass. For a photon, the polarization state with a magnetic quantum number $M=0$
(the quantization axis is directed along the particle momentum) is missing. At the same time, such a state exists for a particle with mass.

Let us proceed to  consideration of the refraction of particles
with  spin $S \ge 1$.

We know already (see (\ref{I.1})) that the refractive index of particles with spin
$S$ can be written as follows
\begin{equation}
 \hat{N}=1+\frac{2\pi\rho}{k^{2}}\hat{f}(0)\,,
 \nonumber
\end{equation}
where $\hat{f}(0)=\textrm{Tr}\hat{\rho}_{J}\hat{F}(0)$;
$\hat{\rho}_{J}$ is the spin density matrix of the scatterer;
$\hat{F}(0)$ is the operator amplitude of forward scattering that
acts in the spin space of the particle and the scatterer with spin
$J$.

If at entering the target the particle wave
function is $\psi_{0}$, then after passing the path length  $z$,  it will be
$\psi=\exp[ik\hat{N}z]\psi_{0}$.

The explicit form of the
amplitude $\hat{f}(0)$ for  particles with spin $1/2$ was obtained above.
In the case in question  we have at our disposal three parameters for forward
scattering: $\vec{S}$, $\vec{J}$, and $\vec{n}=\vec{k}/k$;
$\vec{k}$ is the particle wave vector.

It is known \cite{Landau_3} that the spin matrix of dimensionality
$(2S+1)(2S+1)$ can be expanded in terms of a complete set of
$(2S+1)^{2}$ matrices, in particular in terms of a set of
polarization operators $\hat{T}_{LM}(S)$, where $0\ll L\ll 2S$,
$-L\ll M\ll L$. The polarization operator is an irreducible tensor
of rank $L$. The maximal rank of $\hat{T}_{LM}$ is $2S$. The same
matrix can be expanded in terms of a set of Cartesian tensors of
maximal rank $2S$, i.e., in terms of a set of  products ${S_{i}
	S_{k} S_{l}\ldots S_{m}}$ with the maximal number of factors in
this product equal to $2S$. These Cartesian tensors are reducible
and may be represented as a sum of irreducible tensors. The matrix
$\hat{F}$ operating in the spin space of $\vec{S}$ and $\vec{J}$
can be expanded  into a set of  various products of $S_{i}$ and
$J_{i}$ in the same manner.

The most general form of such expansion allowing for the fact that
$\hat{F}$ should be scalar with respect to rotations is as follows \cite{110,232,Nuclear_optics,A2}
\begin{eqnarray}
	\label{18.17}
	\hat{F}&=& A+A_{1}S_{i}J_{i}+A_{2}S_{i}J_{k}n_{i}n_{k}+A_{3}J_{i}J_{k}n_{i}n_{k}\nonumber\\
	&+& A_{4}S_{i}S_{k}n_{i}n_{k}+A_{5}S_{i}S_{k}J_{i}J_{k}+A_{6}S_{i}S_{k}n_{i}n_{k}J_{l}J_{m}n_{l}n_{m}+\ldots \nonumber\\
	& & \ldots + BS_{i}n_{i}+B_{1}S_{i}I_{m}e_{iml}n_{l}+B_{2}S_{i}n_{i}S_{l}J_{l}+B_{3}S_{i}S_{l}n_{i}n_{l}J_{m}n_{m}\nonumber\\
	&+&B_{4}J_{i}n_{i}+B_{5}S_{i}J_{m}e_{iml}n_{l}S_{p}n_{p}+\ldots,
\end{eqnarray}
where the three dots stand for the terms containing the
products of $S_{i}$ and $J_{i}$ up to $2S$ and $2J$,
the terms containing $A$ and $d$ are caused by strong P--,T--even interactions,
those with $B$, $B_2,~B_3,~B_4$ --- by P-odd T-even, $B_1$ --- by
P--odd T--odd and $B_5$ corresponds to P--even T--odd
interactions, 
$\vec{S}$ is the spin of the particle incident on the target.

Upon averaging $\hat{F}$ using the spin density matrix of the
target nuclei, we find the explicit form of a coherent elastic
zero--angle scattering amplitude, and hence the refractive index
and the particle wave function in the target. According to
(\ref{18.17}), for particles with spin $S>1/2$, there appear
additional terms involving spin operators in the second and higher
powers.

Let us find out what these terms lead to. We shall first pay
attention to the fact that even in the case of a nonpolarized
target, the amplitude $\hat{f}(0)$ depends on the spin operator of
the incident particle and, when the quantization axis $z$ is
directed along $\vec{n}$,  can be written in the form
\begin{equation}
	\label{18.18} \hat{f}(0)=d+d_{1}S_{z}^{2}+d_{2}S_{z}^{4}\ldots
	+d_{s}S_{z}^{2s}\,.
\end{equation}

We consider a
specific case of strong interactions, invariant with respect to
time and space reflections; for this reason, the terms containing the
odd powers of $S$ are dropped. According to (\ref{18.15}), the refractive index
is
\begin{equation}
	\label{18.19}
	\hat{N}=1+\frac{2\pi\rho}{k^{2}}(d+d_{1}S_{z}^{2}+d_{2}S_{z}^{4}\ldots
	+d_{s}S_{z}^{2s})\,.
\end{equation}
(\ref{18.19}) yields an important conclusion that the refractive
index of a particle with spin $S>1/2$  depends on the spin
orientation with respect to the momentum direction. Write $m$ for
a magnetic quantum number, then the refractive index of a particle
in the state which is the eigenstate of the operator $S_{z}$ of
the spin projection on the $z$-axis is
\begin{equation}
	\label{18.20}
	N(m)=1+\frac{2\pi\rho}{k^{2}}(d+d_{1}m^{2}+d_{2}m^{4}+ \ldots
	+d_{s}m^{2s})\,.
\end{equation}
According to (\ref{18.20}), the states of a particle with quantum
numbers $m$ and ($-m$) have the same refractive indices. For a
particle with spin $1$ (for example, a $J/ \psi$-particle,
deuteron)  and for a  particle with spin $3/2$ (for example, $^{21}Ne$ nucleus,
$\Omega^{-}$-hyperon)
\begin{equation}
	\label{18.21} N(m)=1+\frac{2\pi\rho}{k^{2}}(d+d_{1}m^{2})\,.
\end{equation}
As is seen, $\texttt{Re} N(\pm 1)\neq \texttt{Re} N(0)$;
$\texttt{Im} N(\pm 1)\neq \texttt{Im} n(0)$; $\texttt{Re} N(\pm
3/2)\neq \texttt{Re} N(\pm1/2)$; $\texttt{Im} N(\pm 3/2)\neq
\texttt{Im} N(\pm 1/2)$.

From this follows that for particles with spin $S>1/2$, even a
nonpolarized target causes spin dichroism: due to different
absorption, the initially nonpolarized beam passing through matter
acquires polarization, or  more precisely, alignment
\cite{232,110}.

In view of the above analysis 
from
(\ref{18.19})--(\ref{18.21}) follows that in a medium, a moving
particle with spin $S \ge 1$ has a potential energy:
\begin{eqnarray*}
	\hat{U}=-\frac{2\pi \hbar^{2} \rho}{M}(d+d_{1} S_{z}^{2}+d_{2}S_{z}^{4}+\ldots)\,,\\
	U(m)=-\frac{2\pi \hbar^{2}
		\rho}{M}(d+d_{1}m^{2}+d_{2}m^{4}+\ldots)\,.
\end{eqnarray*}
The energy of interaction $\hat{U}$ between
the particle and matter  is similar to that between the atom of
spin $S\ge 1$ and the electric field.
As a result, in the medium, the spin levels of the particle split in
a way similar to Stark splitting of atomic levels in the electric field.
Hence, we may say that a particle of spin $S \ge 1$, moving in matter,
experiences the influence of a certain pseudoelectric field.

Since we have obtained the explicit spin structure of the
refractive index, then we know the wave function $\psi$, and for
every particular case we can find all spin characteristics of the
beam in a target at  depth $z$.

\subsection
{Rotation and Oscillation of Deuteron Spin in Nonpolarized Matter and Spin Dichroism (Birefringence Phenomenon)}
\label{cosy_sec:1.1}

We shall further dwell on the passage of deuterons through matter.

According to (\ref{18.21}), the refractive indices for the states
with ${m=+1}$ and ${m=-1}$ are the same, while those for for the
states with $m=\pm 1$ and $m=0$ are different
($\texttt{Re}{\textit{N}(\pm 1)} \neq \texttt{Re}{\textit{N}(0)}$
and $\texttt{Im}{\textit{N}(\pm 1)} \neq
\texttt{Im}{\textit{N}(0)}$).

\begin{figure}[htbp]
	\epsfxsize = 10 cm \centerline{\epsfbox{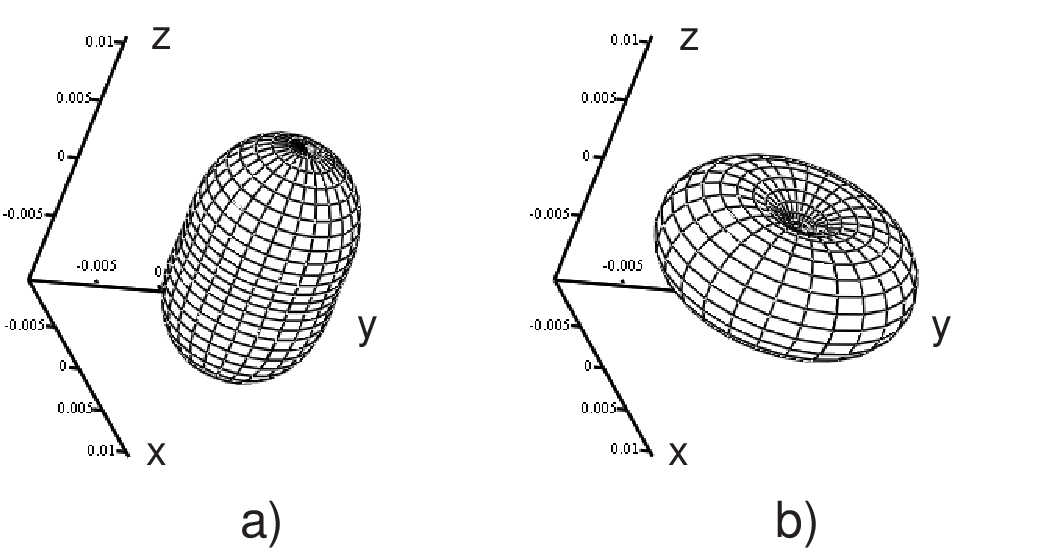}}
	\caption{Squared module for deuteron ground state wave functions
		for the distance of 1.8 fm between its nucleons in the states a)
		$m=\pm 1$; b) $m=0$ }
	\label{VKB_fig1}
\end{figure}

\begin{figure}[htbp]
	\epsfxsize =8 cm \centerline{\epsfbox{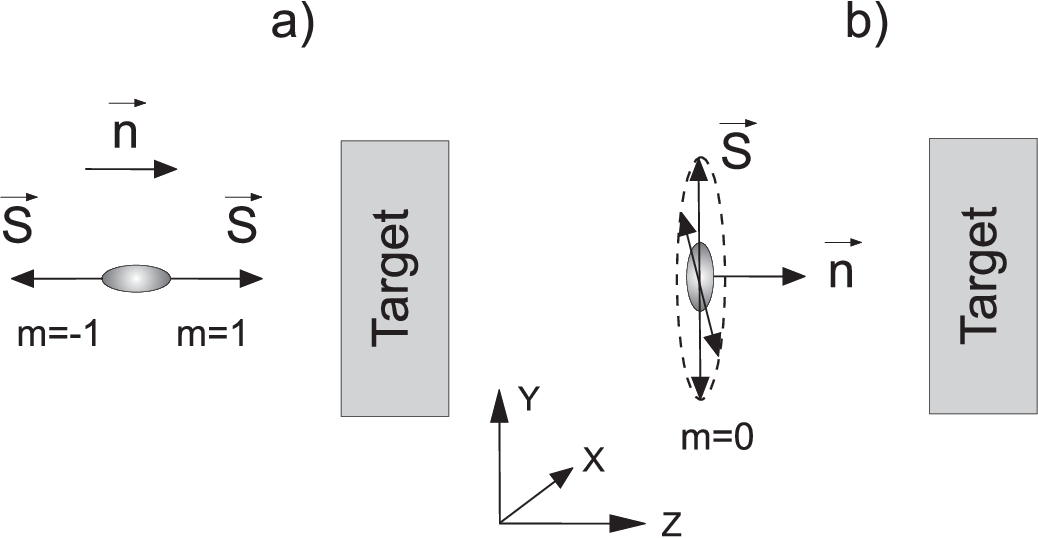}}
	\caption{Two possible orientation of vectors $\vec {S}$ and $\vec
		{n}=\frac{\vec{k}}{k}$: a) $m=\pm 1$; b) $m=0$}
	\label{cosy_sigma}
\end{figure}

This can be explained as follows (see Fig. \ref{VKB_fig1}, \ref{cosy_sigma}):
the shape of a deuteron in the ground state is non--spherical.

Therefore the scattering cross section $\sigma_{\pm 1}$ for a
deuteron with $m= \pm 1$ (deuteron spin is parallel (antiparallel) to its
momentum $\vec{k}$) differs from the scattering cross section
$\sigma_{0}$ for a deuteron with $m=0$:
\begin{equation}
	\label{cosy_eq7} \sigma_{\pm 1} \ne \sigma_{0} ~\Rightarrow ~
	\texttt{Im}\textit{f}_{\pm 1}(0)=\frac{k}{4\pi} \sigma_{\pm 1}\neq
	\texttt{Im}\textit{f}_ 0 (0)=\frac{k}{4\pi}\sigma_0\,.
\end{equation}
According to the dispersion relation, $\texttt{Re}{\textit{f}(0)}
\sim\Phi (\texttt{Im}{\textit{f}(0))}$, hence
${\texttt{Re}{\textit{f}_{0}(0)} \neq \texttt{Re}{\textit{f}_{\pm
			1}(0)}}$

From the above follows that deuteron spin dichroism appears even
when a deuteron passes through an nonpolarized target: owing to
the fact that absorption depends on the orientation of the
deuteron spin, the initially nonpolarized beam acquires alignment.

Let us consider the deuteron spin state in a target.
The spin state of the deuteron is described by its vector and
tensor  polarizations $\vec {p} = \langle \vec {S}\rangle$ and
$p_{ik} =
\langle Q_{ik} \rangle$, respectively.
As the deuteron  moves in matter, its vector and tensor
polarizations  change.
To calculate $\vec {p}$ and $p_{ik}$, one needs to know the explicit
form of the deuteron spin wave function $\psi$.

The wave function of the deuteron that has passed the distance $z$
inside the target is:
\begin{equation}
	\label{cosy_psiz} \psi \left( {z} \right) = \exp\left(
	{ik\hat{N}z} \right)\psi _{0}\,,
\end{equation}
where $\psi_{0}$ is the wave function of the deuteron before
entering the target.
The wave function $\psi$ can be expressed as a superposition of
the basic spin functions $\chi_{m}$, which are the eigenfunctions
of the operators $\hat{S}^{2}$ and $\hat{S}_{z}$ ($\hat {S}_{z}
\chi _{m} = m\chi _{m}$):
\begin{equation}
	\label{cosy_psi} \psi = \sum\limits_{m = \pm 1,0} {a^{m}\chi
		_{m}}\,.
\end{equation}
Therefore
\begin{equation}
	\label{cosy_psidepth}
	\begin{array}{l}
		\psi = \left( \begin{array}{*{20}c}
			{a^{1}} \hfill \\
			{a^{0}} \hfill \\
			{a^{ - 1}} \hfill \\
		\end{array}  \right) = \left( {{\begin{array}{*{20}c}
					{ae^{i\delta _{1}} e^{ikN_{1} z}} \hfill \\
					{be^{i\delta _{0}} e^{ikN_{0} z}} \hfill \\
					{ce^{i\delta _{ - 1}} e^{ikN_{ - 1} z}} \hfill \\
		\end{array}} } \right) = 
	%
	\left( {{\begin{array}{*{20}c}
					{ae^{i\delta _{1}} e^{ikN_{1} z}} \hfill \\
					{be^{i\delta _{0}} e^{ikN_{0} z}} \hfill \\
					{ce^{i\delta _{ - 1}} e^{ikN_{1} z}} \hfill \\
		\end{array}}} \right)\,,
	\end{array}
\end{equation}
according to the above, $N_{1}=N_{-1}$.

Suppose that the plane $(yz)$ coincides with the plane formed by the
initial vector polarization $\vec {p}_0 \neq 0$ and the
momentum $\vec{k}$ of the deuteron. In this case
\[
\delta
_{1}-\delta _{0}= \delta _{0}-\delta _{-1}=\frac{{\pi}}{{2}},
\]
and the components of the polarization vector at $z = 0$ are $p_x =
0,p_y\neq 0$, and $p_z\neq 0$.

The components of the vector polarization
\[
\vec{p}=\langle \vec
{S}\rangle = \frac{{\langle \Psi \left| {\vec {S}} \right|\Psi
		\rangle} }{{\left\langle {{\Psi} } \mathrel{\left| {\vphantom
				{{\Psi} {\Psi} }} \right. \kern-\nulldelimiterspace} {{\Psi} }
		\right\rangle} }
\]
inside the target are:
\\
\begin{eqnarray}
	p_x&=& \frac{{ \sqrt {2} e^{ - \frac{{1}}{{2}}\rho \left( {\sigma
					_{0} + \sigma _{1}} \right)z}b\left( {a - c} \right)\sin\left(
			{\frac{{2\pi \rho} }{{k}}\texttt{Re} d_{1} z}
			\right)}}{{\left\langle {{\Psi} } \mathrel{\left| {\vphantom
					{{\Psi}  {\Psi} }} \right. \kern-\nulldelimiterspace}
			{{\Psi} } \right\rangle} } ,\nonumber\\
	p_y&=&\frac{{\sqrt {2} e^{ - \frac{{1}}{{2}}\rho \left( {\sigma
					_{0} + \sigma _{1}} \right)z}b\left( {a + c} \right)\cos\left(
			{\frac{{2\pi \rho} }{{k}}\texttt{Re} d_{1} z}
			\right)}}{{\left\langle {{\Psi} } \mathrel{\left| {\vphantom
					{{\Psi}  {\Psi} }} \right. \kern-\nulldelimiterspace}
			{{\Psi} } \right\rangle} } ,\nonumber \\
	p_z& =&\frac{{e^{\rho \sigma _{1} z}\left( {a^{2} - c^{2}}
			\right)}}{{\left\langle {{\Psi} } \mathrel{\left| {\vphantom
					{{\Psi}  {\Psi }}} \right. \kern-\nulldelimiterspace} {{\Psi} }
			\right\rangle} }\,.\label{cosy_rot1}
	\\ \nonumber
\end{eqnarray}
Similarly, the components of the tensor polarization
\[
\hat
{Q}_{ij} = \frac{{3}}{{2}}\left( {\hat {S}_{i} \hat {S}_{j} + \hat
	{S}_{j} \hat {S}_{i} - \frac{{4}}{{3}}\delta _{ij}} \right)
\]
are expressed as:
\\
\begin{eqnarray}
	p_{xx}&=&\frac{{ - \frac{{1}}{{2}}\left( {a^{2} + c^{2}}
			\right)e^{ - \rho \sigma _{1} z} + b^{2}e^{ - \rho \sigma _{0} z}
			- 3ace^{ - \rho \sigma _{1} z}}}{{\left\langle {{\Psi} }
			\mathrel{\left| {\vphantom {{\Psi}  {\Psi} }} \right.
				\kern-\nulldelimiterspace} {{\Psi} } \right\rangle} }
	\,,\nonumber
	\\
	p_{yy}&=&\frac{{ - \frac{{1}}{{2}}\left( {a^{2} + c^{2}}
			\right)e^{ - \rho \sigma _{1} z} + b^{2}e^{ - \rho \sigma _{0} z}
			+ 3ace^{ - \rho \sigma _{1} z}}}{{\left\langle {{\Psi} }
			\mathrel{\left| {\vphantom {{\Psi}  {\Psi} }} \right.
				\kern-\nulldelimiterspace} {{\Psi} } \right\rangle} }  \,,
	\nonumber
	\\
	p_{zz}&=&\frac{{\left( {a^{2} + c^{2}}
			\right)e^{ - \rho \sigma _{1} z} - 2b^{2}e^{ - \rho \sigma _{0}
				z}}}{{\left\langle {{\Psi} } \mathrel{\left| {\vphantom {{\Psi}
						{\Psi} }} \right. \kern-\nulldelimiterspace} {{\Psi} }
			\right\rangle} } ,\nonumber\\
	p_{xy}&=& 0 \,,\nonumber
	\\
	p_{xz}&=&\frac{{ \frac{{3}}{{\sqrt {2}} }e^{ - \frac{{1}}{{2}}\rho
				\left( {\sigma _{0} + \sigma _{1}} \right)z}b\left( {a + c}
			\right)\sin\left( {\frac{{2\pi \rho} }{{k}}\texttt{Re} d_{1} z}
			\right)}}{{\left\langle {{\Psi} } \mathrel{\left| {\vphantom
					{{\Psi}  {\Psi }}} \right. \kern-\nulldelimiterspace} {{\Psi} }
			\right\rangle} }  \,,\nonumber
	\\
	p_{yz}&=&\frac{{\frac{{3}}{{\sqrt {2}} }e^{ - \frac{{1}}{{2}}\rho
				\left( {\sigma _{0} + \sigma _{1}} \right)z}b\left( {a - c}
			\right)\cos\left( {\frac{{2\pi \rho} }{{k}}\texttt{Re} d_{1} z}
			\right)}}{{\left\langle {{\Psi} } \mathrel{\left| {\vphantom
					{{\Psi}  {\Psi }}} \right. \kern-\nulldelimiterspace}
			{{\Psi} } \right\rangle} } \,,\nonumber\\
	p_{xx}&+&p_{yy}+p_{zz}=0\,, \label{cosy_rot2}
\end{eqnarray}
\\
\noindent  where
\begin{eqnarray}
	\label{cosy_rot2+}
	\left\langle {{\Psi} } \mathrel{\left| {\vphantom {{\Psi}
				{\Psi} }} \right. \kern-\nulldelimiterspace} {{\Psi} }
	\right\rangle &=& \left( {a^{2} + c^{2}} \right)e^{ - \rho \sigma
		_{1} z} + b^{2}e^{ - \rho \sigma _{0} z},\nonumber\\
	\sigma_{0}&=&\frac{{4\pi} }{{k}}\texttt{Im} f_0,\, \sigma _{1} = \frac{{4\pi}}{{k}}\texttt{Im} f_1,\\
	f_0&=&d,\, f_1=d+d_1\,.\nonumber
\end{eqnarray}

According to (\ref{cosy_rot1}), (\ref{cosy_rot2}), spin rotation and oscillation occur when
the angle between the polarization vector $\vec{p}$ and the momentum
$\vec{k}$ of the particle differs from $\pi/2$. The magnitude of the effect is determined by the phase
\begin{equation}
	\label{cosy_ins} \varphi=\frac{2\pi\rho}{k}\texttt{Re} d_1 z\,.
\end{equation}

For example, let  $\texttt{Re} d_1 >0$. If the angle between the
polarization vector and momentum is acute, then the spin rotates
anticlockwise about the momentum direction, whereas the obtuse
angle between the polarization vector and the momentum gives rise
to a clockwise  spin rotation.

When the polarization vector and momentum are perpendicular
(transversely polarized particle), the components of
the vector polarization at $z = 0$ are: $p_x= 0$, $p_y\ne0$, and $p_z=0$.
In this case $a=c$ and the dependence of the vector polarization
on $z$ can be expressed as:
\begin{eqnarray}
	p_x&=&0,\nonumber\\
	p_y&=&\frac{{\sqrt {2} e^{ - \frac{{1}}{{2}}\rho \left( {\sigma
					_{0} + \sigma _{1}} \right)z}2ba\cos\left( {\frac{{2\pi \rho
				}}{{k}}\texttt{Re} d_{1} z} \right)}}{{\left\langle {{\Psi} }
			\mathrel{\left| {\vphantom {{\Psi}  {\Psi} }} \right.
				\kern-\nulldelimiterspace} {{\Psi} } \right\rangle} }\,,\nonumber\\
	p_z&=&0\,,\nonumber\\
	p_{xx}&=&\frac{{ - 4a^{2}e^{ - \rho \sigma _{1} z}
			+ b^{2}e^{ - \rho \sigma _{0} z}}}{{\left\langle {{\Psi} }
			\mathrel{\left| {\vphantom {{\Psi}  {\Psi} }} \right.
				\kern-\nulldelimiterspace} {{\Psi} } \right\rangle} }\,, \label{cosy_11}\\
	p_{yy}&=&\frac{{2a^{2}e^{ - \rho \sigma _{1} z} + b^{2}e^{ - \rho
				\sigma _{0} z}}}{{\left\langle {{\Psi} } \mathrel{\left|
				{\vphantom {{\Psi}  {\Psi} }} \right.
				\kern-\nulldelimiterspace} {{\Psi} } \right\rangle} }\,,\nonumber\\
	p_{zz}&=&\frac{{2a^{2}e^{ - \rho \sigma_{1} z} - 2b^{2}e^{ - \rho
				\sigma _{0} z}}}{{\left\langle {{\Psi} } \mathrel{\left|
				{\vphantom {{\Psi}  {\Psi} }} \right. \kern-\nulldelimiterspace}
			{{\Psi} } \right\rangle} }\,, \nonumber
	\\
	p_{xz}&=&\frac{{  \frac{{3}}{{\sqrt {2}} }e^{ -
				\frac{{1}}{{2}}\rho \left( {\sigma _{0} + \sigma _{1}}
				\right)z}2ab\sin\left( {\frac{{2\pi \rho} }{{k}}\texttt{Re} d_{1}
				z} \right)}}{{\left\langle {{\Psi} } \mathrel{\left| {\vphantom
					{{\Psi}  {\Psi }}} \right. \kern-\nulldelimiterspace} {{\Psi} }
			\right\rangle} }\,,\nonumber
\end{eqnarray}

\begin{eqnarray}
	p_{yz}&=&0\,, \nonumber\\
	p_{xx}&+&p_{yy}+p_{zz}=0\,.
	\label{cosy_ins11}
\end{eqnarray}
According to (\ref{cosy_ins11}), no rotation occurs in this case; the vector and tensor polarization
oscillate when a transversely polarized deuteron passes through
matter.

Oscillations of $\langle S_{x}\rangle$ and $Q_{yz}$ are opposite in phase.

Thus, if
the difference $\delta_{1}-\delta_{0}$ is
zero when the particle enters the target, and consequently
${\langle S_{x}\rangle=1}$ and $Q_{yz}=0$, then after it has
passed the path length  $l_{1/4}=k(4\rho \texttt{Re} d_{1})^{-1}$,
the polarization vector vanishes while $Q_{yz}$  attains its
maximum value. With further increasing path length, the
polarization vector  changes sign, and $Q_{yz}$  starts
diminishing so that after this quarter of the period is over, just
$\langle S_{x}\rangle$  attains the maximum value, while $Q_{yz}$
vanishes, and so on. After the particle has passed four quarters
of the period, the situation becomes the same as it was in the
beginning when the particle entered the target. Thus, the
transitions between vector and tensor polarizations of the
particle appear. This effect is quite analogous to a well known
optical phenomenon of birefringence in Iceland spar. For example,
after passing a quarter-wave plate, light with right-hand circular
polarization  becomes linearly polarized. With increasing
thickness, the degree of linear polarization decreases, and a
photon with left circular polarization appears. After passing this
quarter of the period, the  left circular polarization of the
photon attains its maximum value, while its linear polarization
vanishes.

Deuteron  birefringence effect in matter is
kinematically analogous to the oscillations of atomic spin in the
electric field, which arise due to the quadratic Stark splitting
of atomic levels. 
As mentioned above, this enables us
to state that the birefringence phenomenon is caused by a
pseudoelectric field acting on a particle in a medium.

\subsection[The Effect of Tensor Polarization Emerging in Nonpolarized Beams Moving in Nonpolarized Matter]
{The Effect of Tensor Polarization Emerging in Nonpolarized Beams
	Moving in Nonpolarized Matter} \label{cosy_sec:1.2}

The effect of birefringence of particles, in particular,
the appearance of tensor polarization of the initially nonpolarized beam
in a real experiment can be most clearly described using the spin density matrix \cite{A2,spinorb_ex2}.

For deuterons (spin $S=1)$, the spin density matrix for the  beam before the target can be
written as follows:
\begin{equation}
	\hat{\rho}_{0}=\frac{1}{3}\hat{I}+\frac{1}{2}\vec{P}_{0}\hat{\vec{S}}+\frac{1}{9}P^{(0)}_{ik}\hat{Q}_{ik}\,,
	\label{cosy_ins+}
\end{equation}
where $\hat{I}$ is the identity (unit) matrix, $\vec{P}_{0}$ is the polarization vector of the beam,
$P^{(0)}_{ik}$  is the polarization vector of the beam incident on the target.
Using (\ref{cosy_psiz}), one can express the density matrix of the
deuteron beam in the target as:
\begin{equation}
	\hat {\rho}  = e^{ik\hat{N}z}\hat{\rho}_{0}e^{-ik\hat{N}^{\ast}z}\,.
	\label{cosy_ins+1}
\end{equation}
As a result, we have
\begin{equation}
	\vec{p}=\langle \vec{S}\rangle = \frac{{{\textrm{Tr}} \left( \hat
			{\rho} \hat {\vec{ S}}\right)}}{{ \textrm{Tr}\left(\hat {\rho}
			\right)}},\quad p_{ik}= \langle Q_{ik} \rangle =
	\frac{{\textrm{Tr}} \left( \hat {\rho}\hat{Q}_{ik} \right)}{
		\textrm{Tr}\left(\hat {\rho} \right)}\,,
	\label{cosy_ins+2}
\end{equation}
where $i,k=x,y,z$.

In the case of thin targets, the vector and tensor polarization of
the deuteron inside the target can be expressed using the
first-order approximation for $e^{ik( \hat{N}-1)z} \approx
1+ik(\hat{N}-1 )z$ as follows:
\begin{eqnarray}
\label{cosy_polpar1}
	p_{x}&=& \frac{{\left[ {1 - \frac{{1}}{{2}}\rho z\left( {\sigma
					_{0} + \sigma _{1}}  \right)} \right] p_{x,0}+
			\frac{{4}}{{3}}\frac{{\pi \rho z}}{k}\texttt{Re} d_{1}
			p_{zy,0}}}{{\textrm{Tr}\hat {\rho }\hat {I}}}\,,\nonumber
	\\
	p_{y} &=& \frac{{\left[ {1 - \frac{{1}}{{2}}\rho z\left( {\sigma
					_{0} + \sigma _{1}}  \right)} \right] p_{y,0}  -
			\frac{{4}}{{3}}\frac{{\pi \rho z}}{k} \texttt{Re} d_{1}
			p_{zx,0}}}{{\textrm{Tr}\hat {\rho }\hat {I}}}\,, \nonumber
	\\
	p_{z}& =& \frac{{\left( {1 - \rho \sigma _{1} z} \right)p_{z,0}} }
	{{\textrm{Tr}\hat {\rho} \hat {I}}}\,,\nonumber
	\\
	p_{xx}&=& \frac{{\left( {1 - \rho \sigma _{1} z} \right)p_{xx,0} +
			\frac{{1}}{{3}}\rho z\left( {\sigma _{1} - \sigma _{0}} \right) -
			\frac{{1}}{{3}}\rho z\left( {\sigma _{1} - \sigma _{0}}
			\right)p_{zz,0}} }{{\textrm{Tr}\hat {\rho} \hat {I}}}\,, \nonumber
	\\
	p_{yy} &=& \frac{{\left( {1 - \rho \sigma _{1} z} \right)p_{yy,0}
			+ \frac{{1}}{{3}}\rho z\left( {\sigma _{1} - \sigma _{0}}  \right)
			- \frac{{1}}{{3}}\rho z\left( {\sigma _{1} - \sigma _{0}}
			\right)p_{zz,0}} }{{\textrm{Tr}\hat {\rho} \hat {I}}}\,, \nonumber
	\\
	p_{zz} &=& \frac{{\left[ {1 - \frac{{1}}{{3}}\rho z\left( {2\sigma
					_{0} + \sigma _{1}}  \right)} \right]p_{zz,0} -
			\frac{{2}}{{3}}\rho z\left( {\sigma _{1} - \sigma _{0}}
			\right)}}{{\textrm{Tr}\hat {\rho} \hat {I}}}\,, \nonumber
	\\
	p_{xy} &=& \frac{{\left( {1 - \rho \sigma _{1} z} \right)p_{xy,0}}
	} {{\textrm{Tr}\hat {\rho} \hat {I}}}\,,
	\\
	p_{xz} &=& \frac{{\left[ {1 - \frac{{1}}{{2}}\rho z\left( {\sigma
					_{0} + \sigma _{1}}  \right)} \right] p_{xz,0} + 3\frac{{\pi \rho
					z}}{{k}}\texttt{Re} d_{1} p_{y,0} } }{{\textrm{Tr}\hat {\rho }\hat
			{I}}}\,,\nonumber
	\\
	p_{yz} &=& \frac{{\left[ {1 - \frac{{1}}{{2}}\rho z\left( {\sigma
					_{0} + \sigma _{1}} \right)} \right]p_{yz,0} - 3\frac{{\pi \rho
					z}}{{k}}\texttt{Re} d_{1} p_{x,0}} }{{\textrm{Tr}\hat {\rho }\hat
			{I}}}\,, \nonumber
\end{eqnarray}
\noindent where
\[
\textrm{Tr}\hat {\rho} \hat {I} = 1 - \frac{{\rho z}}{{3}}\left(
{2\sigma _{1} + \sigma _{0}}  \right) - \frac{{\rho z}}{{3}}\left(
{\sigma _{1} - \sigma _{0}} \right)p_{zz,0}\, .
\]

If the beam is initially nonpolarized ($p_{x,0} = p_{y,0} =
p_{z,0} = p_{xx,0} = p_{yy,0} =$ $p_{zz,0} = p_{xy,0} = p_{xz,0} =
p_{yz,0} = 0$), then after passing through the nonpolarized target
of thickness $z$, the deuteron beam acquires  tensor polarization:
\begin{eqnarray}
	p_{zz}&\approx& - \frac{{2}}{{3}}\rho z\left( {\sigma _{1} -
		\sigma _{0}}  \right) = - \frac{{2}}{{3}} \rho \sigma z  \frac{\Delta \sigma}{\sigma},\nonumber\\
	p_{xx} &=& p_{yy} ~ \approx ~ \frac{{1}}{{3}}\rho \sigma z  \frac{\Delta \sigma}{\sigma}\,. \label{cosy_polpar2}
		\label{eq:pzz23}
\end{eqnarray}
Vector polarization remains equal to zero.


%
The expression for tensor polarization (\ref{cosy_polpar2}) may also be obtained from another viewpoint.

Let a deuteron beam in spin state with $m=1$ pass through a
target. The beam intensity changes as
$I_1(z)=I_{1}^0e^{-\sigma_1\rho z}$, where $I_{1}^0$ is the beam
intensity before entering the target. Similarly, for states
$m={-1}$ and $m=0$, the intensity changes as
$I_{-1}(z)=I_{-1}^0e^{-\sigma_{-1}\rho z}$ and
$I_0(z)=I_{0}^0e^{-\sigma_0\rho z}$, where $I_{-1}^0$ and
$I_{0}^0$ are the beam intensities before entering the target,
respectively.

Let us consider the transmission of an nonpolarized deuteron beam
through an nonpolarized target.

The nonpolarized deuteron beam can be described as a composition
of three polarized beams with equal intensities
$I=I_{1}^0+I_{-1}^0+I_{0}^0$, {$I_{\pm1}^0=I_{0}^0=I/3$}.

In a real experiment
$\sigma_{\pm1,0}\rho z\ll1$ and the change in the intensity of each
beam can be expressed as
$I_{\pm1}(z)=I_{\pm1}^0(1-\sigma_{\pm1}\rho z)$ and
$I_{0}(z)=I_{0}^0(1-\sigma_{0}\rho z)$.
%

%
%

According to \cite{VKB_Ohlsen}, the tensor polarization of the beam can
be expressed as
\[
p_{zz}=\frac{I_{-1}+I_{1}-2I_0}{I_{-1}+I_{1}+I_0}\,.
\]

The tensor polarization of the initially nonpolarized deuteron
beam transmitting through the target of thickness $L$, which
arises from deuteron spin dichroism reads as follows:
\begin{eqnarray}
	p_{zz}(L)&=&\frac{I_{-1}(L)+I_{1}(L)-2I_{0}(L)}{I_{-1}(L)+I_{1}(L)+I_{0}(L)}\nonumber\\
	&\approx&\frac{2N_aL\left(\sigma_{0}-\sigma_{\pm1}\right)}{3M_r}=-\frac{8\pi
		N_aL \texttt{Im}(d_1)}{3kM_r}\,, \label{VKB_pzzdef}
\end{eqnarray}
where $N_a$ is the Avogadro number, $L$ is the target thickness in
g/cm$^2$, $M_r$ is the molar mass of the target matter.

%
%
Note that a deuteron passing through a target loses energy
by ionization of matter, then, taking into account the energy
change, we can write the tensor polarization as

\begin{eqnarray}
	p_{zz}(L)&=&\frac{2N_a}{3M_r}\int_0^L
	\left(\sigma_{0}\left(E\left(L'\right)\right)-\sigma_{\pm1}\left(E\left(L'\right)\right)\right)dL'\nonumber\\
	&=&-\frac{8\pi N_a}{3M_r}
	\int_0^L\frac{\texttt{Im}(d_1\left(E\left(L'\right)\right))}{k(L')}dL'\,.
	\label{VKB_pzzrot}
\end{eqnarray}
According to (\ref{VKB_pzzrot}), the imaginary part of the
spin-dependent forward scattering amplitude can be measured
directly in a transmission experiment
by means of deuteron beam tensor polarization, which arises due to
deuteron spin dichroism.

Thus, theoretical studies of the deuteron beam transmission
through the nonpolarized target predict the appearance of tensor
polarization in a transmitted beam due to deuteron spin dichroism.

\subsection[The Amplitude of Zero--Angle Elastic Scattering of a Deuteron by a Nucleus]
{The Amplitude of Zero--Angle Elastic Scattering of a Deuteron by
	a Nucleus} \label{eikanal3_sec:1.1}

Let us discuss the possible magnitude of the deuteron birefringence
effect in detail. According to (\ref{cosy_rot1}), (\ref{cosy_rot2}), (\ref{cosy_11}), (\ref{cosy_ins11}),
the birefringence effect depends on the amplitudes of zero-angle elastic coherent
scattering of a deuteron by a nucleus $f(m=\pm1)$ and $f(m=0)$.

In order to find the amplitude $f(0)$, one should start with considering the Hamiltonian $H$
describing the interaction of the deuteron with the nucleus \cite{110,232}.

The  Hamiltonian $H$ can be written as
\begin{equation}
	H=H_D(\vec{r_p},\vec{r_n})+H_N(\{\xi_i\})+V_{DN}(\vec{r_p},\vec{r_n},\{\xi_i\})\,,
	\label{eikanal3_ham1}
\end{equation}
where $H_D$ is the deuteron Hamiltonian; $H_N$ is the nuclear
Hamiltonian; $V_{DN}$ stands for the energy of deuteron--nucleus
nuclear and Coulomb interaction; $r_p$ and $r_n$ are the coordinates of
the proton and the neutron composing the deuteron, $\{\xi_i\}$ is the
set of coordinates of the nucleons.

Having introduced the deuteron
center-of-mass coordinate $\vec{R}$ and the relative distance
between the proton and the neutron in the deuteron
$\vec{r}=\vec{r_p}-\vec{r_n}$, we recast (\ref{eikanal3_ham1}) as
\begin{equation}
	H=-\frac{\hbar^2}{2m_D}\Delta(\vec{R})+H_D(\vec{r})+H_N(\{\xi_i\})+V_{DN}^N(\vec{R},\vec{r},\{\xi_i\})
	+ V_{DN}^C(\vec{R},\vec{r},\{\xi_i\})\,, \label{eikanal3_ham2}
\end{equation}
where $H_{D}(\vec{r})$ is the Hamiltonian describing the internal state of the deuteron, $m_{D}$ is the deuteron mass.

In view of (\ref{eikanal3_ham2}), the deuteron--nucleus scattering
is determined by two interactions: nuclear and Coulomb. In this
section we shall content ourselves with finding the amplitude of
forward elastic scattering of a deuteron with energy of hundreds
of megaelectronvolts by a light nucleus due to nuclear interaction
(the term $V^{C}_{DN}$ in (\ref{eikanal3_ham2}) will be ignored).
At lower energies, taking account of the Coulomb interaction is
essential \cite{rins_65}.

In further consideration we shall pay attention to the fact that
for deuterons, for example,  with energy of several tens of
megaelectronvolts, appreciably exceeding the  binding energy of
deuterons $\varepsilon_d$, the time of nuclear deuteron-nucleus
interaction is $\tau^N\simeq5\cdot10^{-22}$ s,  whereas the
characteristic period of oscillation of nucleus in the deuteron is
$\tau\simeq2\pi\hbar/\varepsilon_d\simeq2\cdot10^{-21}$\,s. So we
can apply the impulse approximation \cite{Goldberger}.  
In this
approximation we can neglect the binding energy of nucleons in the
deuteron, i.e., neglect  $H_D(\vec{r})$ in (\ref{eikanal3_ham2}).
As a result,
\begin{equation}
	H=-\frac{\hbar^2}{2m_D}\Delta(\vec{R})+H_N(\{\xi_i\})+V_{DN}^N(\vec{R},\vec{r},\{\xi_i\})\,.
	\label{eikanal3_ham3}
\end{equation}

As is seen, in the impulse approximation the problem of
determining the scattering amplitude reduces to the problem of
scattering by a nucleus of a structureless particle having
the same mass as the deuteron. In this case the
coordinate $\vec{r}$ is a parameter. Therefore, the relations
obtained for the cross section and the forward scattering
amplitude should be averaged over the stated parameter. To
estimate the magnitude of the effect, we shall also neglect the
spin-dependence of internucleonic interaction. This enables using
eikonal approximation for analyzing the magnitude of the amplitude
for fast deuterons \cite{quark_8}.

In this approximation the amplitude of coherent zero--angle
scattering can be written as follows:

\begin{eqnarray}
	f(0)=\frac{k}{2\pi~i}\int \left( e^{i\chi _{D}\left( \vec{b},\vec{r}%
		\right) }-1\right) d^{2}b\left| \varphi \left( \vec{r}\right)\,.
	\right| ^{2}d^{3}r, \label{eikanal3_amp}
\end{eqnarray}
where $k$ is the deuteron wave number, $\vec{b}$ is the impact
parameter, $\varphi(\vec{r})$ is the wave function of the deuteron
ground state. The phase shift due to the deuteron scattering by
carbon is
\begin{equation}
	\chi _{D}=-\frac{1}{\hbar v}
	\int_{-\infty }^{+\infty }V_{DN}\left( \vec{b},z^{^{\prime }},%
	\vec{r}_{\perp }\right) dz^{^{\prime }}\,,
\end{equation}
$\vec{r}_{\perp}$ is the component of $\vec{r}$, which is perpendicular to the
momentum of incident deuteron, $v$ is the deuteron speed. The
phase shift $\chi _{D}=\chi _{1}+\chi _{2}$, where $\chi_{1}$ and
$\chi _{2}$ are the phase shifts caused by proton-nucleus and
neutron-nucleus interactions, respectively.

For the deuteron, the probability $\left| \varphi \left(
\vec{r}\right)\right| ^{2}$ differs
for different spin states. Thus, for
states with magnetic quantum number $m=\pm 1$, the probability is
$\left| \varphi_{\pm 1} \left( \vec{r}\right)\right| ^{2}$,
whereas for $m=0$, it is $\left| \varphi_{0} \left(
\vec{r}\right)\right| ^{2}$.

Owing to the additivity of phase
shifts, equation (\ref{eikanal3_amp}) can be rewritten as
\begin{eqnarray}
	f\left( 0\right)=& &\frac{k}{\pi }\int \left\{
	t_{1} \left(
	\vec{b}-\frac{\vec{r}_{\perp }}{2} \right) +t_{2} \left(
	\vec{b}+\frac{\vec{r}_{\perp }}{2} \right) + 2it_{1} \left( \vec{b
	}-\frac{\vec{r}_{\perp }}{2} \right) t_{2} \left(
	\vec{b}
	+\frac{\vec{r}_{\perp }}{2} \right)
	\right\} \nonumber\\
	&\times&\left| \varphi \left( \vec{r} \right) \right| ^{2}
	d^{2}bd^{3}r\,,
	\label{eikanal3_42}
\end{eqnarray}
where
\[
t_{1(2)}=\frac{e^{i\chi _{1\left( 2\right) }}-1}{2i}\,.
\]

From
(\ref{eikanal3_42}) follows
\begin{equation}
	f(0)=f_{1}(0)+f_{2}(0)+ \frac{2ik}{\pi}
	\int t_{1}\left( \vec{b}-%
	\frac{\vec{r}_{\perp }}{2}\right)t_{2}\left( \vec{b}+%
	\frac{\vec{r}_{\perp }}{2}\right)\left| \varphi \left(
	\vec{r}_{\perp},z\right) \right| ^{2}d^{2}bd^{2}r_{\perp}dz \,,
	\label{eikanal3_integral}
\end{equation}
where
\[
f_{1(2)}(0)=\frac{k}{\pi} \int t_{1(2)}(\vec{\xi})d^{2}\xi=
\frac{m_D}{m_{p(n)}}~f_{p(n)}(0)
\]
and $f_{p(n)}(0)$ is the amplitude of the proton--
(neutron)--nucleus zero--angle elastic coherent scattering.
(\ref{eikanal3_integral}) can be recast as
\begin{equation}
	f(0)=f_{1}(0)+f_{2}(0)+ \frac{2ik}{\pi}\int t_{1}(\vec{\xi})~
	t_{2}(\vec{\eta}) \left| \varphi
	\left(\vec{\xi}-\vec{\eta},z\right) \right|
	^{2}~d^{2}\xi~d^{2}\eta~dz\,. \label{eikanal3_27}
\end{equation}
Then from (\ref{eikanal3_27}), we get
\begin{eqnarray}
	\texttt{Re}f(0)=& &\texttt{Re}f_{1}(0)+\texttt{Re}f_{2}(0) -\frac{2k}{\pi}\texttt{Im} \int
	t_1(\vec{\xi}) t_{2}(\vec{\eta})\nonumber\\
	\times& &\left| \varphi
	\left(\vec{\xi}-\vec{\eta},z\right) \right|
	^{2}~d^{2}\xi~d^{2}\eta~dz \\
	\texttt{Im}f(0)=& &\texttt{Im}f_{1}(0)+\texttt{Im}f_{2}(0)+\frac{2k}{\pi}\texttt{Re} \int t_1(\vec{\xi})
	t_{2}(\vec{\eta})\nonumber\\
	\times& &\left| \varphi \left(\vec{\xi}-\vec{\eta},z\right)
	\right| ^{2}~d^{2}\xi~d^{2}\eta~dz \,. \nonumber
	\label{eikanal3_28}
\end{eqnarray}

In accordance with (\ref{cosy_rot1}), (\ref{cosy_rot2}), the
polarization state of the deuteron in the target is determined by
the difference of the amplitudes $\texttt{Re}f(m=\pm1)$ and
$\texttt{Re}f(m=0)$, and $\texttt{Im}f(m=\pm1)$ and
$\texttt{Im}f(m=0)$.

From (\ref{eikanal3_42}) follows that \cite{110,232,rins_65}
\begin{eqnarray}
\label{eikanal3_d1}
	\texttt{Re}d_1=-\frac{2k}{\pi}\texttt{Im} \int t_1(\vec{\xi})
	t_{2}(\vec{\eta})\left[ \varphi_{\pm 1}^{+}
	\left(\vec{\xi}-\vec{\eta},z\right) \varphi_{\pm 1}
	\left(\vec{\xi}-\vec{\eta},z\right)\right.\nonumber\\
	\left.- \varphi_{0}^{+}
	\left(\vec{\xi}-\vec{\eta},z\right) \varphi_{0}
	\left(\vec{\xi}-\vec{\eta},z\right) \right]d^{2}\xi~d^{2}\eta~dz \\
	\texttt{Im}d_1=\frac{2k}{\pi}\texttt{Re} \int
	t_1(\vec{\xi}) t_{2}(\vec{\eta})\left[ \varphi_{\pm 1}^{+}
	\left(\vec{\xi}-\vec{\eta},z\right) \varphi_{\pm 1}
	\left(\vec{\xi}-\vec{\eta},z\right)\right.\nonumber\\
	\left.- \varphi_{0}^{+} \left(\vec{\xi}-\vec{\eta},z\right)
	\varphi_{0} \left(\vec{\xi}-\vec{\eta},z\right)
	\right]~d^{2}\xi~d^{2}\eta~dz \,.\nonumber 
\end{eqnarray}

Note that according to (\ref{eikanal3_d1}), the spin-dependent part of the scattering
amplitude $d_{1}$ is determined by the rescattering effects of colliding particles.

When  the deuteron is scattered by a light nucleus, its
characteristic radius  is large as compared with the radius of the
nucleus. For this reason, to estimate the effects, we can suppose
that in integration,
the functions $t_1$ and $t_2$ act on $\varphi$ as a $\delta$-function. 
Then
\begin{eqnarray}
	\texttt{Re} d_1=& &-\frac{4\pi}{k}\texttt{Im} {f_1(0)f_{2}(0)} \int_{0}^{\infty} \left[
	\varphi_{\pm 1}^{+} \left(0,z\right) \varphi_{\pm 1}
	\left(0,z\right)\right.\nonumber\\
	& &\left.- \varphi_{0}^{+} \left(0,z\right) \varphi_{0}
	\left(0,z\right) \right]dz\,, \nonumber\\
	\texttt{Im}d_1=& &\frac{4\pi}{k}\texttt{Re}
	{f_1(0)f_{2}(0)} \int_{0}^{\infty} \left[ \varphi_{\pm 1}^{+}
	\left(0,z\right) \varphi_{\pm 1} \left(0,z\right)\right.\nonumber\\
	& &\left.- \varphi_{0}^{+} \left(0,z\right) \varphi_{0}
	\left(0,z\right) \right]dz\,. \label{eikanal3_0d1}
\end{eqnarray}

The magnitude of the birefringence effect is determined by
the difference
\[
\left[ \varphi_{\pm 1}^{+} \left(0,z\right) \varphi_{\pm 1}
\left(0,z\right)- \varphi_{0}^{+} \left(0,z\right) \varphi_{0}
\left(0,z\right)\right]\,,
\]
i.e., by the difference of distributions of nucleon density in the
deuteron for different deuteron spin orientations.
As noted hereinbefore, the spin dichroism effect (the effect of birefringence of particles, nuclei) is caused by the internal anisotropy of nuclei (particles) with spin $S \ge 1$.  
Therefore, the effect magnitude can always be evaluated as follows: it is sufficient, for example, to multiply the total cross section by the degree of asymmetry caused by the difference in areas (volumes) of spherical and ellipsoidal  nuclei shapes.
%
The structure
of the wave function $\varphi_{\pm 1}$ is well known
\cite{rins_94}:
\begin{equation}
	\varphi_m=\frac{1}{\sqrt{4 \pi}} \left\{
	\frac{u(r)}{r}+\frac{1}{\sqrt 8}\frac{W(r)}{r}\hat{S}_{12}
	\right\} \chi_m\,, \label{eikanal3_phi_m}
\end{equation}
where $u(r)$ is the deuteron radial wave function corresponding to
the $S$-wave; $W(r)$ is the radial function corresponding to the
$D$-wave; the operator $\hat{S}_{12}=6(\hat{\vec{S}}
\vec{n}_{r})^2-2\hat{\vec{S}}^2$; $\vec{n}_{r}=\frac{\vec{r}}{r}$;
$\hat{\vec{S}}=\frac{1}{2}(\vec{\sigma}_1+\vec{\sigma}_2)$, and
$\vec{\sigma}_{1(2)}$ are the Pauli spin matrices describing
proton (neutron) spin.

Use of (\ref{eikanal3_phi_m}) yields
\begin{eqnarray}
	\texttt{Re}d_1=-\frac{6}{k}~\texttt{Im} \left\{ f_{1}(0)f_{2}(0)
	\right\} G = -\frac{24}{k} \texttt{Im} \left\{ f_{p}(0)f_{n}(0)
	\right\} G\,,\
	\nonumber\\
	\texttt{Im}d_1=\frac{6}{k}\texttt{Re} \left\{ f_{1}(0)f_{2}(0)
	\right\} G= \frac{24}{k} \texttt{Re}\left\{ f_{p}(0)f_{n}(0)
	\right\} G\,, \label{eikanal3_13}
\end{eqnarray}
where
\[
G=\int_{0}^{\infty} \left( \frac{1}{\sqrt
	2}\frac{u(r)W(r)}{r^2}-\frac{1}{4} \frac{W^2(r)}{r^2} \right)
dr\,.
\]

According to the optical theorem,
\[
\texttt{Im}~f_{p(n)}(0)=\frac{k_{p(n)}}{4 \pi}\sigma_{p(n)}\,,
\]
where $\sigma_{p(n)}$ is the total scattering
cross section of the proton and the neutron by carbon, respectively, and 
\[
k_{p(n)}=\frac{m_{p(n)}}{m_{D}}k\backsimeq\frac{1}{2}k\,.
\]
As a result, (\ref{eikanal3_13}) can be written as
\begin{equation}
	\texttt{Re}~d_1=-\frac{3}{\pi} \left( \texttt{Re}~f_p(0) \sigma_{n}+\texttt{Re}~f_n(0)
	\sigma_{p}
	\right) G
	\label{eikanal3a_id1}
\end{equation}
\begin{equation}
	\texttt{Im}~d_1=\left( \frac{24}{k}
	\texttt{Re}~f_p(0)~\texttt{Re}~f_n(0)-\frac{3k}{8
		\pi^2}\sigma_{p}\sigma_{n} \right) G\,. \label{eikanal3b_id1}
\end{equation}

In view of (\ref{eikanal3a_id1}), the analysis of the birefringence
phenomenon in this simple approximation gives information about
the relation between the real part of the amplitude of the
proton--nucleus zero--angle scattering (nucleon--nucleon
scattering in the case of interaction between deuterons and a
hydrogen target).
Expressions
(\ref{eikanal3a_id1}), (\ref{eikanal3b_id1}) enable  making first
estimates of the magnitude of the effect and demonstrating the
real possibility of its experimental observation
\cite{110,cosy_bar99,rins_65}. In particular, for phase $\varphi$,
determining the spin rotation angle of the deuteron, the following
value was obtained for energies up to hundreds of megavolts for a
carbon target:
\[
\varphi=\frac{2\pi\rho z}{k}\texttt{Re} d_{1}\simeq10^{-3}z\,,
\]
$z$ is the path length (in centimeters) traveled by the particle
in the target.

The magnitude of dichroism and tensor polarization acquired by an
nonpolarized beam after passing the path length $z$ is
$P_{zz}\approx 10^{-2}z$ \cite{cosy_bar99}.

The estimate of the magnitudes of the effect based on the eikonal
approximation formulas (\ref{eikanal3_42}) ignores the spin
dependence of the nucleon-nucleon interaction. The Glauber
multiple scattering theory generalized to the case when the spin
dependence of nucleon--nucleon interaction is included enables one
to take account of this dependence \cite{nn_gl,nn_sit}. The
analysis including the stated dependence was carried out in
\cite{cosy_bar99}. In view of this analysis,
the spin dependence of the nucleon-nucleon interaction  influences the birefringence effect in the range of
deuteron energies less than several Ge\,V in such a way that the estimates carried out using a simplified approach remains valid.

\subsection
{First Observation of Spin Dichroism with Deuterons in a Carbon
	Target}\label{cosy_ch:1}

In 2003, the first experimental observation of the deuteron spin
dichroism (the effect of the appearance of tensor polarization in
a deuteron beam transmitted through a carbon target) was carried
out at the electrostatic HVEC tandem Van-de-Graaff accelerator
with deuteron energy up to 20\,MeV (Institut f\"{u}r Kernphysik of
Universit\"at zu K\"oln) \cite{spinorb_ex1,spinorb_ex2,rins_63}.
These experiments revealed that in the energy interval 5--20\,MeV,
both the value and the sign of the deuteron tensor polarization in
a carbon target change with changing deuteron energy.

In 2007, spin dichroism and the effect of appearance of tensor
polarization for 5.5\,GeV/\,s deuterons transmitted through carbon
targets were observed in another experiment at the Nuclotron in
JINR (Russia) \cite{VKB_dub}. Taking into account both nuclear
interaction and the Coulomb interaction between the deuteron and
the nucleus enables one  to describe the effect of changing sign
of tensor polarization with the change in the deuteron energy,
which was observed in the experiments with the deuterons of energy
$5\div 20$\,MeV \cite{232,110,rins_65}. However, the magnitude of
the effect has not been explained yet \cite{Sey2011}.

For high energy deuterons the nuclear interaction of a deuteron
with a light nucleus gives the main contribution to the
birefringence effect.
Comparison of the theoretical estimation of \cite{232,110}, which
includes the consideration of nuclear interaction, with the
experimental results of \cite{VKB_dub} demonstrates their qualitative
agreement.

Moreover, the sign of the effect obtained in theoretical
calculations coincides with that measured in the experiment.

It is necessary to add that as follows from the experiment, the
phenomenon of deuteron spin dichroism can be used for obtaining a
source of tensor-polarized deuterons
\cite{spinorb_ex1,spinorb_ex2,spinorb_ex3,VKB_dub,rins_63,Sey2011}.

Thus, theoretically predicted phenomena of spin dichroism and
appearance of tensor polarization for deuterons transmitted
through matter \cite{232,110} have been revealed in the
experiment. It would be recalled that in addition to dichroism,
the effect of deuteron birefringence  is accompanied by spin
oscillations, rotation and conversion of vector polarization to tensor that  and vice versa,  
which are still awaiting experimental observation.

\subsection{Tensor polarization rotation and oscillation between vector and tensor polarizations}

Let us examine  the equations (\ref{cosy_polpar1}) more carefully.  
These equations describe change of spin characteristics of a deuteron (nucleus with spin 1) during its motion in a target.  
The change of  vector polarization With the particle's motion through matter is described by the first three equations for polarization components $p_{x}$, $p_{y}$, and $p_{z}$.  
Let us consider the equations for $p_{x}$ and $p_{y}$:
%
	Hereafter, unless otherwise specified, we consider the target to be thin, so that birefringence effects are small.
%
%
	\begin{eqnarray}
		p_{x}&=& \frac{{\left[ {1 - \frac{{1}}{{2}}\rho z\left( {\sigma
						_{0} + \sigma _{1}}  \right)} \right] p_{x,0}+
				\frac{{4}}{{3}}\frac{{\pi \rho z}}{k}\texttt{Re} d_{1}
				p_{yz,0}}}{{\textrm{Tr}\hat {\rho }\hat {I}}}\,,\nonumber
		\\
		p_{y} &=& \frac{{\left[ {1 - \frac{{1}}{{2}}\rho z\left( {\sigma
						_{0} + \sigma _{1}}  \right)} \right] p_{y,0}  -
				\frac{{4}}{{3}}\frac{{\pi \rho z}}{k} \texttt{Re} d_{1}
				p_{xz,0}}}{{\textrm{Tr}\hat {\rho }\hat {I}}}\,, 
		\label{px+py}
	\end{eqnarray}

\begin{figure}[h]
	\epsfxsize = 12 cm \centerline{\epsfbox{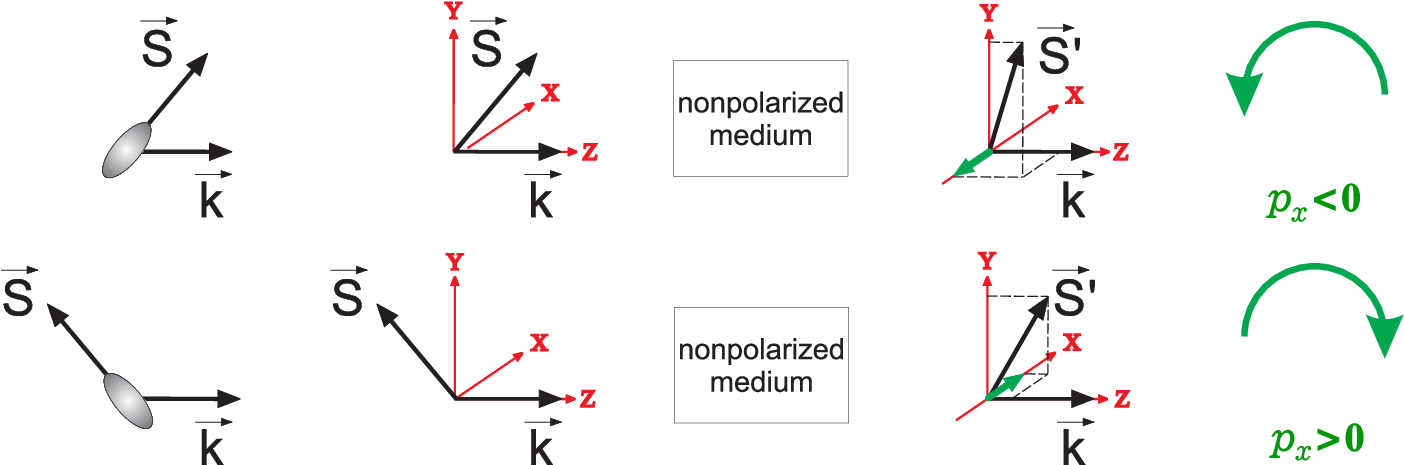}}
	\caption{Rotation for an acute and obtuse angles between spin and momentum occurs in opposite directions} \label{fig:slide35}
\end{figure}

According to (\ref{px+py}), while the particle moves through matter, nonzero initial tensor polarization ($p_{yz,0}$ and $p_{xz,0}$) leads to rotation of the polarization vector.
%
%
%
%
Let us consider a pair of equations (\ref{px+py}) for vector polarization components  $p_x$ and $p_y$. 
Suppose, that component $p_x$ is equal to zero when particle enters the target  ($p_{x,0}=0$). 
In this case tensor polarization component $p_{xz}$ presenting in (\ref{px+py})  is also equal to zero ($p_{xz,0}=0$). 
Polarization vector lays in plane 
 $(y,z)$ (see Fig.~\ref{fig:slide35}).
Equations (\ref{px+py}) for this case can be expressed as follows:
\begin{eqnarray}
	p_{x}&=& \frac{{4}}{{3}}\frac{{\pi \rho z}}{k}\texttt{Re} d_{1}
			p_{yz,0}\,,\nonumber
	\\
p_{y} &=& \frac{{\left[ {1 - \frac{{1}}{{2}}\rho z\left( {\sigma
				_{0} + \sigma _{1}}  \right)} \right] }}{{\textrm{Tr}\hat {\rho }\hat {I}}}\, p_{y,0} \,, 
	\label{px+py_1}
\end{eqnarray}
It is clear, that appearance of nonzero component $p_x$ means rotation of polarization vector. The angle of rotation  can be evaluated as follows:
$$
\vartheta \approx \frac{p_x}{p_y}= \frac{{4 \pi}}{{3}}\frac{{ \rho z}}{k}\texttt{Re} d_{1} \frac{	p_{yz,0}}{p_{y,0}}.
$$
In the specific case, when in the deuteron wavefunction state with  $m=-1$ is absent (amplitude $c$ in (\ref{cosy_psidepth}) is equal to zero)
the ratio $\frac{p_{yz,0}}{p_{y,0}}=\frac{3}{2}$, then the angle of rotation reads as follows:
%
%
$$
\vartheta \approx  2 \pi \lambda_c \, \frac{{1}}{\gamma} \, \texttt{Re} d_{1} z.
$$
%
%

Suppose the nucleus entering the target does not possesses vector polarization, but has tensor polarization.  
In this case, tensor polarization  causes the appearance of vector polarization.

And vice versa, according to expressions
(\ref{cosy_polpar1}) for tensor polarization:
\begin{eqnarray}
	p_{xz} &=& \frac{{\left[ {1 - \frac{{1}}{{2}}\rho z\left( {\sigma
					_{0} + \sigma _{1}}  \right)} \right] p_{xz,0} + 3\frac{{\pi \rho
					z}}{{k}}\texttt{Re} d_{1} p_{y,0} } }{{\textrm{Tr}\hat {\rho }\hat
			{I}}}\,,\nonumber
	\\
	p_{yz} &=& \frac{{\left[ {1 - \frac{{1}}{{2}}\rho z\left( {\sigma
					_{0} + \sigma _{1}} \right)} \right]p_{yz,0} - 3\frac{{\pi \rho
					z}}{{k}}\texttt{Re} d_{1} p_{x,0}} }{{\textrm{Tr}\hat {\rho }\hat
			{I}}}\,.
	\label{pxz+pyz}
\end{eqnarray}
Vector polarization leads to the appearance of tensor polarization components $p_{yz}$ and $p_{xz}$, if the particle entering the target did not possess them.  
At the same time, the vector polarization component $p_{x,0}$ leads to the appearance of $p_{yz}$, and the component $p_{y,0}$ to the appearance of $p_{xz}$.  
If both tensor polarization components $p_{yz,0}$ and $p_{xz,0}$ (or one of the components either $p_{yz,0}$ or $p_{x,0}$) are nonzero upon entering the target, rotation also arises.
%
	Evolution of vector and tensor polarization is similar to axis precession for a gyroscope rotating and moving over complicated trajectory.  A separate paper will be dedicated to exploring this issue in detail.
%
%

From (\ref{px+py_1}) it follows that angle of rotation for polarization vector for a deuteron beam passing distance  $l$  in a thin target with density $\rho$  (the number of nuclei per cm$^3$) reads:
\[
\vartheta= 2\pi {\rho l \lambda_c}\,\frac{1}{\gamma}\, \texttt{Re} d_{1}\simeq 10^{-3} \, l\,,
\]
where $\lambda_c = \frac{\hbar}{Mc}$ is the Compton wavelength.

Analysis shows that for deuterons of several GeV energies passing through a carbon target with thickness of several nuclear interaction lengths angle of polarization  vector (polarization tensor) rotation is about $\vartheta \sim 1 \cdot 10^{-2}$~rad.

For an acute and obtuse angles between spin and momentum rotation  occurs in opposite directions (see Fig.~\ref{fig:slide35}). 

%
Evaluation of spin polarization rotation effect for a deuteron beam in a carbon target is given by Table~\ref{tab:slide35} for conditions similar to those for deuteron dichroism observation \cite{azhgirei2010}, namely: deuteron momentum 5~GeV/c and carbon target thickness $\rho l=83 $\,g~cm$^{-2}$. In experiment \cite{azhgirei2010} tensor polarization component  $p_{zz}$ was measured to be as high as  $p_{zz} \approx 0.1$ that corresponded to $\Delta\sigma/\sigma \approx 0.06$. Rotation  of polarization  vector  reveals by acquiring components  $p_x$  
{and rotation angle} is evaluated in Table~\ref{tab:slide35} for number of deuterons in a bunch $N_{b}$.
Evaluations show that time required to observe the discussed phenomena for beams of polarized deuterons  at NuclotronM is about 10--30 hours.

\begin{table}[ht]
	\setlength{\extrarowheight}{2pt}
	\renewcommand{\arraystretch}{1.6}
	\caption{Evaluation of deuteron spin 
		rotation effect for carbon target}
	\label{tab:slide35}	
	\centering
	\begin{tabular}{|c|c|c|c|c|c|c|c|}
		\hline
		
		~~$N_{b}$~~ &
		$~~{\sigma}$, b~~ &
		${\rho l}$, g$\cdot$cm$^{-2}$ &
		$\Delta\sigma/\sigma$ &
		$\lvert \frac{\mathrm{Re}(d_1)}{\mathrm{Im}(d_1)} \rvert$ &
		$ \vartheta \approx \frac{|p_x|}{|p_y|}$
		\\		\hline
		$10^{10}$ &
		0.6 &
		83 &
		0.06 &
		0.1 &
		$4\cdot10^{-3}$ 
		\\		\hline
	\end{tabular}
\end{table}
%


\subsection{Dichroism effect, rotation of spin and tensor polarization for heavy nuclei} 
\label{sec:heavy0}

All the above considered phenomena, which are caused by the birefringence effect (spin dichroism, spin rotation and tensor polarization, transition of vector polarization into tensor and tensor into vector), exist for all particles with spin $S \ge 1$. 
It is very important to note that, as mentioned hereinabove, the birefringence effect associated
with particles motion through nonpolarized matter 
is caused by particle's internal anisotropy. 
For nuclei heavier than deuteron, this anisotropy could significantly exceed that for deuterons. 
For example, for the $^{21}$Ne nucleus, the deviation of the nuclear shape from spherical is about 30 percent.  
Many other nuclei possess a noticeable rate of anisotropy.

Therefore, initially nonpolarized beam of heavy nuclei with spin $S \ge 1$ passing through a nonpolarized external target at Nuclotron would acquire tensor polarization due to spin dichroism. 
%

\begin{figure}[h]
	\begin{center}
		\resizebox{!}{50mm}{\includegraphics{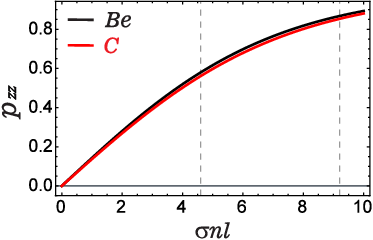}}
	\end{center}
	\caption{Dependence of tensor polarization  $p_{zz}$ acquired by $^{21}$Ne nuclei passing through carbon and beryllium targets. Vertical dashed lines in plot mark target thickness $\sigma n l=4.6$ and $9.2$ corresponding beam attenuation 100 and $10^4$ times, respectively }
	\label{fig:neon}
\end{figure}

For example, a beam of $^{21}$Ne nuclei, which possess spin $3/2$ and large quadrupole deformation ($\beta_2$=0.463), acquire tensor polarization, which value is the higher, the longer is  the nuclei path in the target.
If the beam passes through the external target 
at Nuclotron  distance $l=4.6 \cdot L_{nuc}$, it is attenuated 100 times ($N/N_0=10^{-2}$), attenuation $10^4$ times corresponds to $l=9.2 \cdot L_{nuc}$, here $L_{nuc}=\frac{1}{n \sigma}$, $n$ is the number of scatterers in the target per cm$^3$, $\sigma$ is the total cross-section of $Ne$ scattering on the target nuclei, $\sigma n l$ has the meaning of target thickness in nuclear lengths.
According to evaluations \cite{school2025_1,school2025_2,SPD_seminar}  tensor polarization of $^{21}$Ne beam in carbon target appears as high as $p_{zz}=0.56$ at $\sigma n l =4.6$  and $p_{zz}=0.85$ at $\sigma n l =9.2$.  
Evaluations carried by Lebedev and Shimansky  for $^{21}$Ne beam passing through 20cm thick beryllium  target give similar results:  tensor polarization $p_{zz} \sim $0.4 can be obtained \cite{SPIN2025}. 
All the results are gathered in 
Table~\ref{tab:Ne-C} and Fig.~\ref{fig:neon}, in calculations $\Delta \sigma / \sigma$ was supposed to be $\Delta \sigma / \sigma = 0.133$ and $0.128$ for beryllium   and  carbon  targets, respectively.

\begin{table}[ht]
	\setlength{\extrarowheight}{2pt}
	\renewcommand{\arraystretch}{1.6}
	\caption{Evaluation of tensor polarization acquired by $^{21}$Ne nuclei in carbon and beryllium targets for different beam attenuation values}
	\label{tab:Ne-C}	
	\centering
	\begin{tabular}{|c|c|c|c|}
		\hline
		
		~~$N/N_0$~~ & ~~~$\sigma n l$ ~~~
		& ~~~$p_{zz}~(Be)~~~$ &
		~~~$p_{zz}~(C)~~~$ 
		\\		\hline
		$10^{-2}$ &
		4.6 
		& 0.58  &
		0.56  
		\\		\hline
		$10^{-4}$ &
		9.2 
		&  0.87&
		0.85  
		\\		\hline
	\end{tabular}
\end{table}
%


Great value of tensor polarization appearing due to spin dichroism for  many nuclei, gave possibility to V.Lebedev and C.Shimansky to formulate an idea for application of nuclei, which acquire tensor polarization at low energies, for further acceleration and bringing into NICA ring. Therefore, one gains possibility to study reactions induced at collisions of tensor polarized nuclei beams.

Certainly, it is also necessary to consider experiments with heavy nuclei enabling to observe the effect of tensor polarization rotation and that of  tensor polarization conversion to vector polarization.


\subsection{Possibilities to observe spin dichroism at the internal target  of NuclotronM}
\label{sec:heavy}

Let us consider deuteron beam motion in a storage ring in the presence of external magnetic fields.
The spin precession of the particle, caused by the interaction of the particle magnetic moment with the external magnetic field, is described by the Bargmann-Michel-Telegdi equation \cite{6,23}
\begin{equation}
	\frac{d\vec{p}}{dt}=[\vec{p}\times\vec{\Omega}_{0}],
	\label{2.1}
\end{equation}
where $t$ is the time in the laboratory frame,
\begin{equation}
	\vec{\Omega}_{0}=\frac{e}{mc}\left[\left(a+\frac{1}{\gamma}\right)\vec{B}
	-a\frac{\gamma}{\gamma+1}\left(\vec{\beta}\cdot\vec{B}\right)\vec{\beta}
	\right]
	,
	\label{2.2}
\end{equation}
$m$ is the particle mass, $e$ is its charge, $\vec{p}$ is the polarization vector, $\gamma$ is the Lorentz factor,
$\vec{\beta}=\vec{v}/c$, $\vec{v}$ is the particle velocity, $a=(g-2)/2$, $g$ is the gyromagnetic ratio,
$\vec{B}$ is the magnetic field at the particle location.

Thus, evolution of the deuteron spin is described by the following equation:
\begin{eqnarray}
	\frac{d\vec{p}}{dt}=
	\frac{e}{mc}\left[\vec{p}\times\left\{\left(a+\frac{1}{\gamma}\right)\vec{B}
	-a\frac{\gamma}{\gamma+1}\left(\vec{\beta}\cdot\vec{B}\right)\vec{\beta}
	\right\}
	\right].
	\label{2.4}
\end{eqnarray}

However, the equation (\ref{2.4}) alone is not sufficient to describe spin evolution in the Nuclotron with an internal target: it is necessary to supplement it with a contribution
caused by interaction of the deuteron with the internal target.
This interaction is described by effective potential energy $\hat{V}$, which a particle in matter possesses
\cite{232,110,A2}:
\begin{equation}
	\hat{V} =- \frac{2 \pi {\hbar}^2}{M} {\rho} \hat{f}(0),
	\label{U1}
\end{equation}
where $\hat{f}(0)$ is the amplitude of elastic coherent forward scattering, the explicit expression of $\hat{f}(0)$ for a particle with  spin $S \ge 1$ (see (\ref{18.17})) was obtained in   \cite{232,110,A2}.
In case if one neglect weak interactions it converts to the following form:
\begin{equation}
	\hat{f(0)}=d+d_{1}\left(\hat{\vec{S}}\vec{n}\right)^{2},
	\label{1.2}
\end{equation}
where  $\vec{n}$ is the unit vector in the direction of particle momentum.

The density matrix of a system ''deuteron beam + target''  can be expressed as \cite{rins_98}:
\begin{eqnarray}
	\hat{\rho}=\hat{\rho}_{d}\otimes \hat{\rho}_{t},
	\label{2.6}
\end{eqnarray}
where $\hat{\rho}_{d}$ is the density matrix of a deuteron beam, $\hat{\rho}_{t}$ density matrix of a target.
\noindent The density matrix of a deuteron beam
\begin{eqnarray}
	\hat{\rho}_{d}=I(\vec{k})\left(\frac{1}{3}\hat{\texttt{I}}
	+\frac{1}{2}\vec{p}(\vec{k})\hat{\vec{S}}+\frac{1}{9}p_{ik}(\vec{k})\hat{Q}_{ik}\right),
	\label{2.7}
\end{eqnarray}
$I(\vec{k})$ is the beam intensity, $\vec{p}$ is the polarization vector, $p_{ik}$ is the polarization tensor for the deuteron beam.
%

Equation for density matrix of a deuteron beam reads as: 
\begin{eqnarray}
	\frac{d\hat{\rho}_{d}}{dt}=-\frac{i}{\hbar}\left[\hat{H},\hat{\rho}_{d}\right]{+\left(\frac{\partial\hat{\rho}_{d}}{\partial
			t}\right)_{col}},
	\label{2.9}
\end{eqnarray}
where $\hat{H}=\hat{H}_{0}+ {\hat{V}}$.

The term responsible for collisions   
$\left(\frac{\partial\hat{\rho}_{d}}{\partial t}\right)_{col}$ can be obtained with the use of method described in \cite{rins_98} and section~\ref{sec:protonspin}:
\begin{eqnarray}
	\left(\frac{\partial\hat{\rho}_{d}}{\partial
		t}\right)_{col}
	=vN\emph{Sp}_{t}\left[\frac{2\pi
		i}{k}\left[\hat{F}(\theta=0)\hat{\rho}-\hat{\rho} \hat{F}^{+}(\theta=0)\right] +\int
	d\Omega \hat{F}(\vec{k}^{'})\hat{\rho}(\vec{k}^{'})\hat{F}^{+}(\vec{k}^{'})\right],
	\label{2.11}
\end{eqnarray}
where $\vec{k}^{'}=\vec{k}+\vec{q}$, $\vec{q}$ is the momentum, transferred from the incident particle  to matter,
$v$ is the velocity of the incident particle, $N$ is the number of atoms in cm$^3$ of matter,
$\hat{F}$ is the scattering amplitude, which depends on spin operators of deuterons and nuclei (atoms) of matter, $\hat{F}^+$ is the operator Hermitian conjugate to operator $\hat{F}$.
The first term in (\ref{2.11}) describes coherent scattering of the particle by the nuclei of matter, while the second one is responsible on the multiple scattering.

Let us consider the first term in (\ref{2.11}) in more details:
\begin{eqnarray}
	\left(\frac{\partial\hat{\rho}_{d}}{\partial
		t}\right)_{col}^{(1)}=vN\frac{2\pi i}{k}
	\left[
	\hat{f}(0)\hat{\rho}_d-\hat{\rho}_d \hat{f}(0)^{+}
	\right] .
	\label{2.12}
\end{eqnarray}
Amplitude $\hat{f}(0)$ of forward scattering of a deuteron  in a nonpolarized target  can be expressed as:
\begin{eqnarray}
	\hat{f}(0)=\emph{Sp}_{t} \hat{F}(0) \hat{\rho}_{t}.
	\label{2.13}
\end{eqnarray}
In accordance with (\ref{1.2}) the amplitude reads: 
\begin{eqnarray}
	\hat{f}(0)=d+d_{1}(\hat{\vec{S}}\vec{n})^{2},
	\label{2.14}
\end{eqnarray}
where $\vec{n}=\vec{k}/k$, $\vec{k}$ is the deuteron momentum.
As a result expression (\ref{2.12}) reads as follows:
\begin{eqnarray}
	\left(\frac{\partial\hat{\rho}_{d}}{\partial t}\right)_{col}^{(1)} =
	-\frac
	i\hbar\left(\hat{V} {\hat{\rho}_d}-{\hat{\rho}_d} \hat{V}^{+}\right).
	\label{2.15}
\end{eqnarray}
where $\hat{V}$ is defined by (\ref{U1}).

\noindent And finally, equation   (\ref{2.9}) can be presented as follows:
\begin{eqnarray}
	\frac{d\hat{\rho}_{d}}{dt}=-\frac{i}{\hbar}\left[\hat{H},\hat{\rho}_{d}\right]
	-\frac
	i\hbar\left(\hat{V} {\hat{\rho}_d}-{\hat{\rho}_d} \hat{V}^{+}\right){+
		vN
		\emph{Sp}_{t} \int d\Omega
		F(\vec{k}^{'})\hat{\rho}(\vec{k}^{'})F^{+}(\vec{k}^{'})}.
	\label{2.9_new}
\end{eqnarray}
%
%
%
The last term, proportional to $\emph{Sp}_{t}$, describes multiple scattering and resulting depolarization (in more details see section~\ref{sec:protonspin} and \cite{mil,nik1,nik2,nik}).
Hereinafter the target thickness enabling to neglect this term is used.

Beam intensity reads as
\begin{eqnarray}
	I (t)=\emph{Sp}_{d}\hat{\rho}_{d}.
	\label{2.16}
\end{eqnarray}
Therefore, the rate of intensity change is determined by scattering amplitude $\hat{f}(0)$ as follows:
\begin{eqnarray}
	\frac{dI}{dt}=vN\frac{2\pi
		i}{k}\emph{Sp}_{d}\left[\hat{f}(0)\hat{\rho}_{d}-\hat{\rho}_{d} \hat{f}^{+}(0)\right].
	\label{2.17}
\end{eqnarray}
Substituting (\ref{2.7}) and (\ref{2.14}) to (\ref{2.17}) one can get 
the rate of intensity change caused by the tensor polarization components as follows
\begin{eqnarray}
	\frac{dI}{dt}=\frac{\chi}{3}\left[2+p_{ik}n_{i}n_{k}\right]I(t)+\alpha
	I(t),
	\label{2.18}
\end{eqnarray}
where parameters  $\chi=-\frac{4\pi
	vN}{k}\texttt{Im}d_{1}=-vN(\sigma_{\pm 1}-\sigma_0)$ and
$\alpha=-\frac{4\pi vN}{k}\texttt{Im}d=-v N \sigma_0$ depend on total scattering cross sections $\sigma_{\pm 1}$
and $\sigma_0$  for quantum numbers   $m= \pm 1$
and $m=0$, respectively.

Vector polarization $\vec{p}$ of a deuteron beam reads as follows:
\begin{eqnarray}
	\vec{p}=\frac{\emph{Sp}_{d}\hat{\rho}_{d} \hat{\vec{S}}}{\emph{Sp}_{d}\hat{\rho}_{d}}=\frac{\emph{Sp}_{d}\hat{\rho}_{d} \hat{\vec{S}}}{I(t)}.
	\label{2.19}
\end{eqnarray}
Differential equations describing vector polarization can be obtained from (\ref{2.19}):
\begin{eqnarray}
	\frac{d\vec{p}}{dt}=\frac{\emph{Sp}_{d}(d\hat{\rho}_{d}/dt) \hat{\vec{S}}}{I(t)}-
	\vec{p} \,\frac{\emph{Sp}_{d} (d\hat{\rho}_{d}/dt)}{I(t)}.
	\label{2.20}
\end{eqnarray}
The tensor polarization components $p_{ik}$  are defined as:
\begin{eqnarray}
	p_{ik}=\frac{\emph{Sp}_{d}\hat{\rho}_{d} \hat{Q}_{ik}}{\emph{Sp}_{d}\hat{\rho}_{d}}=\frac{\emph{Sp}_{d}\hat{\rho}_{d} \hat{Q}_{ik}}{I(t)},
	\label{2.21}
\end{eqnarray}
where quadrupolarization operator $\hat{Q}_{ik}$ reads:
$\hat{Q}_{ik}=\frac{3}{2}\left({\hat{S}_{i}}{\hat{S}_{k}}+{\hat{S}_{k}}{\hat{S}_{i}}-\frac{4}{3}\delta_{ik}\hat{\texttt{I}}\right)$.
%
%
Similar to the vector polarization, the following equation describes evolution of the tensor polarization
\begin{eqnarray}
	\frac{dp_{ik}}{dt}=\frac{\emph{Sp}_{d}(d\hat{\rho}_{d}/dt) \hat{Q}_{ik}}{I(t)}-
	p_{ik}\frac{\emph{Sp}_{d} (d\hat{\rho}_{d}/dt)}{I(t)}.
	\label{2.22}
\end{eqnarray}
%

\noindent Combining equations (\ref{2.7}) and (\ref{2.1}), (\ref{2.20}) and (\ref{2.22}), along with condition  $p_{xx}+p_{yy}+p_{zz}=0$, yields a system \cite{rins_98,Baryshevsky2005} describing the evolution of both vector and tensor polarization components for a deutron:

\begin{eqnarray}
	\left\{
	\begin{array}{l}
		\frac{d\vec{p}}{dt}=
		\frac{e}{mc}\left[\vec{p}\times\left\{\left(a+\frac{1}{\gamma}\right)\vec{B}
		-a\frac{\gamma}{\gamma+1}\left(\vec{\beta}\cdot\vec{B}\right)\vec{\beta}
		\right\}\right]+\\
		+\frac{\chi}{2}(\vec{n}(\vec{n}\cdot\vec{p})+\vec{p}) + \\
		+  \frac{\eta}{3}[\vec{n}\times\vec{n}^{'}]
		-\frac{2\chi}{3}\vec{p}-\frac{\chi}{3}(\vec{n}\cdot\vec{n}^{'})\vec{p}
		,\\
		{} \\
		\frac{dp_{ik}}{dt}  =  -\left(\varepsilon_{jkr}p_{ij}\Omega_{r}+\varepsilon_{jir}p_{kj}\Omega_{r}\right) + \\
		+
		\chi\left\{-\frac{1}{3}+n_{i}n_{k}+\frac{1}{3}p_{ik}-\frac{1}{2}(n_{i}^{'}n_{k}+n_{i}n_{k}^{'})
		+\frac{1}{3}(\vec{n}\cdot\vec{n}^{'})\delta_{ik}\right\} + \\
		+
		\frac{3\eta}{4}\left([\vec{n}\times\vec{p}]_{i}n_{k}+n_{i}[\vec{n}\times\vec{p}]_{k}\right)
		-\frac{\chi}{3}(\vec{n}\cdot\vec{n}^{'})p_{ik},
		\\
	\end{array}
	\right.
	\label{2.23}
\end{eqnarray}
where $\vec{n}=\vec{k}/k$,
$\eta=-\frac{4 \pi N}{k} \texttt{Re}d_{1}$,
$n_{i}^{'}=p_{ik}n_{k}$, 
$\Omega_{r}$ are the components of  $\vec{\Omega}$
($r=1,2,3$ corresponds to $x,y,z$):
\begin{eqnarray}
	\vec{\Omega} & = &
	\frac{e}{mc}\left\{\left(a+\frac{1}{\gamma}\right)\vec{B}
	-a\frac{\gamma}{\gamma+1}\left(\vec{\beta}\cdot\vec{B}\right)\vec{\beta}
	\right\} .
	\label{2.24}
\end{eqnarray}

Now let us take into account that the target is located in the ring section, where the magnetic field is absent. As a result, the system of equations (\ref{2.23}) is conveniently split into two systems: one (see (\ref{2.23_a})) describes the behavior of deuteron beam spin characteristics in that section and during the time interval, where the magnetic field $\vec{B}$ is present, but the target is absent, and the second (see (\ref{2.23_b})) describes spin characteristics inside the target, where magnetic field is not applied:
\begin{eqnarray}
	\label{2.23_a}
	\left\{
	\begin{array}{l}
		\frac{d\vec{p}}{dt}=
		\left[\vec{p} \times \vec{\Omega} \right]
		,\\
		{} \\
		\frac{dp_{ik}}{dt}  =  -\left(\varepsilon_{jkr}p_{ij}\Omega_{r}+\varepsilon_{jir}p_{kj}\Omega_{r}\right)
		\\
	\end{array}
	\right.
\end{eqnarray}

\begin{eqnarray}
	\label{2.23_b}
	\left\{
	\begin{array}{l}
		\frac{d\vec{p}}{dt}=
		\frac{\chi}{2}(\vec{n}(\vec{n}\cdot\vec{p})+\vec{p}) + 
		\frac{\eta}{3}[\vec{n}\times\vec{n}^{'}]
		-\frac{2\chi}{3}\vec{p}-\frac{\chi}{3}(\vec{n}\cdot\vec{n}^{'})\vec{p}
		,\\
		{} \\
		\frac{dp_{ik}}{dt}  = 
		\chi\left\{-\frac{1}{3}+n_{i}n_{k}+\frac{1}{3}p_{ik}-\frac{1}{2}(n_{i}^{'}n_{k}+n_{i}n_{k}^{'})
		+\frac{1}{3}(\vec{n}\cdot\vec{n}^{'})\delta_{ik}\right\} + \\
		+
		\frac{3\eta}{4}\left([\vec{n}\times\vec{p}]_{i}n_{k}+n_{i}[\vec{n}\times\vec{p}]_{k}\right)
		-\frac{\chi}{3}(\vec{n}\cdot\vec{n}^{'})p_{ik},
		\\
	\end{array}
	\right.
\end{eqnarray}
where $\vec{n}=\vec{k}/k$,
$\eta=-\frac{4 \pi N}{k} \texttt{Re}d_{1}$,
$n_{i}^{'}=p_{ik}n_{k}$, 
%
%
$\chi=-\frac{4\pi
	vN}{k}\texttt{Im}d_{1}=-vN(\sigma_1-\sigma_0)$.

Suppose that at instant $t_0$
a target of thickness $L$ is inserted into the beam’s path.
At this instant, particles possessing  polarization vector $\vec{p}_0$ and polarization tensor $p_{ik}^{(0)}$ pass through the boundary of the target.
After beam entering the target spin characteristics  $\vec{p}_0$ and $p_{ik}^{(0)}$ change due to interaction with the target in accordance with formulas (\ref{2.23_b}).
If the target is thin enough to make changes of vector and tensor polarization for a particle small, equations (\ref{2.23_b}) can be solved using the perturbations theory.
Therefore, from (\ref{2.23_a}) and (\ref{2.23_b})  for spin characteristics of a particle leaving the target  $\vec{p}(t_0 + \tau)$  and  
$p_{ik}(t_0 + \tau)$ one can write:
\begin{eqnarray}
	\label{2.25_b}
	\vec{p}\,(t_0+\tau)=\vec{p}_0 + 
	\frac{\chi}{2}(\vec{n}(\vec{n}\cdot\vec{p}_0)+\vec{p}_0)\tau
	+\frac{\eta}{3}[\vec{n}\times\vec{n}^{'}_{0 }]\tau 
	-\frac{2\chi}{3}\vec{p}_0\tau
	-\frac{\chi}{3}(\vec{n}\cdot\vec{n}^{'}_{0})\vec{p}_0\tau 
	,
\end{eqnarray}
\begin{eqnarray}
	\label{2.26_b}
	\lefteqn
	p_{\,~ik}(t_0+\tau) & =&
	p_{ik}^{(0)}
	+ 
	\chi\left[-\frac{1}{3}+n_{i}n_{k}+\frac{1}{3}p_{ik}^{(0)}
	-\frac{1}{2}(n_{i0}^{'}n_{k}+n_{i}n_{k0 }^{'})
	+\frac{1}{3}(\vec{n}\cdot\vec{n}^{'}_{0 })\delta_{ik}\right]\tau + \nonumber \\
	& + &  \frac{3\eta}{4}\left([\vec{n}\times\vec{p}_0]_{i}n_{k}
	+n_{i}[\vec{n}\times\vec{p}_0]_{k}\right)\tau
	-\frac{\chi}{3}(\vec{n}\cdot\vec{n}^{'}_{0})p_{ik}^{\,(0)}\tau ,
\end{eqnarray}
where $\vec{p}_0$ is the beam polarization  at instant $t_{0}$,
$n_{i0}^{'}=p_{ik}^{\,(0)}n_{k}$,
$p_{ik}^{\,(0)}$ are the components of polarization tensor at the same instant, $\tau$ is the time interval, which the particle spends in the target.

The further evolution of $\vec{p}$ and $p_{ik}$ is again determined by the equations (\ref{2.23}).
After one revolution period $T$, a particle enters the target again possessing spin parameters $\vec{p}\,(t_0 + \tau + T)$ and
$p_{ik}(t_0 + \tau + T)$, which have changed compared to their values at the time $(t_0 + \tau)$ due to the spin rotation in the magnetic field in Nuclotron ring.
These new values can be used as the initial conditions, when solving the equations (\ref{2.23_b}), i.e. one can use the solutions of (\ref{2.25_b}) and (\ref{2.26_b}) with the replacement of $\tau$ by $\tau + T$.
This iterative process can be continued further.
%
%
Solutions (\ref{2.25_b})  and (\ref{2.26_b}) are given for a single turn of a deuteron in a ring.  
The frequency of spin precession and that of cyclotron motion over the orbit are different (their ratio i.e. spin tune is unconstrained). 
Therefore, after each particle turn in a ring those components of vector and tensor polarization, which lay in the orbit plane, differ from their initial values in the point, in which the target is supposed to be located. 
That is why, the polarization vector component lying in the orbital plane of a particle rotating in a ring, when averaged over a time interval exceeding the rotation period, vanishes.
However, the situation is entirely different for tensor polarization: the diagonal components of the polarization tensor, which arise due to the spin dichroism effect in an initially nonpolarized beam, do not vanish after averaging over time! (see section~\ref{sec:deuteron}).
Consequently, the arising spin dichroism effect  does not dependent on the ratio of the spin precession frequency to the cyclotron frequency (i.e. spin tune).  
%
%

Hereinafter it is more convenient to consider evolution of polarization characteristics of a particle beam in an internal target of Nuclotron in a different way.

\subsection{ Evolution of polarization characteristics of a particle beam in an internal target of Nuclotron}
\label{sec:evolution}


General dynamics of tensor polarization of particles in external fields
is given in \cite{Huang,Silenko1,Silenko2}, the general theory describing motion of  particles possessing tensor polarization in a target in presence of electromagnetic fields is described in \cite{A2,rins_98}.

To find the explicit expressions for the quantum-mechanical evolution operators, let us suppose that  $y$-axis is directed along the direction of magnetic field $\vec{B}$, and $z$-axis is parallel to the  momentum of a particle at the instant it enters a target.
Then two parts of the Hamiltonian, which are responsible for the spin dynamics of the particle in the target ($\hat V$)  can be written as follows~\cite{rins_98,A2}:
\begin{equation}
	\hat V = -\frac{2\pi \hbar^2 N}{M \gamma} \hat f(0),
\end{equation}
and the Larmor precession in the storage ring ($\hat H$) reads as~\cite{2005Mane}:
\begin{equation}
	\hat H = -\frac{e \hbar}{M c} \left( \frac{g-2}{2} + \frac{1}{\gamma} \right) B_y \hat S_y.
\end{equation}
Here $N$ is  the number of scatterers in cm$^3$ of the target, $M$ is the mass of the incident particle, $\gamma$ is its Lorentz factor, $e$ is the particle charge  and $g$ is $g$-factor (for the deuteron $g \approx 0.86$), $\hat f(0) = d_0 + d_1 \hat S_z^2$ is the amplitude of coherent elastic forward scattering in the reference frame in which the target rests.
Note that operator $\hat V$ is non-Hermitian ($\hat V \neq \hat V^+$) due to presence of nonzero imaginary parts for parameters $d_0$ and $d_1$.

Since $\tau$ is the time interval for a particle to pass through the target once, and $T - \tau \approx T$ is the time interval, when the particle moves in the storage ring beyond the target  ($\tau \ll T$), the evolution operators after passing each of two sections at a single turn are
\begin{equation}
	\hat U_V = \,{\texttt{e}}^{-i \hat V \tau / \hbar}
\end{equation}
and
\begin{equation}
	\hat U_B = \,{\texttt{e}}^{-i \hat H T / \hbar}.
\end{equation}
Then, the evolution operator for one turn in the storage ring is the product $\hat U_1 = \hat U_B \hat U_V$, and after $n$ turns it is defined as  $\hat U^{(n)} = \hat U_1^n$.

Using the following equalities valid for particles with spin $S=1$, namely: $\hat S_z^4 = \hat S_z^2$ and $\hat S_y^3 = \hat S_y$, the evolution operators $\hat U_{B(V)}$ can be transformed as follows:
\begin{equation}
	\hat U_B = \hat{\texttt{I}} + \left(\cos(\phi) - 1 \right) \hat S_y^2 + i \sin(\phi) \hat S_y
\end{equation}
and
\begin{equation}
	\hat U_V = \,{\texttt{e}}^{i \alpha} \left( \hat{\texttt{I}} + (\,{\texttt{e}}^{i \zeta} - 1) \hat S_z^2 \right),
\end{equation}
where
$$\phi=\frac{e}{mc\hbar}\left(\frac{g-2}{2}+ \frac{1}{\gamma} \right)B_y T$$
is the spin rotation angle around the magnetic field direction per single turn in the storage ring,

\begin{equation}
	\label{eq:alpha}
	\alpha=\frac{2\pi \hbar N}{M\gamma}d_0\tau
\end{equation}
is the complex quantity responsible for spin-independent beam attenuation, 
and
\begin{equation}
	\label{eq:beta}
	\zeta=\frac{2\pi \hbar N}{M\gamma}d_1\tau
\end{equation}
is responsible  for spin dichroism and spin rotation after a single pass through the target.


%
	
	Let the initial state of the particle beam in the storage ring be described by the density matrix $\hat{\rho}_0$. 
	Then the average values $p_{ij}$ of the Cartesian components of the quadrupolarization operator~\cite{rins_98,Landau_3}
	\begin{equation}
		\hat Q_{ij} = \frac{3}{2 S (S - 1)} \left( \hat S_i \hat S_j + \hat S_j \hat S_i - \frac{2}{3} \hat{\vec{S}}^{\,2} \delta_{ij} \right)
	\end{equation}
	and the average values $p_i$ of the spin operator components $\hat S_i$ read as follows:
	\begin{equation}
		\label{eq:pij}
		p_{ij} = \frac{\mathrm{Sp}(\hat \rho^{(n)} \hat Q_{ij})}{\mathrm{Sp}(\hat \rho^{(n)})}
	\end{equation}
	and
	\begin{equation}
		p_i = \frac{\mathrm{Sp}(\hat \rho^{(n)} \hat S_i)}{\mathrm{Sp}(\hat \rho^{(n)})},
	\end{equation}
	where 
	$\hat \rho^{(n)}=\hat U^{(n)}\hat\rho_0\hat U^{(n)+}$
	is the density matrix after $n$ turns of the particle in the accelerator. 
	Since the relation
	$\hat \rho^{(n)}=\hat U_1\hat\rho^{(n-1)}\hat U_1^{+}$
	holds, the matrices $\hat \rho^{(n)}$ form an explicitly defined iterative sequence, which can be constructed using mathematical software packages.
	
	In case of insignificant change of the state for an initially nonpolarized deuteron beam due to birefringence in a target
	($|\zeta n|\ll1$) the linear approximation over $\zeta$ can be used, evaluation  $\hat U_V\approx \,{\texttt{e}}^{i\alpha }
	\hat{\texttt{I}}+i\,{\texttt{e}}^{i\alpha }\zeta \hat S_z^2$ is valid and evolution operator   $\hat U$ can be approximately expressed as follows:
	\begin{equation}
		\label{eq:Ulin}
		\hat U\approx \,{\texttt{e}}^{i\alpha n}\hat U_B^n+i\,{\texttt{e}}^{i\alpha n}\Big(\sum_{j=0}^{j=n-1}{\hat U_B^{j}\zeta \hat S_z^2\hat U_B^{-j}}\Big)\hat U_B^{n-1}.
	\end{equation}
	Then at $n \gg 1$ component  $p_{zz}$ of quadrupolarization tensor  reads as follows:
	\begin{equation}
		p_{zz}\approx-  \frac{1}{3} \, n \, \texttt{Im}\zeta.
		\label{eq:pzz}
	\end{equation}
	Discarding the rapidly oscillating term in \eqref{eq:pzz}, which comprises ratio of two sines, and using  \eqref{cosy_rot2+} and \eqref{eq:beta} one can get for $p_{zz}$ the following: 
	%
	\begin{equation}
		p_{zz}\approx\frac{\Delta\sigma}{\sigma}\frac{N \sigma z}{6}.
		\label{eq:pzzlin}
	\end{equation}
	where $\Delta\sigma=\sigma_0-\sigma_{\pm 1}$ and $\sigma=\frac{2}{3}\sigma_{\pm 1}+\frac{1}{3}\sigma_0$ (for deuterons $\Delta\sigma>0$), $N$ is  the number of atoms in cm$^3$ of the target.
	Note that for an internal target $p_{zz}$ value is four times lower than for an external target at the same path length $z$ for the particle in the target (see \eqref{eq:pzzlin}).

	Thus, deuterons passing through the Nuclotron internal target acquire tensor polarization. 
	Quadrupolarization tensor component $p_{zz}$ appears to be proportional to  length $z$ of the path, which a particle passes in the target.
	%
	It should be noted that in case of polarization measurements for deuterons interacting with the internal target,  either deuterons scattered in the target or the products of their interactions with the nuclei in the internal target are detected rather than polarization of the transmitted beam.
	For measurement with an external target, just polarization of the transmitted beam is investigated.
	Particles, which came into collisions, are scattered and leave the beam.
	The average value of the path length $z$ for a particle in the target is equal to mean free path $1/N\sigma$.
	Therefore, the average value of $p_{zz}$ for a single cycle of Nuclotron: 
	\begin{equation}
		\bar p_{zz}\approx\frac{\Delta\sigma}{6\sigma}.
		\label{eq:pzzav}
	\end{equation}
	Further discussion is based on the application of formulas~\eqref{eq:pij} and ~\eqref{eq:pzzav} to the description of spin dichroism phenomenon for deuterons in the internal target of the storage ring.

	\subsection{Dichroism effect for a deuteron beam moving in Nuclotron with internal target}
	\label{sec:deuteron}

	When conducting experiments with an internal target in the storage ring, it is necessary to consider the presence of a magnetic field, which leads to Larmor precession of the deuteron spin. 
	{This phenomenon leads to averaging the physical quantities and makes change in the relationship between the diagonal components of tensor polarization of the beam. }
	Due to precession, the average value of component $p_{zz}$ becomes equal to the average value of component $p_{xx}$, and due to relation $p_{xx} + p_{yy} + p_{zz} = 0$, the average value of $p_{yy}$ appears to be twice as large in magnitude as the average value of $p_{zz}$. 
	While the signs of the average values of the components $p_{zz}$ and $p_{yy}$ are opposite.
	As it was already mentioned, in experiments with an internal target the detector counts scattered particles, which is in contrast to an experiment with an extracted beam, where particles transmitted through the target are detected. 
	In contrast to the experiments with an external target, where a beam passes through the target only once, multiple beam turns $N_{turns}$ in Nuclotron contribute to  path length  $z$ for a particle in the internal target, $z=N_{turns} l$, where $l$ is the target thickness.
	
\begin{figure}[ht]
		\begin{center}
			\resizebox{80mm}{!}{\includegraphics{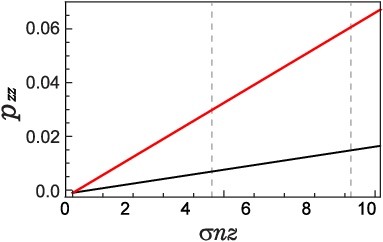}}
		\end{center}
		\caption{
			Dependence of tensor polarization component $p_{zz}$  on the path length for the particle in the target  at $\Delta\sigma / \sigma = 0.01$. Black curve corresponds to the tensor polarization of the beam in the experiments with an internal target, and red curve is for experiments with an external target. Vertical lines on the graphs correspond to beam attenuation by two and four orders of magnitude} \label{fig:deuteronsPzz}
	\end{figure}

	Figure~\ref{fig:deuteronsPzz} shows the dependence of the component $p_{zz}$ on the path length in the target, which is expressed in the terms of $L_{nuc}$ similar section~\ref{sec:heavy0}.
	Here, the value of $\Delta\sigma / \sigma$ is taken to be $0.01$
	\footnote{Analysis of deuteron dichroism experiments with momentum 5 GeV$/c$~\cite{VKB_dub,azhgirei2010} shows that this value can be considerably higher ($\Delta\sigma / \sigma \approx 0.06$).}. 
	For comparison, the same graph also shows the dependence of $p_{zz}$ for the case of an external target, which is 6 times higher (compare (\ref{eq:pzz23}) with (\ref{eq:pzzlin}) and	(\ref{eq:pzzav})). 
	
	
	As an example let us consider measurements with the use of polarimeter developed by~\cite{2011Kurilkin}, which comprises polyethylene~CH$_2$ target of 10$\mu$m thickness and the deuteron beam with 270~MeV energy (see also \cite{Ladygin_1990,Ladygin_1995,Ladygin_1998}).  
	It is noteworthy that during a single cycle of Nuclotron operation, which duration is about several seconds, the particle beam is completely absorbed in the target.

	Several operation cycles are required to 	 detect  the effect.
	Preliminary analysis of possibility to 
	observe  the spin dichroism effect at the polarimeter developed in JINR shows that    accelerator operation during  20 to 30 hours enables effect acquisition  \cite{arxiv.2025}.
	%

	The  birefringence phenomena, which is described above and reveals itself as diverse effects, namely: spin  and tensor polarization  rotation around the momentum direction, spin oscillations, vector polarization conversion to tensor that and vice versa, as well as spin dichroism, 
	should be taken into account when conducting precision experiments with either nonpolarized and polarized particle beams, since they lead to changes in the components of vector and tensor polarization of the beam and thus introduce systematic errors into the measurement results.

	Evaluations of deuteron dichroism effect (acquired $\langle p_{zz} \rangle$ value) for carbon and polyethylene (CH$_2$) targets are given in Table~\ref{tab:slide33}. Evaluations are made for two $\Delta \sigma / \sigma$ values: the lowest expected $\Delta \sigma / \sigma =0.01$ and that derived from measurements \cite{azhgirei2010} $\Delta \sigma / \sigma =0.06$.

	\bigskip
	
	\begin{table}[h]
		\setlength{\extrarowheight}{2pt}
		\renewcommand{\arraystretch}{1.6}
		\centering
		\caption{Evaluation of deuteron dichroism  for carbon and polyethylene (CH$_2$) targets}	
		\label{tab:slide33}
		\begin{tabular}{|c|c|c|c|c|c|c|c|c|}
			\hline
			{Particle} & {~~Target~~} & {~~E, MeV~~} & ~~$N_0$~~ & ~~$\sigma$, b~~ & ~~$\Delta \sigma / \sigma$~~ & ~~$\langle p_{zz} \rangle$~~ 
			\\ 		\hline
			d & CH$_2$ & 270 & $10^{10}$ & 0.8 & 0.06 & $1 \cdot 10^{-2}$ 
			\\ 		\hline
			d & CH$_2$ & 270 & $10^{10}$ & 0.8 & 0.01 & $2 \cdot 10^{-3}$ 
			\\ 		\hline
			d & C & 3700 & $10^{10}$ & 0.6 & 0.06 & $1 \cdot 10^{-2}$ 
			\\ 		\hline
			d & C & 270 & $10^{10}$ & 0.6 & 0.06 & $1 \cdot 10^{-2}$ 
			\\ 		\hline
		\end{tabular}
	\end{table}

	
	\subsection{About Possible Influence of Birefringence Effect on the Processes of Production (Photoproduction, Electroproduction) of Vector Mesons  in Nuclei} \label{spinorb_ch:1}

	Collision of high energy particles (proton, electron,
	$\gamma$-quanta, nucleus) with a nucleus yields a lot of hadronic
	processes inside nuclei, which are accompanied by the appearance of
	secondary particles with spin $S\ge 1$ (vector mesons, $\Omega^-$ hyperons, and so
	on).
	In particular, the processes of photoproduction
	(electroproduction) of vector mesons by nuclei have been studied
	because the photoproduction vertex for hadronic probes inside
	nuclei is well known and the analysis of the results is simple and
	more reliable.

	Moreover, experiments demonstrate the production of both
	longitudinally (L) and transversally (T) polarized vector mesons
	and L/T-ratios depend on $Q^2$ \cite{spinorb_dorohov,130_add}.

	According to \cite{spinorb_01}, treating photoproduction of vector
	mesons inside nuclei and their rescattering via strong
	interactions within the framework of the Glauber multiple
	scattering theory allows one to consider many peculiarities of the
	process, which are important for understanding unconventional
	effects, such as color transparency.
	According to the analysis given in \cite{rins_76}, theoretical consideration of
	the processes of rescattering of produced particles inside the nucleus
	must take into account the possible influence of the spin of produced
	particle on rescattering inside the nucleus
	and, therefore, the influence of birefringence effect (spin dichroisn and rotation) on characteristics of the produced vector mesons.
	%

	In particular, as is known from the previous sections, birefringence phenomenon
	appears when a particle with spin $S \ge 1$ passes
	through matter. Specifically, the effect of spin dichroism arises.

	For a particle, produced inside a nucleus and moving through the
	nuclear matter, the conception of the refraction index can be
	applied too \cite{stod66,Goldberger}.

	As a result, the particle produced inside the nucleus undergoes
	refraction described by the spin--dependent index of refraction \cite{rins_76}  (see also (\ref{I.1})):
	\begin{equation}
		\hat{N}=1+\frac{2 \pi \rho(\vec{r})}{k^2} \hat{f} (0)\,,
\nonumber
	\end{equation}
	where $\rho(\vec{r})$ is the density of scatterers in matter (the number of
	scatterers in 1 cm$^3$), $k$ is the particle wave number,
	$\hat{f}(0)$ is used to denote the amplitude of zero-angle elastic
	coherent scattering of a particle by a scattering center, this
	amplitude is an operator acting in the particle spin space.
	
	Refraction of particles in matter (either conventional or nuclear)
	implies the existence of an optical pseudopotential depending on
	particle spin:
	\begin{equation}
		\hat{V}_{\mathrm{eff}}= - \frac{2 \pi \hbar^2\rho(\vec{r})}{m
			\gamma} \hat{f}(0)\,,
		\label{spinorb_Vopt}
	\end{equation}
	where $m$ is the particle mass and $\gamma$ is its Lorentz factor.

	In the case we are concerned with, for example, of production of
	particles with spin 1 (vector mesons), the amplitude of
	zero--angle elastic coherent scattering by nonpolarized nuclear
	nucleons can be expressed in the following general form:
	\begin{equation}
		\hat{f} (0)=d+d_1 (\vec{S} \vec{n})^2\,,
		\label{spinorb_f}
	\end{equation}
	where $\vec{S}$ is the operator of the particle spin,
	$\vec{n}=\frac{\vec{k}}{k}$ is the unit vector along the particle
	momentum.
	The angle of spin rotation is determined by
	$\texttt{Re}~d_1$, while $\texttt{Im}~d_1$ describes dichroism.

	The occurrence of spin dichroism means that spin
	features of a particle produced in nuclear matter (another
	particle with  spin  $\ge 1$) will differ from spin properties
	of the particle produced by a stand--alone nucleon (which does not
	compose the nucleus), which brings about, in particular, the change in the $L/T$-ratio \cite{rins_76}.

	To describe rescattering processes in the energy range, where
	${\texttt{Re} f(0)<< \texttt{Im} f(0)}$, the expressions obtained
	in \cite{spinorb_01} can be used.
	But the total cross-sections of vector meson production given there
	should be replaced by $\sigma_{M=\pm 1}$ or $\sigma_{M=0}$.
	
	In the case when  $\texttt{Re} f(0)$ is comparable with
	$\texttt{Im} f(0)$ or larger, the additional analysis is required.
	
	The detailed analysis of dichroism effect in case of vector meson production by nuclei was recently presented in \cite{Guskov}.
	%
	%

	It should be mentioned that in the general case,  two correlations
	are present in photoproduction. Correlation $[\vec{e}\,^{*}
	\vec{e}] \vec{S}$ (where $\vec{S}$ is the spin operator of a
	produced particle)
	is sensitive to circular polarization of photons and
	the produced particle has vector polarization.
	Correlation $(\vec{e} \vec{S})^2$, which is sensitive to the linear
	polarization of a photons, corresponds to production of particle with
	tensor polarization.
	Due to the birefringence effect, the produced particles are absorbed differently by the
	nucleus.
	Therefore, the yield of vector-mesons  depends on the photon polarization,
	i.e., the production cross-sections are different for different
	polarizations of the incident photons $\sigma_{\mathrm{circ}} \ne \sigma_{\mathrm{lin}}$  \cite{rins_76}.


	\section{"Optical" spin rotation and birefringence of  particles  at NICA}
	\label{sec:NICA}

	%
	%
	%
	

	Recall now that matter (gas, jet target) can be contained in a
	storage ring.
	As has been shown in the previous sections, a particle moving in
	matter undergoes the action of a pseudomagnetic and a
	pseudoelectric fields, which leads to "optical" spin rotation and
	birefringence effects. As a result, the presence of a target
	influences the particle spin motion in a storage ring.
	It would be recalled that the particle refractive index in matter has the form (see also (\ref{I.1})):
	\begin{equation}
		n=1+\frac{2\pi N }{k^{2}}f\left( 0\right)\,,
		\nonumber
		\label{journ_refr_ind}
	\end{equation}
	where $N $ is the number of particles per cm$^{3}$ and $k$ is the
	wave number of the particle incident on the target, $f(0)$ is the
	coherent elastic zero--angle scattering amplitude.

	Let us consider particle refraction on the vacuum--medium boundary
	(see Fig.~{\ref{journ_fig1}).
		
		\begin{figure}[htbp]
			\epsfxsize = 5 cm \centerline{\epsfbox{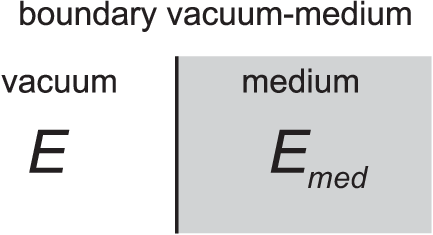}}
			\caption{Particle energy $E$ in vacuum is not equal to particle
				energy $E_{\mathrm{med}}$ in  a medium.} \label{journ_fig1}
		\end{figure}
		The wave number of the particle in the vacuum is denoted $k$.
		The wave number of the particle in the  medium is $k^{\, \prime} =
		k n$.
		As is evident, the particle momentum in the vacuum $p=\hbar k~$ is
		not equal to the particle momentum in the medium.
		Therefore, the particle energy in the vacuum $E=\sqrt{\hbar^2 k^2
			c^2+m^2 c^4}$ is not equal to the particle energy in the medium
		$E_{\mathrm{med}}=\sqrt{\hbar^2 k^2 n^2 c^2+m^2 c^4}$.

		As has been stated above, the energy conservation law immediately
		necessitates the particle in the medium to have an effective
		potential energy $V_{\mathrm{eff}}$. This energy can be easily found  from the relation
		\[
		E={E_{\mathrm{med}}}+V_{\mathrm{eff}}\,,
		\]
		i.e.,
		\begin{equation}
			V_{\mathrm{eff}}=E-E_{\mathrm{med}}=- \frac{2 \pi {\hbar}^2}{m
				\gamma} {N} \, f(E,0) = (2 \pi)^3 N \, \mathcal{T}(E)\,, \label{journ_U1}
		\end{equation}
		where  $\mathcal{T}(E)$ is the $\mathcal{T}$-matrix \cite{Goldberger} and
		\[
		f(E,0)=- (2 \pi)^2 ~\frac{E}{c^2 \hbar^2} ~\mathcal{T}(E)= - (2 \pi)^2
		~\frac{m \gamma }{\hbar^2}~ \mathcal{T}(E)\,.
		\]
		
		\subsection{Mutual refraction of colliding beams	}
		
		Until now we have discussed the rest target.
		But a bunch, which moves in a storage ring, can also be considered
		as a target.
		Therefore, (\ref{journ_refr_ind}) and (\ref{journ_U1}) should be
		generalized for this case \cite{A2,rins_98}.

		Let $E_1$ and $\gamma_1$ denote the energy and Lorentz-factor of
		the particles of the first beam in the rest frame of the storage
		ring and $E_2$ and $\gamma_2$ denote those of the particles of the
		second beam.

		Bearing in mind that the wave phase in the medium is
		Lorentz-invariant one can find it in the following way.
		
		Let the
		second beam rests in the reference frame, then the refraction index
		can be expressed in the conventional form (\ref{journ_refr_ind}):
		\begin{equation}
			n_1^{\, \prime }=1+\frac{2\pi N_2^{\, \prime } }{{k_1^{\, \prime
				}}^{2}}f\left(E_1^{\, \prime}, 0\right)\,, \label{journ_refr_ind3}
		\end{equation}
		where $N_2^{\, \prime }=\gamma_2^{-1} N_2$ is the density of bunch
		2 in its rest frame and $N_2$ is the density of the second bunch
		in the storage ring frame, $k_1^{\, \prime}$ and $E_1^{\, \prime}$
		are the wave number and energy of particles of the first bunch in
		the rest frame of  bunch 2, respectively.
		Let $L$ denotes the length of  bunch 2 in its rest frame, then
		$L=\gamma_2 ~ l$, where $l$ is the length of this bunch in the
		storage ring frame.
		
		The interaction of a particle from  bunch 1 (particle 1) with
		particles of bunch~2 causes a change in the phase of the wave:
		\begin{equation}
			\phi=k_1^{\, \prime}(n_1^{\, \prime}-1)L=\frac{2 \pi N_2^{\,
					\prime}}{k_1^{\, \prime}} f(E_1^{\, \prime},0)~L =\frac{2 \pi
				N_2}{k_1^{\, \prime}}{f(E_1^{\, \prime},0)}{k_1^{\, \prime}}~l\,.
			\label{journ_phase}
		\end{equation}
		It is known \cite{Goldberger} that the ratio
		\[
		\frac{f(E_1^{\,
				\prime},0)}{k_1^{\, \prime}}
		\]
		is invariant, i.e.,
		\[
		\frac{f(E_1^{\,
				\prime},0)}{k_1^{\, \prime}}=\frac{f(E_1,0)}{k_1}\,,
		\]
		where
		$f(E_1,0)$ is the amplitude of elastic coherent forward scattering
		of  particle 1 by moving particles of e bunch 2 in the rest frame
		of the storage ring.
		
		As a result
		\begin{equation}
			\phi=\frac{2 \pi N_2}{k_1} f(E_1,0) \cdot l=\frac{2 \pi N_2}{k_1}
			f(E_1,0) \cdot v_{\mathrm{rel}} \cdot t\,, \label{journ_phase1}
		\end{equation}
		where $v_{\mathrm{rel}}$ is the velocity of the relative motion of
		particle 1 and bunch~ 2 (for opposing motion
		$v_{\mathrm{rel}}=(v_1+v_2)(1+\frac{v_1 v_2}{c^2})^{-1})
		$
		and $t$ is the time of interaction of  particle~ 1 with  bunch~ 2 in the rest
		frame of the storage ring.
		The particle with  velocity
		$
		v_1=\frac{\hbar k_1 c^2}{E_1}
		$
		passes the distance $z=v_1 \cdot t$ over  time~$t$. Note that the
		distance $z$ differs from the length of  bunch 2, because it
		moves. Expression (\ref{journ_phase1}) can be rewritten as:
		\begin{equation}
			\phi=\frac{2 \pi N_2}{k_1}
			f(E_1,0)~\frac{v_{\mathrm{rel}}}{v_1}~z=k_1 (n_1-1)z\,,
			\label{journ_phase_z}
		\end{equation}
		where the index of refraction of  particle 1 by the beam of moving
		particles 2 is:
		\begin{equation}
			n_1=1+\frac{2 \pi
				{N}_2}{{k_1}^2}~\frac{v_{\mathrm{rel}}}{v_1}f(E_1,0)\,.
			\label{journ_r_ind_z}
		\end{equation}
		When $v_2=0$, expression (\ref{journ_r_ind_z}) converts to the conventional
		view (\ref{journ_refr_ind}).

		Thus, the effective potential energy $V_{\mathrm{eff}}$ being
		acquired by  particle 1 when it collides with the particles of
		bunch 2, can be written as follows:
		\begin{eqnarray}
			\label{journ_Veff_new} V_{\mathrm{eff}} & =&
			E_1-E_{1~\mathrm{med}}=E_1-\sqrt{p_1^2 c^2
				n_1^2 + m_1^2 c^4} \nonumber \\
			& =&  -2 \pi \hbar^2 N_2  v_{\mathrm{rel}} \frac{f(E_1,0)}{p_1}=-2
			\pi \hbar^2 N_2 v_{\mathrm{rel}} \frac{f(E_1^{\,
					\prime},0)}{p_1^{\, \prime}}\,.
		\end{eqnarray}
		Therefore,
		\begin{equation}
			V_{\mathrm{eff}}=- \frac{2 \pi \hbar^2 N_2 }{m_1 \gamma_1
				\gamma_2} f(E_1^{\, \prime},0)= (2 \pi)^3 N_2 \mathcal{T} (E_1^{\,
				\prime})\,, \label{journ_Veff_new1}
		\end{equation}
		where $E_1^{\, \prime}=m_1 c^2 \gamma_1 \gamma_2$ is the energy of
		particle 1 in the rest frame of bunch~2, $p_1$ denotes the
		momentum of particle 1 in the storage ring frame, while $p_1^{\,
			\prime}$ is the momentum of particle 1 in the rest frame of
		bunch~2
		$ (p_1^{\, \prime}=\frac{E_1^{\, \prime} v_1^{\,
				\prime}}{c^2})
		$
		and $v_1^{\, \prime}=v_{\mathrm{rel}}$.
		
		When obtaining (\ref{journ_Veff_new}) it was used $|n_1-1| \ll 1$.

		Let us consider now a particle with the nonzero spin.
		In this case the amplitude of the zero-angle scattering depends on
		the particle spin and, as a consequence, the index of refraction
		depends on the particle spin and can be written as:
		
		\begin{equation}
			\hat {n}_1=1+\frac{2 \pi
				{N}_2}{{k_1}^2}~\frac{v_{\mathrm{rel}}}{v_1}\hat {f}(E_1,0)\,,
			\label{journ_hatn1}
		\end{equation}
		where $\hat {f}\left(E_1,{0} \right) = Tr \hat {\rho} _{J}
		\hat {F}\left( {0} \right)$,  the operator of the forward
		scattering amplitude  $\hat {F}\left( {0} \right)$ acts in the
		combined spin space of the particle and scatterer spin, $\hat
		{\rho} _{J} $ is the spin density matrix of the scatterers.

		According to the above,  a particle
		in matter possesses some effective potential energy
		$V_{\mathrm{eff}}$.
		If the amplitude $ \hat {f}(0)$ of particle scattering depends on
		the particle spin, then the effective energy depends on the spin
		orientation:
		\begin{equation}
			\hat{V}_{\mathrm{eff}} =- \frac{2 \pi \hbar^2 N_2 }{m_1 \gamma_1
				\gamma_2} \hat{f}(E_1^{\, \prime},0)\,.  \label{journ_1.1}
		\end{equation}
		Therefore, all the above considered quasi-optical phenomena (spin rotation and spin dichroism) arise at collisions of particles and nuclei beams.
		For example (see section \ref{cosy_ch:1}), scattering amplitude $\hat
		{f}(0)$ of a particle with spin $S=1$ (for example, deuteron) in
		a nonpolarized target depends on the particle spin and can be
		written as:
		\begin{equation}
			\label{journ_hatf} \hat{f}(E_1^{\, \prime},0) = d + d_{1}
			(\hat{\vec{S}} \cdot  \vec{n})^{2}\,,
		\end{equation}
		where $\hat{\vec{S}}$ is the deuteron spin {operator} and
		$\vec{n}$ is the unit vector along the deuteron momentum
		$\vec{k}$.

		Substituting  (\ref{journ_hatf}) to (\ref{journ_1.1}), one can
		obtain for a particle with  spin
		$S=1$:
		\begin{equation}
			\hat{V}_{\mathrm{eff}}=-\frac{2\pi \hbar^{2}}{m_1 \gamma_1
				\gamma_2} N_2 \left(d+d_{1}\left(\hat{\vec{S}} \cdot
			\vec{n}\right)^{2}\right)\,. \label{journ_1.2}
		\end{equation}
		Let the quantization axis z is directed along $\vec{n}$ and $M$
		denotes the magnetic quantum number.
		Then, for a particle in the eigenstate of the operator  of spin
		projection onto the z-axis $\hat{S}_{z}$, the effective potential
		energy can be written as:
		\begin{equation}
			\hat{V}_{\mathrm{eff}}=-\frac{2\pi \hbar^{2}}{m_1 \gamma_1
				\gamma_2}N_2 \left(d+d_{1}M^{2}\right)\,. \label{journ_1.3}
		\end{equation}
		According to (\ref{journ_1.3}), splitting of the deuteron energy levels
		in matter is similar to splitting of the atom energy levels in an
		electric field due to the quadratic Stark effect.
		Therefore,
		the above effect could be considered as caused by splitting of the
		particle spin levels in the pseudoelectric nuclear field of
		matter.
		
		As a result, the
		spin of a particle with $S \ge 1$, which 
		participates in collisions of  counter-propagating beams,
		rotates and oscillates (birefringence effect) (see section
		\ref{cosy_ch:1})
		due to interaction with the  pseudoelectric
		nuclear field of bunches.
		
		\subsection{Evaluations of spin dichroism effect at beam collisions at NICA}

		At collision of two initially nonpolarized deuteron beams at NICA both beams acquire tensor polarization $p_{zz}=-(p_{xx}+p_{yy}) \ne 0$ (see Fig.~\ref{fig:slide30-}). 
		In case if siberian snakes are not available, rotation of polarization occurs at each turn. 
		\begin{figure}[h]
			\epsfxsize = 12 cm \centerline{\epsfbox{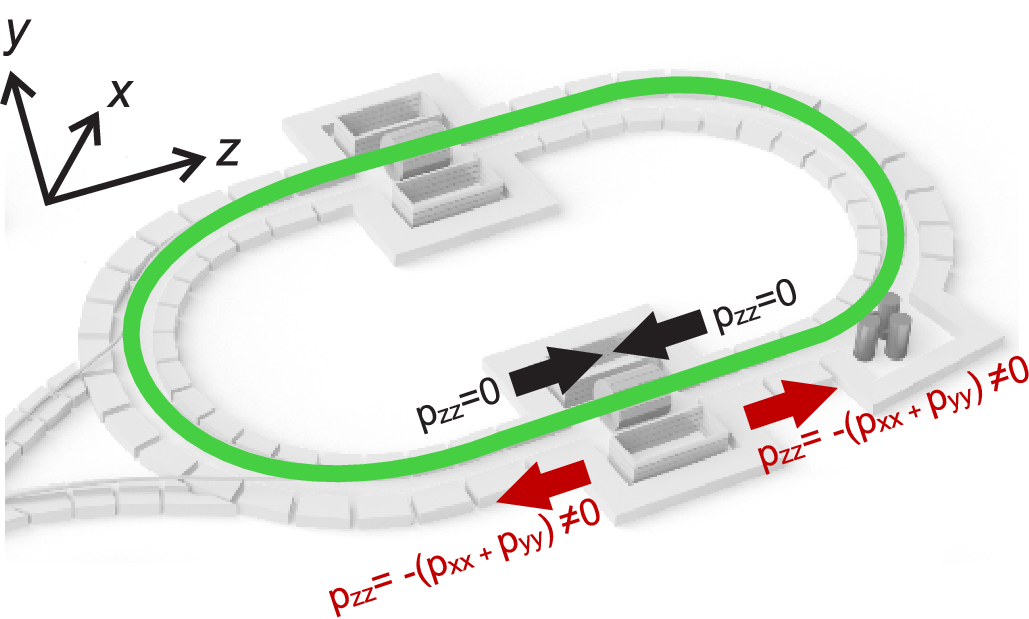}}
			\caption{Spin dichroism effect at beams collisions at NICA without siberian snakes} \label{fig:slide30-}
		\end{figure}

		Let us evaluate the acquired tensor polarization $p_{zz}$ at typical NICA parameters.
		According to \cite{SPD_design_report} the cycle of NICA operation is described as follows:
		''After the beginning of the beam collisions at a maximal luminosity $L_0$, the data taking continues for
		a period $T_1$ that is lower than the luminosity decay time $\tau_L$ (which, in turn, is much longer than
		the polarization decay time $\tau_P$). During this time the RF system of the collider provides an acceptable
		longitudinal size of the colliding bunches. After that, the existing beams are dumped and the accelerator
		complex spends time $T_2$ to accelerate and store a new portion of particles. Assuming $T_1 = $2~hours, 
		$T_2 = 1$~hour and $\tau_L = 6$~hours, the effective luminosity $L_{eff}$ averaged over the cycle is 
		about $0.6 \times L_0$.''
		
		
		Tensor polarization acquired by a deuteron beam during period $T_1$ is given by:
		\begin{equation}
			p_{zz} = \frac{1}{6} \frac{\Delta \sigma}{\sigma} \frac{\sigma L T_1}{N_b},
			\label{eq:pzz}
		\end{equation}
		where $L$ is the luminosity, $N_b$ is the number of deuterons per beam, $\sigma$ is the total cross-section for dd collisions  $\sigma=\frac{2}{3} \sigma_{\pm 1} + \frac{1}{3} \sigma_{0}$ and $\Delta \sigma$ is the difference between scattering cross-sections for spin projections with $m=0$ and $m=\pm 1$ i.e. $\Delta \sigma = \sigma_{0} - \sigma_{\pm 1}$.
		Typical parameters for deuteron beams collision at NICA \cite{SPD_design_report} are used: number of particles per beam  $N_b= 1.3 \cdot 10^{13}$, 
		luminosity $L=4 \cdot 10^{31}$ cm$^{-2} $s$^{-1}$ and data taking period $T_1=2$~hours. Total cross-section of dd collisions is supposed to be as high as $\sigma=0.2$~barn.
		Two values of  $\Delta \sigma/\sigma$ ratio are used for calculation: 0.01 as the lowest expected and 0.06 as reported in \cite{azhgirei2010}.
		Therefore, the acquired tensor polarization $p_{zz}$ is expected to be within interval $~0.75 \cdot 10^{-5}$ and $~0.5 \cdot 10^{-4}$.
		To understand whether the expected $p_{zz}$ value is observable, one should analyze statistics on the number of events during a data-taking period
		(the systematic errors, which also have action on possibility to observe the effect, should be analyzed at real experiment planning). 
		Several operation cycles are required to detect the effects.
		{Analysis presented in 
 \cite{SPD_seminar,school2025_1,school2025_2}
			demonstrate that 
		dozens and hundreds of hours 
			are required to observe the effect - the specific experiment duration depends on particular cross-sections and analyzing powers for the certain energy of colliding beams.}
		
Change in number of beam particles  occurs
		due to elastic and inelastic collisions (reactions) (other loss mechanisms also exist, for example, intra-beam scattering \cite{SPD_design_report}). 
		Let us henceforward study inelastic processes.
		Influence of elastic processes is hardly accountable since part of elastically scattered particles are scattered at a small angle and do not leave the orbit, thus,  making no contribution to change of beam polarization. For all evaluations beam parameters are taken from \cite{SPD_design_report}.

	Suppose $\sigma_r$ to be the sum of cross-sections for all reactions sensitive to  tensor polarization, which value is of the order of the total cross-section ($\sigma_r \sim \sigma$). The analyzing powers are denoted by $A_r$. The number of reactions during period $T_1$  reads as follows:
	\begin{equation}
		\label{eq:Nr}
		N_r = \sigma_r LT_1\,.
	\end{equation}
	Since some deuterons in the beam moving in Nuclotron
	acquire tensor polarization, the actual number of reactions changes with time.

	{The difference in numbers of events $N_{diff}$, which can be observed at $p_{zz}=0$ and $p_{zz} \ne 0$, i.e. the part of events caused by nonzero  $p_{zz}$ reads as follows: }
	\begin{equation}
		N_{diff} \sim A_r N_r p_{zz}= A_r \sigma_r L T_1 p_{zz}.
	\end{equation}
	%

	Statistical error for number of reactions is proportional to the root of  number of reactions as follows:
	\begin{equation}
		\Delta N_r \sim \sqrt{\sigma_r LT_1}.
	\end{equation}
	One could improve signal-to-noise ratio by increasing the observation time $T$, which is composed of periods $T_1$  and $T_2$ repeated as many times $n_c$ as many operation cycles are made:
	\begin{equation}
		T=n_c (T_1+T_2).
	\end{equation} 
	%
	%
	The number of reactions $N_r^T$ during period $T=n_c (T_1+T_2)$ reads as follows:
	\begin{equation}
		N_r^T=\sigma_r LT_1 n_c = \sigma_r LT_1 \frac{T}{(T_1+T_2)}= \sigma_r L \frac{T_1}{(T_1+T_2)}T.
	\end{equation}
	Therefore, 
	\begin{equation}
		N_{diff}^T  \sim A_r  \sigma_r p_{zz} L \frac{T_1}{(T_1+T_2)}T.
	\end{equation}
	Statistical error can be evaluated as
	\begin{equation}
		\Delta	N_{r}^T  \sim  \sqrt {\sigma_r  L \frac{T_1}{(T_1+T_2)}T}.
	\end{equation}
	Signal-to-noise ratio reads as follows:
	\begin{equation}
		\frac{N_{diff}^T}
		{\Delta	N_{r}^T}  = A_r p_{zz} \sqrt {\sigma_r  L \frac{T_1}{(T_1+T_2)}T}.
	\end{equation}
	%
	%
	%
	For example, 1~month observation time at $\frac{\Delta \sigma}{\sigma}=0.01$ gives
	\begin{equation}
		\frac{ N_{diff}}{\Delta N_r} \sim \frac{1  }{2 } A_r \cdot 10^{3}
		\label{eq:221}
	\end{equation}
	and makes even more promises at $\frac{\Delta \sigma}{\sigma}=0.06$:
	\begin{equation}
		\frac{ N_{diff}}{\Delta N_r} \sim  A_r \cdot 10^{2}
		\label{eq:221}
	\end{equation}

	\noindent Assumption that analyzing power is as high as $A_r=0.1$ assures effect measurement, though premise $A_r=0.01$ does not confine this possibility.
	Evaluations of the acquired tensor polarization $p_{zz}$ at dd collisions are given by Table~\ref{tab:slide30--}.

	\begin{table}[h!]
		\setlength{\extrarowheight}{2pt}
		\renewcommand{\arraystretch}{1.6}
		\centering
		\caption{Spin dichroism effect: evaluations for dd collisions at NICA without Siberian snake at luminosity as high as  $L=4 \cdot 10^{31}$~cm$^{-2}$s$^{-1}$) and $T=T_1 =2$~hours} 
		\label{tab:slide30--}
		\begin{tabular}{|c|c|c|c|c|c|c|}
			\hline
			\textbf{Part.} & $\sigma$,b & $\Delta\sigma/\sigma$ & $N_b$ & $N_{\text{diff}}$ & $\Delta N_r$ & {$p_{zz}$} 
			\\
			\hline
			d-d & 0.2 & 0.01 & $1.3 \cdot 10^{13}$ & $2 \cdot 10^{6} A_r$ & $2 \cdot 10^{5}$ & {$0.75 \cdot 10^{-5}$} 
			\\ \hline
			d-d & 0.2 & 0.06 & $1.3 \cdot 10^{13}$ & $10^{7} A_r$          & $2 \cdot 10^{5}$ & {$0.5 \cdot 10^{-4}$} 
			\\
			\hline
		\end{tabular}
	\end{table}

Note that presence of Siberian snakes or operation in the spin transparency (ST) mode  both increases the magnitude of dichroism effect and enables to observe  the effects of coherent spin rotation and conversion of vector  polarization to tensor one and vice versa, as well as spin rotation in pseudomagnetic field in case of polarized beams.


	\section{Conclusion}	
	\label{sec:conclusion}

	Quasi-optical phenomenon of nuclear spin precession of particles (nuclei) in the pseudomagnetic field of matter with polarized spins and the phenomenon of birefringence of particles (nuclei) with spin $S \ge 1$ can be observed at the Nuclotron\,M-NICA complex. 
	Studies of these phenomena allow measurement of the spin-dependent part of the amplitude of elastic coherent forward scattering.  
	The above mentioned phenomena are not caused by strong interactions only. 
	According to sections~\ref{sec:high-energy} and \ref{sec:birefringence1} the T-odd P-odd, T-odd P-even, T-even P-odd interactions also contribute.
	Limits for the value of these contributions at energies available at complex NuclotronM-NICA can be obtained by investigating all these phenomena.
	When studying polarized particles collisions, it is necessary to consider possible influences of quasi-optical phenomena of spin rotation and spin dichroism caused by nuclear precession and birefringence.

In the nearest future it appears feasible to prepare and carry out the following experiments	at the Nuclotron-M NICA complex:
	\newline
	1. Investigation of spin dichroism effect for deuterons at Nuclotron with the internal target.
	\newline
	2. Observation and study of rotation of spin and tensor polarization, as well as conversion of vector polarization to tensor	for deuterons passing through a nonpolarized target with the use of the extracted beam at Nuclotron.
	\newline
	3. Observation of tensor polarization acquiring by heavy nuclei (for example, Ne) when passing through an external target at Nuclotron.
	\newline
	4. Start preparing of such experiments  at NICA.


\section*{Acknowledgements}

\noindent Special thanks to Dr. Sergei Anischenko and Dr. Alexandra Gurinovich for fruitful discussions and assistance in preparing this paper.



\section*{Funding}

\noindent  Research was carried in the framework of fundumental research activities in the Institute of Nuclear Problems of Belarusian State University.

\section*{Conflicts of Interest}

\noindent  The author declare no conflicts of interest.






\end{document}